\documentclass[aps,pre,twocolumn,amsmath,amssymb,superscriptaddress,floatfix]{revtex4}
\usepackage{graphicx,float}
\usepackage{subfigure}
\usepackage{bm}
\usepackage[usenames]{color}
\usepackage{colordvi}
\usepackage{units}
\usepackage{bbm}
\usepackage{enumitem}
\usepackage{calc}
\usepackage{verbatim}
\usepackage{multirow}
\usepackage[utf8]{inputenc}
\usepackage{amsmath}
\usepackage{amssymb}
\usepackage{epsfig}
\usepackage{rotating}

\usepackage{algorithm}

\usepackage{algpseudocode}

\usepackage{amsfonts}
\usepackage{subscript}
\usepackage{placeins}
\usepackage{braket}

\usepackage{float}

\newcommand{\nnn}{\nonumber\\}

\newcommand{\mb}[1]{\mathbf{#1}}

\newcommand{\pd}[1]{\frac{\partial}{\partial #1}}

\newcommand{\kr}[1]{\left({#1}\right)}
\newcommand{\pdou}[2]{\frac{\partial {#1}}{\partial {#2}}}

\newcommand{\thetab}{\vartheta_\text{abso}}

\DeclareMathOperator*{\maxi}{\text{max}}

\newcommand{\hs}[1]{\hspace*{#1mm}} 

\newcommand{\non}{\nonumber}

\newcommand{\skaltime}{\mathbb{T}}

\newcommand{\incgraptw}[2]{\includegraphics[width={#1}\textwidth]{#2}}
\newcommand{\incgrapcw}[2]{\includegraphics[width={#1}\columnwidth]{#2}}
\newcommand{\incgrapw}[2]{\includegraphics[scale={#1}]{#2}}

\begin{document}

\title{Numerical analysis of homogeneous and inhomogeneous intermittent search strategies}

\author{Karsten Schwarz}
\email{kschwarz@lusi.uni-sb.de}
\author{Yannick Schröder}
\email{yannick@lusi.uni-sb.de}
\author{Heiko Rieger}
\email{h.rieger@mx.uni-saarland.de}
\affiliation{Theoretische Physik, Universit\"at des Saarlandes, 66123
Saarbr\"ucken, Germany}
\date{\today}

\begin{abstract}
A random search is a stochastic process representing the random motion of a particle (denoted as the searcher) that is terminated when it reaches (detects) 
a target particle or area the first time. In intermittent search the random motion alternates between two or more motility modes, one of which is non-detecting. 
An example is the slow diffusive motion as the detecting mode and fast, directed ballistic motion as the non-detecting mode, which can lead to much faster detection 
than a purely diffusive search. The transition rate between the diffusive and the ballistic mode (and back) together with the probability distribution of directions 
for the ballistic motion defines a search strategy. If these transition rates and/or probability distributions depend on the spatial coordinates within the search 
domain it is a spatially inhomogeneous search strategy, if both are constant, it is a homogeneous one. Here we study the efficiency, measured in terms of the mean 
first-passage time, of spatially homogeneous and inhomogeneous search strategies for three paradigmatic search problems: 1) the narrow escape problem, 
where the searcher has to find a small area on the boundary of the search domain, 2) reaction kinetics, which involves the detection of an immobile target 
in the interior of a search domain, and 3) the reaction-escape problem, where the searcher first needs to find a diffusive target before it can escape through a 
narrow region on the boundary.  Using families of spatially inhomogeneous search strategies, partially motivated by the spatial organization of the cytoskeleton in living 
cells with a centrosome, we show that they can be made almost always more efficient than homogeneous strategies. 
\end{abstract}


\maketitle

\section{Introduction}
The successful usage of efficient search strategies is one of the most important needs in biology and human behavior. It 
can be observed on all length scales of life and in all kinds of complexity. Just to mention a few examples, humans use them for pattern recognition 
\cite{Najemnik2008}. Predators apply certain strategies for hunting their moving prey \cite{Humphries2010}. Ants use special 
techniques to find each other after being separated while being on a tandem run \cite{Franks2010}. Some eukaryotic cells improve their 
chance to find a target by performing  random walks with characteristic persistent time and persistent lengths, even in the absence of 
external signals \cite{Li2008}. And there are many more observed examples in biological literature. \\ 
Although all these examples are quite different and seem to have nothing in common, they can commonly be described by first-passage processes \cite{Redner2001book},
which are stochastic processes that end if a certain event happens for the first time $t_{f}$. 
The probability density $\rho_f$ for the time $t_f$ contains all temporal information about the efficiency of the search strategy. In 
\cite{Mattos2012} it is shown, that one sometimes has to be careful with the reduction of this information to only one value, 
the so called mean first-passage time (MFPT)
\begin{equation}
T=\braket{\,t_f\,}=\int_0^\infty t\;\rho_f(t)\; dt\;.\nonumber
\end{equation}
Nevertheless, this is in most cases the only property which is used to classify the efficiency of the search strategy. 
Apart from the obvious reason of simplification for comparison, there is a second reason for this reduction: Often it is very hard or even
impossible to calculate the whole first-passage time density function $\rho_f(t)$ as a function of the initial conditions, but it 
is much easier to solve the time-independent differential equation system for its first moment, which is derived with the help of corresponding 
backward equations \cite{Risken1996book,Redner2001book}.\\
The MFPT $T$ is a function of the tunable and the non-tunable parameters of the stochastic first-passage process. Typical tunable parameters are for 
example the persistence length in random walks \cite{Tejedor2012}, the desorption rate in surface mediated diffusion \cite{Benichou2010, Calandre2014} 
or the resetting rate in random motion with stochastic resetting \cite{Evans2011, Kusmierz2014}. Typical non-tunable parameters of the search problem  
are for example the target size, the detection rate, the size and shape of the searching domain and constants of motion (velocity, diffusivity).
A complete set of tunable parameters defines a search strategy for the problem which is defined via the non-tunable parameters.
Hence, the best strategy is the set of tunable parameters which minimizes the MFPT $T$.\\

A frequently used way of modeling real search is a so called intermittent search \cite{Benichou2005, Benichou2005a, Benichou2007,
Benichou2011, Loverdo2008,Bressloff2009, Loverdo2009, Smith2001, Bressloff2012}. The searcher switches between phases of fast directed ballistic 
motion, during which it cannot recognize a target and phases of slow diffusion for detecting a target. \\

For a given size and shape of the search domain and the target, the efficiency, i.e. the MFPT $T$, of an intermittent search still depends on a 
number of parameters. Since increasing the diffusion constant for the diffusive mode or increasing the velocity modulus $v$ for the ballistic mode 
always decreases the MFPT, even if done only locally, both are assumed to be fixed in the following. Then the MFPT $T$ is a function of the switching 
rates between both motility modes and a functional of distribution of the directions into which the searcher moves after a switch to the ballistic mode. 
If the searcher does not have a knowledge about his position in the search domain and the search domain is homogeneous such that at no position in the search 
domain certain directions for ballistic motion are preferred the directional distribution can be assumed to be uniform over all solid angles - as was done 
in \cite{Benichou2005, Benichou2005a, Benichou2007, Benichou2011, Loverdo2008, Loverdo2009}. This we denote as a spatially homogeneous (and isotropic) intermittent 
search strategy. \\
If on the other hand ballistic motion is only possible along predefined tracks, like in molecular motor assisted intracellular transport along the filaments of 
the cytoskeleton \cite{Alberts2014book}, or in cases the searcher utilizes any other transport network, the directional distribution for the ballistic motion should be 
described by a spatially inhomogeneous direction distribution, which then must represent the spatial organization of the tracks. Also in cases when the searcher does have 
knowledge about its position in the search domain and about its shape it might be more efficient to move in certain regions of the search domain preferentially into other 
directions than in other regions. An intermittent search strategy with a spatially varying direction distribution we denote as a spatially inhomogeneous (and non-isotropic) search 
strategy. In a recent letter \cite{schwarz2016} we introduced the concept of spatially inhomogeneous intermittent search strategies and we presented results that showed 
that their optimum is in general more efficient than the optimum of homogeneous search strategies. In this paper we elaborate these and more results in detail, 
explain the computational techniques and show all computations explicitly.\\

Thus the goal of this paper is to compare the efficiency of spatially homogeneous 
and inhomogeneous search strategies in spherical domains by determining, numerically, the optimal parameter for different setups: 1) the narrow escape problem 
, where a searcher has to find a small region on the boundary of the search area, 2) the reaction kinetics enhancement by ballistic motion, where the searcher 
has to find a immobile target particle within the search domain, and 3) the reaction-escape problem, which combines 1 and 2 such that a searcher has to 
find a mobile target particle first before it can escape through a narrow region on the boundary of the search domain. The latter example is motivated by a transport process 
within T-cells attached to a target cell that it is supposed to kill: vesicles loaded with cytotoxic proteins first have to attach to another vesicle containing 
receptor proteins before they can dock at the immunological synapse, a small region on the cell membrane in contact with the target cell, and release their content 
there. \\

Since determining the optimum of the MFPT as a functional of a space and angle dependent direction distribution is not feasible we confine ourselves 
to two different families of direction distributions. The first (one-parameter) family is specially designed for solving the narrow escape problem efficiently 
and only investigated in that scenario. The second (two-parameter) family is inspired by the spatial organization of the cytoskeleton of spherical cells with a centrosome.
It will be studied for all the three scenarios.

In order to compare the gain of efficiency for different situations, 
we introduce the dimensionless time  
\begin{eqnarray}
\skaltime=\frac{T}{T_\text{diff}}\; , \label{def_gain}
\end{eqnarray} 
which is the MFPT $T$ of the intermittent search strategy normalized by the MFPT $T_\text{diff}$ for the purely diffusive searcher.
Hence, for $\skaltime<1$ an intermittent searcher is more efficient and for $\skaltime>1$ a purely diffusive search is faster on average.\\

The paper is organized as follows:
Section \ref{modeldef} introduces our model of intermittent search in the general case with space and time dependent transition rates. It explains the meaning
of the occurring parameters exemplarily in the context of intracellular transport. In almost all cases, it is not possible to solve the differential equation system of the model in a 
straight forward way via finite element method (FEM).\\
In consequence,  section \ref{algo} introduces the Green's function method, which is used 
to solve the model stochastically. \\
Section \ref{narrow_escape} faces the classical narrow escape problem, meaning, a particle looks for a 
certain region at the boundary. For the purely diffusive scenario the scaling of the MFPT as a function of the size and the position of the target area is 
understood for quite a large range of problems 
\cite{Krapivsky1996a,Redner2001book,Benichou2005b,Singer2006a,Singer2006b,Singer2006c,Schuss2007, Schuss2012, Benichou2008,Chevalier2011,Cheviakov2012}. 
Even in the absence of analytic or asymptotic expressions, the purely diffusive MFPT problem can be solved fast and easily via FEM calculations.  
For spatial dimensions $d>1$ this is in most cases not possible for the master equation system of intermittent search 
(Eqs. (\ref{diff})-(\ref{ball})) due to the integro type of the partial differential equation. As far as we know, there are no studies on the intermittent 
search narrow escape problem in a sphere available. Hence, we start the numeric study of this problem in the case of a homogeneous velocity direction distribution. 
Afterwards we modify the velocity direction distribution to show, that there are more efficient strategies than a homogeneous one. \\
Section \ref{target_in_sphere} asks for the best search strategy for a target located within the sphere. In the case of a homogeneously distributed velocity 
direction and a target which is centered in the middle of the sphere, there are studies on this problem \cite{Benichou2011, Loverdo2008, Loverdo2009}. 
We numerically confirm their results, including the very weak dependence of the MFPT on the transition rate $\gamma$ from diffusive to ballistic motion, but disprove their
optimality assumption for $\gamma$. Furthermore,  we study less homogeneous cases, for which there are no MFPT expressions available up to now. \\
Section \ref{pred-prey} finally faces a reaction-escape problem for two particles, i.e an intermittent searching predator-particle is looking 
for a mobile prey-particle. After having found the prey, the particle-complex has to find a small escape area at the boundary. Again, there are already some results 
for purely diffusive predators in different domains \cite{Redner2001book,Krapivsky1996,Redner1999}, but not for intermittent searching ones in a spherical domain. \\
Finally, appendix \ref{Samplingappendix} introduces exact and very fast methods to sample the later defined probability densities of the algorithm of section \ref{algo} .
\section{The model}
\label{modeldef}
Intermittent search is generally based on (at least) two different phases for a searcher \cite{Benichou2011}. On the one hand, there is a searching phase of slow (or none) motion, 
in which the searcher is able to detect a target. On the other hand, there is a relocation phase of directed fast motion without the ability of target detection. Commonly, and also in 
our case, the searching phase is modeled by pure diffusion with diffusivity $D$. The probability density for being in the diffusive state at position $\mb{r}\in V$ at time $t$ will be 
called $P_0(\mb{r},t)$ in the following, where $V\subset\mathbbm{R}^d$ denotes the search volume of the particle. The relocation phase is modeled by straight ballistic motion. 
The probability density for being in the ballistic state at position $\mb{r}\in V$ at time $t$ and moving with velocity $\mb{v}_\Omega=v\cdot\mb{e}_\Omega$ is denoted $P_\Omega(\mb{r},t)$ 
in the following, where $\mb{e}_\Omega$ is the unity vector in direction of the solid angle $\Omega$. 

In intracellular transport vesicles (proteins, organelles) switch between diffusion within the cytosol and almost ballistic motion by molecular motor assisted movement along 
cytoskeleton filaments. The density of these filaments in direction of the solid angle $\Omega$ is generally very inhomogeneous in space: for instance in cells with a centrosome 
microtubules emanate radially from the centrosome towards the cell periphery, where the actin cortex, a thin sheet of actin filaments underneath the cell membrane, provides 
transport in random directions. Sometimes the filament density even varies over time (for
instance during cell polarization). In consequence, the likelihood of a switch between the two phases and the choice of the ballistic direction $\mb{e}_\Omega$ generally depends on the 
position of the searcher. Formally we describe a spatially varying distribution of directions by the density  $\rho_\Omega(\mb{r},t)$. It is proportional to the rate of a switch from diffusive to ballistic
motion in direction $\Omega$ at position $\mb{r}$ at time $t$. In the context of intracellular transport it can be interpreted as the filament density of the cytoskeleton in direction $\Omega$.\\

The master equation system of our model for one searching particle is given by the Fokker-Planck equation system:
\begin{eqnarray}
\frac{\partial}{\partial t} P_0(\mb{r},t)
&=&D\Delta P_0(\mb{r},t)
 -\left[\gamma\,\int d\Omega\,\rho_\Omega(\mb{r},t)\right]\,P_0(\mb{r},t)\nonumber\\ 
 && +\gamma'\int d\Omega\, P_\Omega(\mb{r},t)\label{diff}\\
\frac{\partial}{\partial t} P_\Omega(\mb{r},t) 
&=&-\nabla\cdot ({\bf v}_\Omega P_\Omega(\mb{r},t))
 +\gamma\,\rho_\Omega(\mb{r},t)P_0(\mb{r},t)\nonumber\\
 &&-\gamma' P_\Omega(\mb{r},t)\;,\label{ball}
\end{eqnarray}
where $\gamma$ and $\gamma'$ are transition rates from diffusive to ballistic motion and vice versa. In the context of modeling intracellular transport they are the attachment 
and detachment rates (from cytoskeleton filaments). \\
In consequence, the diffusing searcher experiences a total annihilation rate 
\begin{equation}
k(\mb{r},t)=\gamma\int d\Omega\,\rho_\Omega(\mb{r},t)\;, \label{total_diff_ball_rate}
\end{equation}
with which it is transformed into a
ballistically moving particle with a randomly chosen direction
$\Omega$ (and velocity ${\bf v}_\Omega$) with probability
\begin{equation}
\rho_\mb{v}(\Omega|\mb{r},t)=\frac{\rho_\Omega(\mb{r},t)}{\int\rho_{\Omega'}(\mb{r},t) d\Omega'}\; . \label{rhovomega}
\end{equation}
A ballistically moving particle switches back to diffusive motion with rate $\gamma'$. \\

Within this article, a target shall always be detected immediately, when the diffusive searcher reaches the target area for the first time (in reaction kinetics this means 
reaction upon contact). One could also consider detection or reaction with a finite rate $k_\text{det}$ within the target area \cite{Benichou2011}. But we restrict ourselves 
to the case $k_\text{det}\rightarrow \infty$, i.e. target detection is always modeled via the boundary condition 
\begin{eqnarray}
P_0({\bf r},t)=0\;\quad\forall\;\mb{r}\in A\;, \label{abso_cond}  
\end{eqnarray}
where either $A \subset V$ is the detection area within $V$ or $A \subset \partial V$ is the detection area at the surface of $V$ (narrow escape problem). \\

In section \ref{pred-prey}, we consider the problem of two moving particle, which will react immediately if their distance becomes smaller than a 
certain value. Hence their probability distributions are not independent and the solution does not factorize. Consequently, the master equation 
system depends on $6$ spatial coordinates and $4$ coordinates for $\Omega$. As the exact notation of this master equation system
and the corresponding boundary conditions is very lengthy but straightforward, we will skip it here.\\

Apart from the initial conditions $P_0({\bf r},t=0)=\delta(\bf r -\bf r_0)$, $P_\Omega({\bf r},t)=0$ 
the equation system is augmented by boundary conditions at the boundary $\partial V \setminus A$ for $P_0({\bf r},t)$  and boundary conditions at $\partial V$ for $P_\Omega({\bf r},t)$.
Two different boundary conditions, in the following called BB (Ballistic-Ballistic) and BD (Ballistic-Diffusive), will be used within the studies of this article.
\subsection*{BB boundary condition}
For BB boundary conditions we assume that a ballistically moving particle hitting the boundary is simply reflected and stays in the ballistic mode, 
no matter whether this happens at the target area or not (i.e. the target is not detected then). A particle in the diffusive mode is reflected at 
every point of the boundary $\partial V$ , which does not belong to the target area and stays in the diffusive mode. Fig. \ref{BB_BD_explain}a visualizes the BB condition in a sketch. 
Formally this boundary conditions are described by
\begin{eqnarray}
\frac{\partial}{\partial \bf{n}_{\bf{r}}} P_0({\bf r},t)&=&0  \quad\quad \forall\;\mb{r}\in \partial V \setminus A  \nnn
P_\Omega({\bf r},t) &=&  P_{\Omega_\text{refl}}({\bf r},t)  \quad\quad \forall\;\mb{r}\in \partial V\;, \label{BB} 
\end{eqnarray}
where ${\bf{n}_{\bf{r}}}$ denotes the outward pointing unity vector perpendicular to the boundary at position $\mb{r}$ and $\Omega_\text{refl}$ denotes the solid angle which belongs to
the reflection of $\mb{e}_{\Omega}$ at the surface position $\mb{r}$.
\subsection*{BD boundary condition}
\begin{figure} [!htb]
\includegraphics{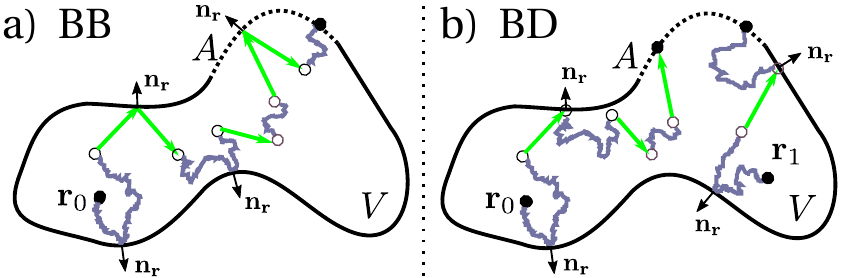} 
\caption{Sketch of boundary conditions in a search volume $V$ with escape area $A$ (dotted line). Grey wiggly lines represent diffusive motion, green lines ballistic motion.  
\textbf{a)} BB: a ballistically moving particle is reflected at the boundary and stays in the ballistic mode, i.e $A$ is only detected if it is reached diffusively. 
\textbf{b)} BD: a ballistically moving particle switches to diffusion at the boundary, i.e $A$ will be detected if it reaches $A$ ballistically 
(trajectory, starting at $\tilde{\bf{r}}_0$) or diffusively (trajectory, starting at $\tilde{\bf{r}}_1$). 
}
\label{BB_BD_explain}
\end{figure}

For BD boundary conditions we assume that a ballistically moving particle hitting the boundary switches to the diffusive motion.  If this part of the boundary 
belongs to the target area, the particle is immediately detected.  A particle in the diffusive mode is reflected at every point of the boundary $\partial V$, 
which does not belong to the target area and stays in the diffusive mode. Fig. \ref{BB_BD_explain}b visualizes the BD condition in a sketch.  Formally this boundary conditions are described by
\begin{eqnarray}
D\frac{\partial}{\partial \bf{n}_{\bf{r}}} P_0({\bf r},t)&=& \int d\Omega\, \kr{{\bf v}_\Omega P_\Omega({\bf r},t)} \cdot \bf{n}_{\bf{r}}\quad \forall\;\mb{r}\in \partial V \setminus A\nnn
P_\Omega({\bf r},t)&=& 0 \quad\quad\forall\;{\bf r}\in\partial V ,\;\Omega \; | \;  {\bf{n}_{\bf{r}}}\cdot{\bf e}_\Omega<0 \label{BD} \;.
\end{eqnarray}

\subsection*{Nondimensionalisation}
In order to reduce the number of parameters to a minimal independent set, characteristic length- and time-scales where 
chosen by introducing the dimensionless spatial and temporal coordinates
\begin{equation}
 \tilde{\mb{r}}=\frac{1}{R}\mb{r}\text{ and } \tilde{t}=\frac{v}{R}t. \label{entdimensionalisierung}
\end{equation}
In consequence, Eqs.  (\ref{diff}) and (\ref{ball}) are always solved in the unit sphere and look the following way:
\begin{eqnarray}
\frac{\partial}{\partial \tilde{t}} P_0
&\hs{-1}=\hs{-1}&\tilde{D}\tilde{\Delta} P_0
 -\tilde{\gamma}\left[ \int\hs{-1}\rho_{\Omega}(\tilde{\mb{r}},t) d\Omega\right] P_0
  +\tilde{\gamma}'\hs{-1}\int\hs{-1} d\Omega\, P_\Omega\hs{5}\label{diff_dimensionless}\\
\frac{\partial}{\partial \tilde{t}} P_\Omega
&\hs{-1}=\hs{-1}&-\mb{e}_{\Omega}\cdot\left(\tilde{\nabla}P_\Omega\right)
 +\tilde{\gamma}\,\rho_{\Omega}(\tilde{\mb{r}},t) P_0
 -\tilde{\gamma}' P_\Omega\;,\label{ball_dimensionless}
\end{eqnarray}
with
\begin{equation}
\tilde{D}=\frac{D}{vR}, \tilde{\gamma}=\frac{R}{v}\gamma \text{ and } \tilde{\gamma}'=\frac{R}{v}\gamma'\;. \label{skaling_of_D_gamma}
\end{equation}
Apart from the sphere radius $R$, the absolute value of the velocity also vanished in the dimensionless coordinates, as $\tilde v=1$ holds. Furthermore,
 $\skaltime$ is not changed by the dimensionless units, i.e. $\skaltime=T/T_\text{diff}=\tilde T/\tilde T_\text{diff}$. 
\subsection*{Models for the direction distribution $\rho_\mb{v}(\Omega|\tilde{\mb{r}},\tilde t)$}
Eq. (\ref{total_diff_ball_rate}) introduced the total transition rate $k(\mb{r},t)$ for a switch from diffusive to ballistic motion 
at position $\mb{r}$ at time $t$. Although it is numerically possible to handle this most general scenario (Algorithm 1 in section \ref{algo}) this rate will 
be constant in time and space in the investigated models, i.e. without further loss of generality we set $\int\rho_{\Omega}(\tilde{\mb{r}}, \tilde t) d\Omega=1$ and in consequence  
Eq. (\ref{rhovomega}) simplifies to  
\begin{equation}
\rho_\mb{v}(\Omega|\tilde{\mb{r}},\tilde t)=\rho_\Omega(\tilde{\mb{r}}, \tilde t)\; .
\end{equation}
Within the studies of this paper, two different families of time-independent inhomogeneous distributions $\rho_\mb{v}(\Omega|\tilde{\mb{r}})$
will be compared to the homogeneous distribution  
\begin{eqnarray}
 \rho_\text{hom}\kr{\Omega}=\frac{1}{4\pi}\; . \label{rho_hom}
\end{eqnarray}

Both will be rotational symmetric, i.e. $\rho_\mb{v}(\Omega|\tilde{\mb{r}})$ depends 
only on the radius $\tilde r=||\tilde{\mb{r}}||$ and the angle  
\begin{eqnarray}
\alpha(\tilde{\mb{r}},\mb{e}_\Omega)=\text{arcos}\left(\frac{\tilde{\mb{r}}\cdot\mb{e}_\Omega}{||\tilde{\mb{r}}||}\right) \label{defalphaangle}
\end{eqnarray}
between the vectors $\tilde{\mb{r}}$ and $\mb{e}_{\Omega}$. \\
This symmetry also holds for the homogeneous case of $\rho_\text{hom}\kr{\Omega}$, where the probability density for the angle $\alpha\in[0;\pi]$  is independent 
of $\tilde{\mb{r}}$ and given by
\begin{eqnarray}
 \rho^\alpha_\text{hom}(\alpha)=\int_0^{2\pi} d\varphi  \; \frac{1}{4\pi} \sin(\alpha)= \frac{1}{2} \sin(\alpha)\; . \label{rho_alpha_hom}
\end{eqnarray}

\subsubsection*{varying Gaussian distribution}
\begin{figure} [htb]
\incgrapw{0.9}{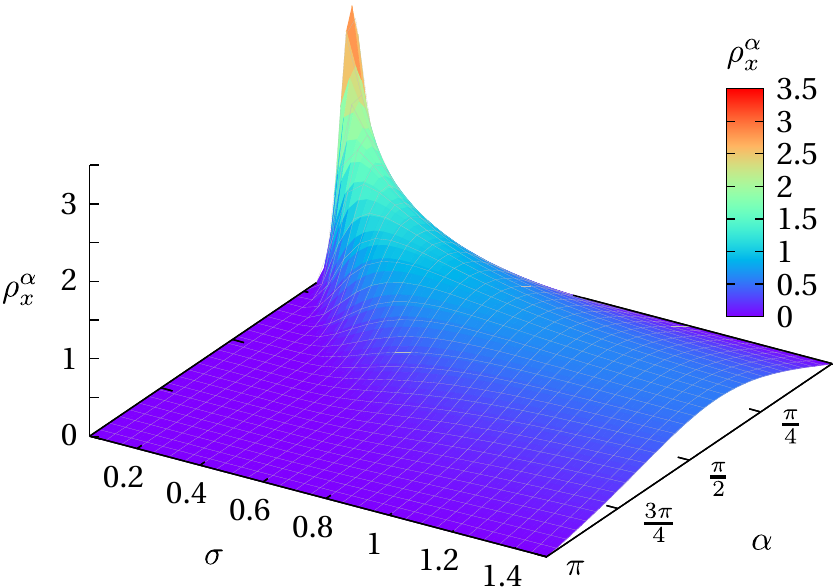} 
\caption{The class of probability densities $\rho^\alpha_x$ (Eq. (\ref{rhoalphainhom})) as a function of the variable $\alpha$ and the spreading parameter $\sigma\kr{x,\tilde r}$.}
\label{rho_alpha_sig}
\end{figure}

The distribution for the angle $\alpha$ (Eq. (\ref{defalphaangle})) , introduced now, will only be applied to the narrow escape problem. 
The principle idea is to find the probability density $\rho^\alpha\kr{\alpha|\tilde{r}}$, which minimizes the MFPT of the narrow escape problem. 
Mathematically, this is a variational problem. In consequence, 
a numeric solution requires an apriori assumption for a class of density functions, which is motivated now:\\
If the particle is close to the center of the simulation sphere, a mainly radially outward pointing velocity direction is
for sure the best strategy, as it is the fastest way to reach the sphere's boundary. At the boundary this distribution is not optimal any more, as there is no 
velocity component in parallel to the boundary. Without this parallel component, the searcher gets stuck at a relatively small part of the boundary. \\
In consequence, the spread of the distribution should increase with $\tilde r$.    
Following this argumentation, the gaussian-like probability density $\rho^\alpha_x\kr{\alpha|\tilde r}$, illustrated in Fig. 
\ref{rho_alpha_sig}, was chosen for our simulations:
\begin{equation}
\rho^\alpha_x\kr{\alpha|\tilde r}\hspace*{-0.05cm}=\hspace*{-0.05cm}2\pi\sin(\alpha)\,N\hspace*{-0.05cm}\big( \hspace*{-0.01cm}\sigma\hspace*{-0.05cm}\kr{x,\tilde r}\hspace*{-0.05cm}\big) \,\exp\hspace*{-0.1cm}\kr{\hspace*{-0.1cm}\frac{-\hspace*{-0.02cm}\big(\hspace*{-0.06cm}\cos(\alpha)\hspace*{-0.06cm}-\hspace*{-0.06cm}1\big)^2}{2\left[\sigma\hspace*{-0.1cm}\kr{x,\tilde r}\right]^2}\hspace*{-0.03cm}}\, ,\label{rhoalphainhom}
\end{equation}
where 
\begin{eqnarray}
\sigma\kr{x,\tilde r}&=&\sqrt{\frac{x}{1-x}} \tilde r \label{defsigmaspread}
\end{eqnarray}
denotes the spreading of the gaussian and
\begin{eqnarray}
N(\sigma)&=&\frac{1}{\pi\sigma\sqrt{2\pi}\text{erf}\kr{\frac{\sqrt{2}}{\sigma}}}
\end{eqnarray}
is the normalization of the distribution. 
The class parameter  $x\in\; ]0;1[$ controls the speed of the increase of the distribution spreading. 
For $x\rightarrow 0^+$, the velocity direction points radially outwards for all $\tilde r\in[0;1]$ as $\sigma\kr{x,\tilde r}$ tends to zero. The 
spread (Eq. (\ref{defsigmaspread})) increases monotonically in $x$ and in $\tilde r$. For $x\rightarrow 1^-$, we are dealing with the totally 
homogeneous velocity direction distribution $\rho^\alpha_\text{hom}(\alpha)$. 

\subsubsection*{radial-peripheral distribution}
\begin{figure}[htb]
\incgrapcw{1}{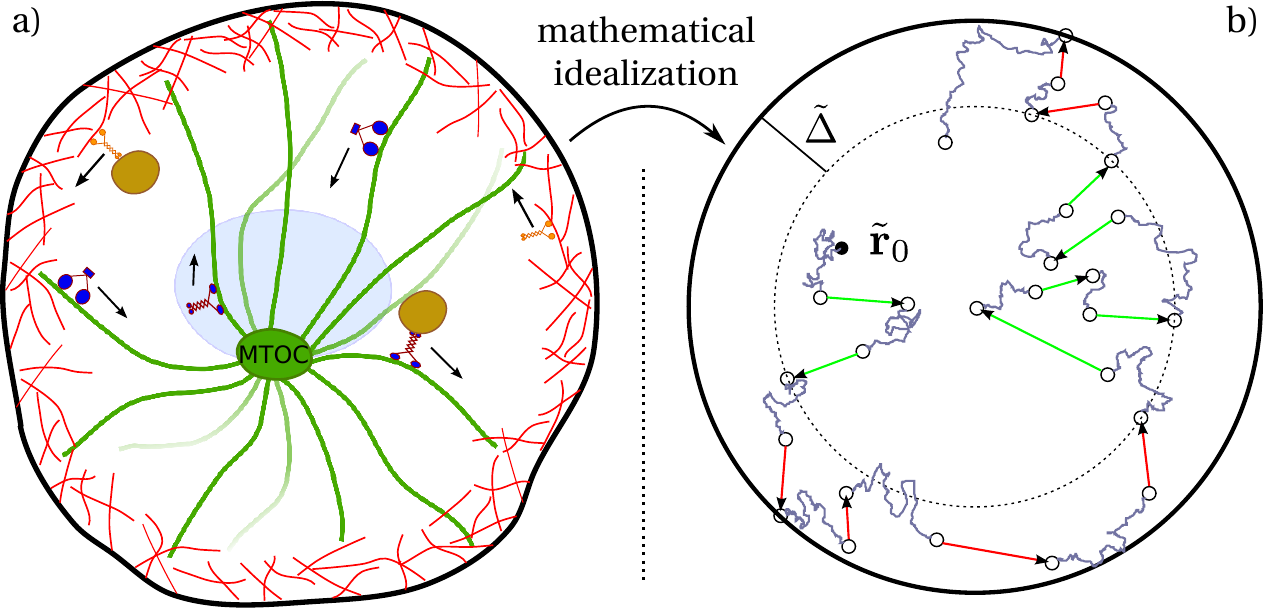} 
\caption{{\bf a)} The cytoskeleton transport network of a spherical cell with a centrosome: Microtubules (green lines), emanating from the
MTOC of the cell close to the nucleus, are orientated predominantly  radially  to  the  cell  membrane. Kinesin and Dynein motor proteins 
transport cargo along them. The actin cortex (red lines) close to membrane is built by isotropically orientated actin filaments. 
Myosin motors transport cargo along. {\bf b)} Sketch of the stochastic process with the direction distribution $\rho_{p,\tilde\Delta}^\alpha$.
$\tilde\Delta$ is the thickness of the outer region. ${\bf{\tilde{r}_0}}$ is the starting point of the particle, Grey wiggly lines represent diffusive motion,
green lines ballistic radial motion (for $|{\bf \tilde{r}}|<1-\tilde{\Delta}$, outward with probability $p$, inward with probability $1-p$), red lines ballistic motion in random directions(for $1-\tilde{\Delta}<|{\bf \tilde{r}}|<1$).
}
\label{sketch_cytoskeleton}
\end{figure}

The second investigated distribution is 
inspired by the spatial organization of the cytoskeleton of spherical cells with a centrosome and was introduced in \cite{schwarz2016},
see Fig. \ref{sketch_cytoskeleton}a for a sketch. It contains two parameters:
\begin{eqnarray}
\rho_{p,\tilde \Delta}^\alpha\hs{-0.5}(\alpha|\tilde r)\hs{-1}=\hs{-1}
\left\{
\begin{array}{ccc}
\hspace{-1mm} p\,\delta(\alpha)\hspace{-0.5mm}+\hspace{-0.5mm}(1\hs{-0.5}-\hs{-0.5}p)\, \delta(\alpha\hs{-0.5}-\hs{-0.5}\pi)&\hs{-2},&0\hs{-0.5}<\hs{-0.5}\tilde r\hs{-0.5}<\hs{-0.5}1\hs{-0.5}-\hs{-0.5}\tilde\Delta\\
\rho^\alpha_\text{hom}(\alpha)&\hs{-2},&1\hs{-0.5}-\hs{-0.5}\tilde\Delta\hs{-0.5}<\hs{-0.5}\tilde r\hs{-0.5}<1 
\end{array}
\right.\hs{-1}.
\end{eqnarray}
The parameter $p\in [0;1]$ is the probability to move radially outwards, and $1-p$ the probability to move inwards inside the inner spherical region with radius
$1-\tilde\Delta$. $\tilde\Delta$ represents the width of the outer shell in which the homogeneous strategy is applied, 
hence $\tilde \Delta=1$ represents the totally homogeneous searching strategy.
A ballistically moving particle switches to the diffusive state when it reaches the radius
$\tilde r=0$ and $\tilde r=1-\tilde\Delta$. The distribution $\rho_{p,\tilde \Delta}^\alpha$ will only be investigated for the boundary condition BD. 
A sketch of the resulting stochastic processes is given in Fig. \ref{sketch_cytoskeleton}b. 
\section{The algorithm}
\label{algo}
Due to the integro type of Eq. (\ref{ball}) and/or the large number of spatial coordinates in two-particle problems, it is not possible to solve the complete Fokker-Planck equation system 
(Eqs. (\ref{diff})-(\ref{ball})) via FEM. Only the purely diffusive case of one particle is always solvable. Hence, the numerical results of this article were mostly derived with Monte Carlo 
techniques, which will be explained in this section. \\     

Green’s function reaction dynamics \cite{Zon2005a,Zon2005b} and first-passage kinetic Monte Carlo methods \cite{oppelstrup2006,oppelstrup2009,donev2010} 
are currently the most powerful tools for simulating diluted reaction-diffusion processes. In contrast to the traditional way of simulating diffusion by 
an enormous number of very small (compared to the system size) random hops, they propagate diffusing particles randomly within so called protective domains over rather long distances. 
The core of these methods are Green's functions, the solution of the initial value diffusion problem within the protective domains. In essence, these methods
work the following way:\\
For a given starting configuration of $N$ interacting diffusing particles within a domain $V$, a protective domain $G_i\subset V$ is assigned to each particle $i$ 
with $G_i\cap G_j=\O{}$ for $i\neq j$. A necessary restriction for the choice of each domain is the knowledge of an analytic expression for the Green's function 
for the initial value diffusion problem according to absorbing boundary conditions at the interior of $V$ and the boundary conditions of $V$ at common boundaries 
of $V$ and $G_i$ (as far as they exist). Based on these Green's functions it is possible to sample for the particle $i$ which will leave its domain first and a 
corresponding time $\tau_i$ 
for this first-passage event. Finally, the exit position $\mb{r}_i\in \partial G_i$ is sampled depending on 
$\tau_i$. If the distance of $\mb{r}_i$ to the protective domains of all other particles is larger than a given threshold, we look for a new protective domain 
for the particle $i$. Otherwise, we have to sample new positions for all particles, whose protection domains are too close to $\mb{r}_i$. In the end, a new 
protective domain has to be assigned to all these particles.\\
In \cite{schwarz2013} we developed an improvement of these routines for a wider range of applications including external space and time depending transition rates. \\    
For a more detailed general explanation of these methods and for proofs of their correctness, the reader is referred to the original articles 
\cite{Zon2005a,Zon2005b,oppelstrup2006,oppelstrup2009,donev2010,schwarz2013}. The rest of this methodical chapter only focuses on describing the concrete 
algorithm for particles in a sphere, switching between ballistic and diffusive motion according to the model definition of section \ref{modeldef}. The method will be explained
in the most general context of spatially and temporally varying rates, see Eqs. (\ref{diff} -\ref{ball}).
\subsection{non-interacting particles in a sphere of radius $R$}
\begin{figure} [!htb]
\includegraphics{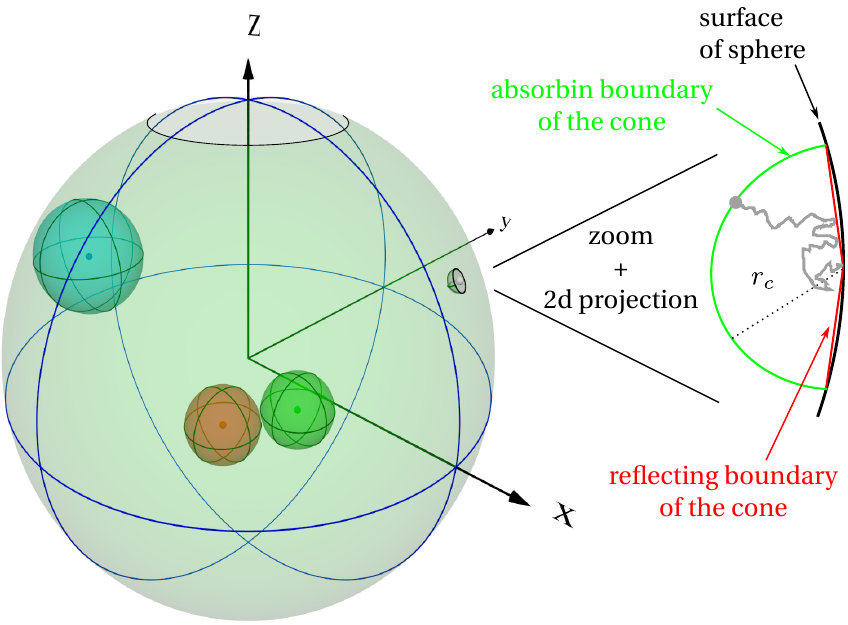} 
\caption{Illustration for the choices of protecting spheres/cones: The blue particle on the left is closer to the boundary of the simulation sphere than to any other
particle, in consequence the radius of its protecting sphere is limited by the distance to the boundary. The green and the red particle in the middle are closer to each 
other than to the cell boundary. If they can react, their radius is limited by their distance. The gray particle has reached the boundary. Hence it is propagated in a cone,
which is illustrated in the 2d projection.}
\label{Illustration_of_protection_boxes}
\end{figure}

Algorithmically, the case of several non-interacting particles is identical to the case of only one particle. Consequently, we restrict the following algorithm description to only 
one particle.\\ 
For a diffusive particle being at position $\mb{r}_0$ at time $t_0$ we use two different types of domains for propagating the particle within the simulation sphere of radius $R$.
If the distance $d_{\partial V}=R-||\mb{r}_0||$ to the boundary of the sphere is larger than a very small threshold value $\epsilon_R$, a sphere $G$ with radius $R_{pro}=d_{\partial V}$, 
centered around $\mb{r}_0$, will be assigned to the particle. An example for such a situation is the blue particle in Fig. \ref{Illustration_of_protection_boxes}. 
Based on the solution of the diffusive initial value problem in $G$ (appendix Eq. (\ref{iniprobsphere})) it is possible to generate stochastically a first-passage 
time $t_b=t_0+\tau_b$ to the boundary of $G$, where $\tau_b$ is sampled according to the corresponding first-passage time probability $\rho_b$ 
(appendix Eq. (\ref{defrho_b})). If the particle does not switch to ballistic movement before time $t_b$, 
a random position update to the boundary of the sphere $G$ is done and a new protective domain must be assigned to the particle afterwards. 
Otherwise a new particle position within the sphere $G$ is sampled by using the radial probability density $\rho_n$ (appendix Eq. (\ref{defrhor}))

Due to the fact, that the boundaries of 
the sphere $V$ and the protecting sphere $G$ have always only one point in common, it does not work to use only spheres for the protecting domains. With probability one, the particle
will touch the boundary of the simulation sphere, i.e. we would end up with an infinite sequence of protecting spheres, whose radii converge to zero. The best possibility 
to overcome this problem would be the usage of protection domains, whose boundaries coincide locally with the boundary of the simulation sphere in an area and not just in 
one point. Due to the missing knowledge of corresponding Green's functions and/or the ability to sample efficiently within these domains, this is not possible. In consequence, 
for $d_{\partial V}\leq\epsilon_R$, we locally approximate the boundary of the simulation sphere by a suitable geometry, which is a spherical cone with a reflecting conical and an 
absorbing spherical boundary  (appendix Eq. (\ref{iniprobcone})). An example for such a situation is the gray particle in Fig. \ref{Illustration_of_protection_boxes}.    
If the distance to the boundary is larger than $\epsilon_R$ after being propagated within the cone, we again go on with a protecting sphere, otherwise, we use again a cone.   
The accuracy of this method is tunable via the two geometry boundary approximation parameters $\epsilon_R$ and the maximum radius $R_{pro}=r_c$ of the protecting cone.
It is important to mention, that $r_c$ is just an upper limit for the cone's radius. If the center position $\mb{x}_c\in\,\partial V \setminus A $ of the cone is closer 
than $r_c$ to the target area $A$, $R_{pro}$ is chosen to be the minimal distance of $\mb{x}_c$ to $A$. \\

\begin{figure} [tb] 
\incgrapw{0.9}{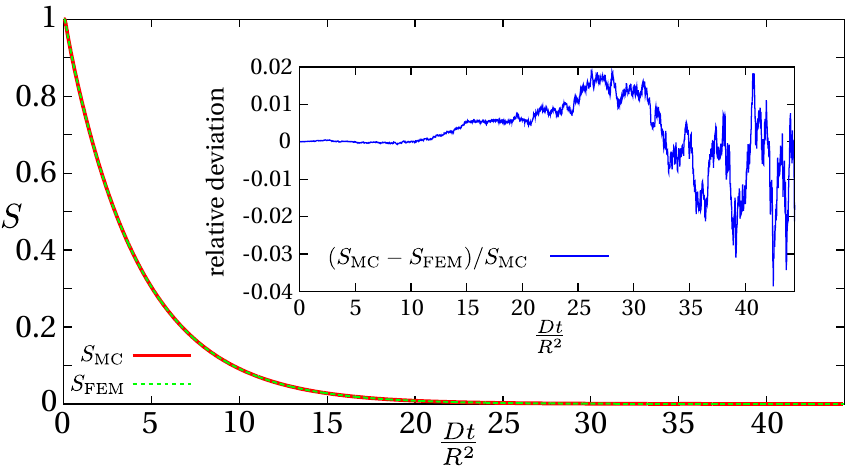} 
\caption{Comparison of the Monte Carlo method and the solution of a FEM solver on the basis of the survival probability $S(t)$ for a purely diffusive narrow escape process with 
$\vartheta_{abso}=\pi/12$ and $r_0=0$. The inset shows the very small relative deviation of these distributions.} 
\label{exact_check}
\end{figure}

In order to demonstrate the high accuracy, we compared our Monte Carlo method with the solution of a commercial FEM 
solver for a purely diffusive 
narrow escape problem. Starting at $r_0=0$, the searcher has the find the escape area with polar angle $\vartheta_{abso}=\pi/12$ 
(bright area at the top in Fig. \ref{Illustration_of_protection_boxes}). The FEM simulation was done on a very fine 
triangulation ($\approx$ 200000 elements) using the rotation symmetry of the problem and yields the expectation value $EW_{FEM}=4.1972R^2/D$ for the needed
search time. The Monte Carlo simulation with $10^7$ samples was done for the geometry approximating parameters $\epsilon_R=10^{-4}\,R$, $r_c=0.04\,R$ and
yields the almost perfectly matching value $EW_{MC}=4.1984R^2/D$. A much stronger criterion than the comparison of expectation values is the equality of the 
survival probability $S(t)$ (probability of not having reached the escape area until $t$) for all $t>0$. The again almost perfectly matching result is shown 
in Fig. \ref{exact_check}. \\       
All numerical results of this article are expected to be in the same numerical exactness (expect the sampling deviation in the case of a smaller number of samples), 
as the values of $\epsilon_R$ and $r_c$ where always chosen to be on the save side according to the smallest occurring length scale. However, the accuracy was successfully checked
wherever this was possible (either by analytic values or FEM values).\\

For a diffusive particle at position $\mb{r}$, the total rate for a switch to a ballistic movement in an arbitrary direction $\Omega$ is given by $k(\mb{r},t)$ (see 
Eq. (\ref{total_diff_ball_rate})). If this rate is spatially inhomogeneous, the methods of \cite{Zon2005a,Zon2005b,oppelstrup2006,oppelstrup2009,donev2010} will 
fail, as there is in general no analytic solution (Green's function) to the diffusion-annihilation equation available. The algorithm, presented in \cite{schwarz2013},
overcomes this problem by using a spatially maximal rate      
\begin{equation}
k_m(t)= {\rm max}_{\mb{r}\in V}\{k(\mb{r},t)\}\;  \label{def_k_m}
\end{equation}
in order to sample a candidate time $t_\text{cand}$ for a switch from diffusive to ballistic motion according to the probability density 
\begin{equation}
\rho_m(t|t_0)=-\frac{d}{dt} \left[ e^{-\int_{t_0}^t \label{def_rho_m}
k_m(t')dt'}\right]\; .
\end{equation}
A new position $\mb{r}$ is assigned to the particle with the help of $\rho_n(\cdot|t_\text{cand}-t_0)$ (appendix Eq. (\ref{defrhor})). 
With probability $1-k(\mb{r},t_\text{cand})/k_m(t_\text{cand})$ the particle moves on diffusively. With probability $k(\mb{r},t_\text{cand})/k_m(t_\text{cand})$
it switches to ballistic motion with velocity $\mb{v}$, sampled according to the probability density $\rho_\mb{v}(\Omega|\mb{r},t)$ (see 
Eq. (\ref{rhovomega})). \\
For a back switch to diffusive motion only a corresponding time must be sampled, as there is a one-to-one relation between time and space in the case of ballistic motion. \\
For a better understanding, the pseudo-code details are shown in Algorithm \ref{Algo1}, exemplarily for the BD boundary condition. \\

\begin{algorithm}
\caption{one particle}
\begin{algorithmic}[1]
\State{{\bf{Input}}: $\mathbf{r}_0\in\mathbbm{R}^3$}
\State{\bf{Output}: $\mb{r}, t$}
\State $t\gets 0$;
\State $t_{\text{cand}}\gets 0$;
\State $diffusive \gets true$;
\Repeat
\If{( $diffusive$ )}
 \If {($t_{\text{cand}}\leq t$)}
 \State $t_{\text{cand}}\gets$ random number according to $\rho_{m}(\cdot|t)$;
 \EndIf
 \State Choose the protecting sphere/cone $P$ with \hspace*{1.1cm} maximal radius $R_{pro}$ as a function of $\mathbf{r}$;
 \State $t_b\gets$ $t$ + random number according to $\rho_b(\cdot)$ \hspace*{1.1cm} for $R_{pro}$;
 \If {($t_b<t_{\text{cand}}$)}
 \State $\mathbf{r}\gets$ rand. position at absorbing part of $\partial P$;  
 \State $t\gets t_b$;
 \Else
 \State $\mathbf{r}\gets$ rand. position update within $P$ according \hspace*{2cm} to $\rho_n(\cdot|t_{\text{cand}}-t)$;  
 \State $t\gets t_{\text{cand}}$;
 \If {$\left(k(\mb{r},t)/k_m(t) \geq ran[0;1]\right)$}
 \State $diffusive \gets false$;
 \State $\mb{v}\gets$ random velocity according to $\rho_\mb{v}(\cdot|\mb{r},t)$;
 \EndIf
 \EndIf
\Else
\State $t_b\gets$ time when ballistic particle hits boundary;  \hspace*{2cm}  $(\;||\mb{r}+(t_b-t)\cdot\mb{v}||=R\,,\,\,t_b>t\; )$
\State $t_{\text{cand}}\gets $ random exponentially distributed number  \hspace*{2cm} with rate $\gamma'$;
\If {($t_b<t_{\text{cand}}$)}
 \State $\mathbf{r}\gets \mb{r}+(t_b-t)\cdot\mb{v}$;
 \State $t\gets t_b$;
 \Else
 \State $\mathbf{r}\gets \mb{r}+(t_\text{cand}-t)\cdot\mb{v}$;  
 \State $t\gets t_{\text{cand}}$;
 \EndIf
 \State $diffusive \gets true$;
\EndIf
\Until{ (distance to absorbing part of sphere $<$ threshold)}
\State\Return $(\mb{r},t)$
\end{algorithmic}
\label{Algo1}
\end{algorithm}

\subsection{interacting particle in a sphere of Radius $R$}
If there are at least two particles in the simulation sphere, which are able to react, the choice of the protection 
boxes does not only depend on the position of the particle, but also on the distance between these reacting particles. 
In general, protecting spheres/cones of reacting particles are not allowed to be closer to each other than the interaction 
distance $d_i$.
An example for such a situation is the red and green particle 
in the middle of Fig. \ref{Illustration_of_protection_boxes}. A similar problem as the boundary approximating problem in the 
subsection before has to be solved here. If we choose the protecting spheres/cones of interacting particles always in a way, 
that the boundaries of these protection boxes have their minimal distance in only one point, we will for sure end up in an 
infinite sequence of protection spheres/cones, whose radii tend to zero. In general there are two ways to overcome this problem. 
The first one is discussed in \cite{Zon2005b} and the problem is solved via a coordinate transformation for the two particle 
positions to the difference vector and the mass point vector. As the problem factorizes in these coordinates, one ends
up with two independent problems. Although a position update takes more time in these situations due to the fact, that 
radial symmetry is lost within the protection boxes in these coordinates, this is a very powerful tool for particles, 
which are far away (compared to their distance) from the boundary of the simulation sphere. But for particles, whose 
distance to the boundary is only a little bit larger than their distance to each other, this method does not work well. 
Hence, we decided to use a second tunable approximation by defining a parameter $\epsilon_{dist}$: If the distance between 
two reactive particles is less than $d_i+\epsilon_{dist}$ these particles react. If we choose  $\epsilon_{dist}=10^{-4}\cdot d_i$, 
we are numerically for sure on the safe side, as all results look totally the same as in the case $\epsilon_{dist}=5\cdot10^{-4}\cdot d_i$.
A comparison to the solution of a FEM solver is not possible anymore, even for purely diffusive particles, due to the high spatial 
dimension ($2\cdot3=6$) of the problem. A pseudo-code description would be quite large and the general idea is the same as 
in Algorithm 1. The interested reader is again referred to \cite{Zon2005a,Zon2005b,oppelstrup2006,oppelstrup2009,donev2010,schwarz2013}.  
\section{Narrow escape problem}
\label{narrow_escape}
The narrow escape problem for a purely diffusive particle in a sphere (and other simple domains) has already been studied in several publications. 
A nice overview, containing analytic asymptotic expressions, is given in \cite{Cheviakov2012} and \cite{Schuss2012}.  
Within this section, we consider the problem of a particle, which moves according to an intermittent search strategy, meaning, the master-equation 
system of its movement is given by the Eqs. (\ref{diff_dimensionless}) and (\ref{ball_dimensionless}) until it reaches the absorbing part of the boundary of the simulation sphere 
for the first time. This escape area is given by a spherical cab with polar angle $\vartheta_\text{abso}$, like it is shown at the north pole of 
Fig. \ref{Illustration_of_protection_boxes}. The position of this cap is of course not known by the particle. \\
The MFPT $\tilde T$ to the absorbing cap is a function of $\vartheta_\text{abso},\tilde D, \tilde \gamma, \tilde\gamma'$ and the velocity direction distribution $\rho_\mb{v}(\Omega|\tilde r)$.
Furthermore, it depends on the initial position of the diffusively starting particle. But in the case of small target areas, the relative influence of the initial position totally 
vanishes. Depending on the diffusivity $\tilde D$ and $\vartheta_\text{abso}$, we study the optimal solution to the escape problem, i.e. we always look for the transition 
rates $(\tilde{\gamma},\tilde{\gamma}')$ which minimize $\skaltime$ (and simultaneously $\tilde T$).\\
\subsubsection*{purely diffusive search}
\begin{figure} [htb]
\incgrapw{0.9}{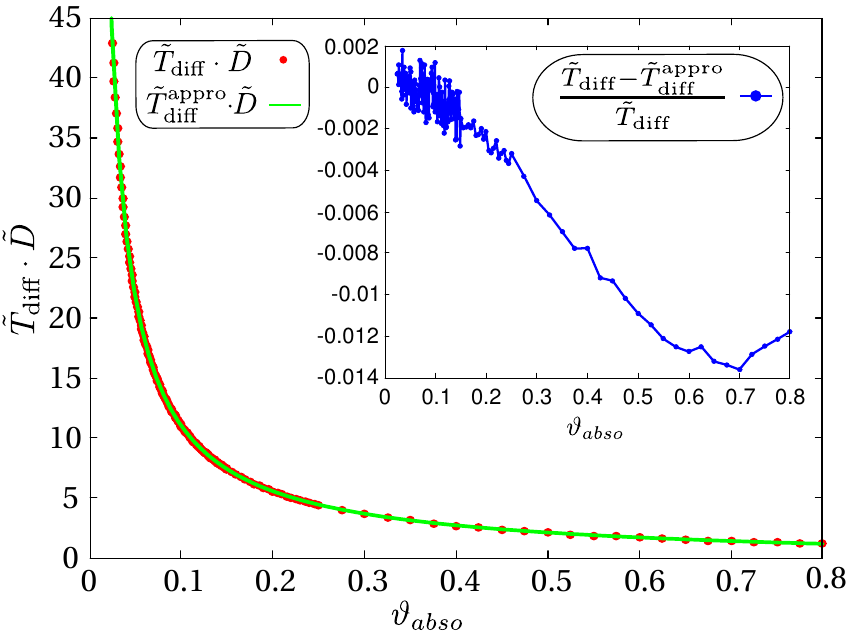}
\caption{$\tilde T_\text{diff}\cdot\tilde D$ and $\tilde T_\text{diff}^\text{appro}\cdot\tilde D$ as a function of $\vartheta_{abso}$ for $\tilde{\mb{r}}_0=0$: 
Each red dot is the average of $2\cdot 10^6 \;-\; 10^7$ Monte Carlo samples. It coincides very well with the analytic 
approximation of \cite{Cheviakov2012} (green line), given in Eq. (\ref{approx_purely_diff_narrow_escape}). The inset shows the 
relative difference between the curves.} 
\label{MFPT_diff_general_plot}
\end{figure}

The reference time $\tilde T_\text{diff}$ for a purely diffusive searcher ($\tilde\gamma'\rightarrow \infty$ and/or $\tilde\gamma=0$) is inversely proportional to the diffusivity $\tilde D$. 
Among others, \cite{Cheviakov2012} has derived a very exact analytic approximation $\tilde T_\text{diff}^\text{appro}(\vartheta_{abso})$  of the problem for small $\vartheta_{abso}$ for arbitrary starting positions 
$\tilde{\mb{r}}_0$. For $\tilde{\mb{r}}_0=0$,
\begin{eqnarray}
\tilde T_\text{diff}^\text{appro}(\vartheta_{abso})=\frac{1}{\tilde D}\kr{\frac{\pi}{3\vartheta_{abso}}-\frac{1}{3}\text{ln}(2\vartheta_{abso})} \label{approx_purely_diff_narrow_escape}
\end{eqnarray}
holds. $\tilde T_\text{diff}$ has been calculated via $2\cdot 10^6$ $(\vartheta_{abso}\leq 0.15)$ - $10^7$ $(\vartheta_{abso}> 0.15)$ Monte Carlo samples for each $\vartheta_{abso}$ and compared to the analytic 
approximation in Eq. (\ref{approx_purely_diff_narrow_escape}). The result is shown in Fig. \ref{MFPT_diff_general_plot}.
For small values of $\vartheta_{abso}$ the relative deviation between the simulated value of $\tilde T_\text{diff}(\vartheta_{abso})$ 
and $\tilde T_\text{diff}^\text{appro}(\vartheta_{abso})$ is extremely small and only based on stochastic fluctuations (inset of Fig. \ref{MFPT_diff_general_plot} ). For larger 
values of  $\vartheta_{abso}$ it slightly increases, which is not based on a drop of exactness in our numerical routines, 
but on the fact that the approximation $\tilde T_\text{diff}^\text{appro}$ becomes worse for larger opening angles.
If the initial position $\tilde{\mb{r}}_0$ of the particle is equally distributed within the sphere, $\tilde T_\text{diff}$ and $\tilde T_\text{diff}^\text{appro}$ exactly decrease by $1/(10\tilde D)$ for all $\vartheta_{abso}$, which  
has also been checked numerically. \\

 \subsubsection*{random velocity model}

  Before studying intermittent strategies, it is insightful to have a look at the
  opposite choice of transitions rates $\tilde\gamma$ and $\tilde\gamma'$, which is a random velocity model, given 
  by the limit $\tilde\gamma\rightarrow \infty$ and $\tilde\gamma'=0$. \\
  For the BB boundary condition the corresponding MFPT $\tilde T_\text{v}$  tends trivially to infinity for all $\vartheta_{abso}\in[0;\pi]$,
  as the ballistically moving particle is reflected at the boundary without target area detection and never switches to diffusive mode (see Fig. \ref{BB_BD_explain}b). 
  
\begin{figure} [htb]
   \incgrapw{0.9}{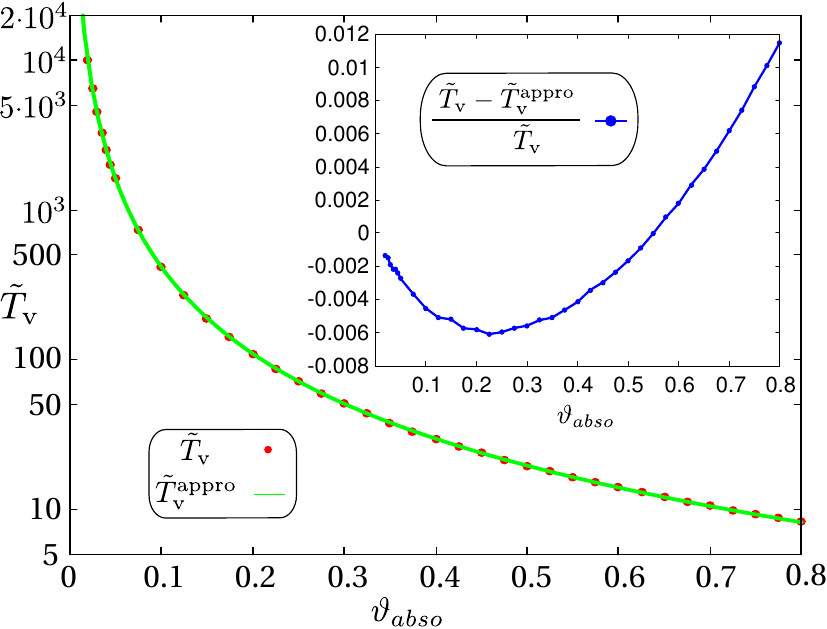} 
   \caption{$\tilde T_\text{v}$ and $\tilde T_\text{v}^\text{appro}$ as a function of $\vartheta_{abso}$ for $\tilde{\mb{r}}_0=0$: Each red dot is the average of $10^8$ 
  Monte Carlo samples. It coincides very well with the analytic approximation $\tilde T_\text{v}^\text{appro}$ (green line), given in Eq. (\ref{hatqv}). The inset shows the 
  relative difference between the curves.} 
   \label{T_v_and_q_v}
  \end{figure}

  For the BD boundary condition, this is not the case. The resulting random velocity model
  is given by a ballistically moving particle, which detects the escape area at the boundary when reaching it and randomly chooses a new direction for the ballistic motion when reaching 
  a part of the sphere's boundary which does not belong to the target area.  
  In consequence, the corresponding MFPT $\tilde T_\text{v}$ depends on the velocity direction distribution $\rho_\mb{v}(\Omega|\tilde{\mb{r}})$, the opening 
  angle $\vartheta_{abso}$ and slightly on the initial position $\tilde{\mb{r}}_0$. For $\tilde{\mb{r}}_0=0$ and the case of a homogeneous velocity direction density (Eq. \ref{rho_hom}) 
  we derived an approximating expression $\tilde T_\text{v}^\text{appro}(\vartheta_{abso})$  
  for $\vartheta_{abso}\in [0;\pi/2]$:
  \begin{eqnarray}
  \tilde T_\text{v}^\text{appro}(\vartheta_{abso})= 1+ \frac{1+\cos(\vartheta_{abso})}{2}\cdot \Psi(\vartheta_{abso}) \label{hatqv}
  \end{eqnarray}
  \vspace*{-0.5cm}
  \begin{eqnarray}
  \text{with }\;\; \Psi(x)=\frac{1+\cos(x)}{1-\cos(x)}\times\hspace*{1.5cm}\nnn
  \frac{1}{\frac{\cos(x)-1}{\sin(x)}\hs{-0.6}+\hs{-0.6}x\sin(x)\hs{-0.6}+\hs{-0.6}4\hs{-0.4}\cos\hs{-0.6}\kr{\frac{x}{2}}\hs{-0.6}-\hs{-0.6}3\hs{-0.6}-\hs{-0.6}4\hs{-0.4}\cos(x)\hs{-0.4}\ln\hs{-0.8}\kr{\hs{-0.8}\frac{\cos(\frac{x}{4})}{\cos(\frac{x}{2})}\hs{-0.8}}  }\; . \non
  \end{eqnarray}
  Fig. \ref{T_v_and_q_v} shows   $\tilde T_\text{v}$ ($10^8$ samples) and $\tilde T_\text{v}^\text{appro}$ in a logscale plot.
  The relative deviation of $\tilde T_\text{v}$ and $\tilde T_\text{v}^\text{appro}$ vanishes for $\vartheta_{abso}\rightarrow 0$, which is shown in the inset. 
  If the initial position $\tilde{\mb{r}}_0$ of the particle is equally distributed within the sphere, $\tilde T_\text{v}$ and $\tilde T_\text{v}^\text{appro}$ exactly 
  decrease by $1/4$ for all $\vartheta_{abso}$.\\

\begin{figure*}
\incgrapw{0.9}{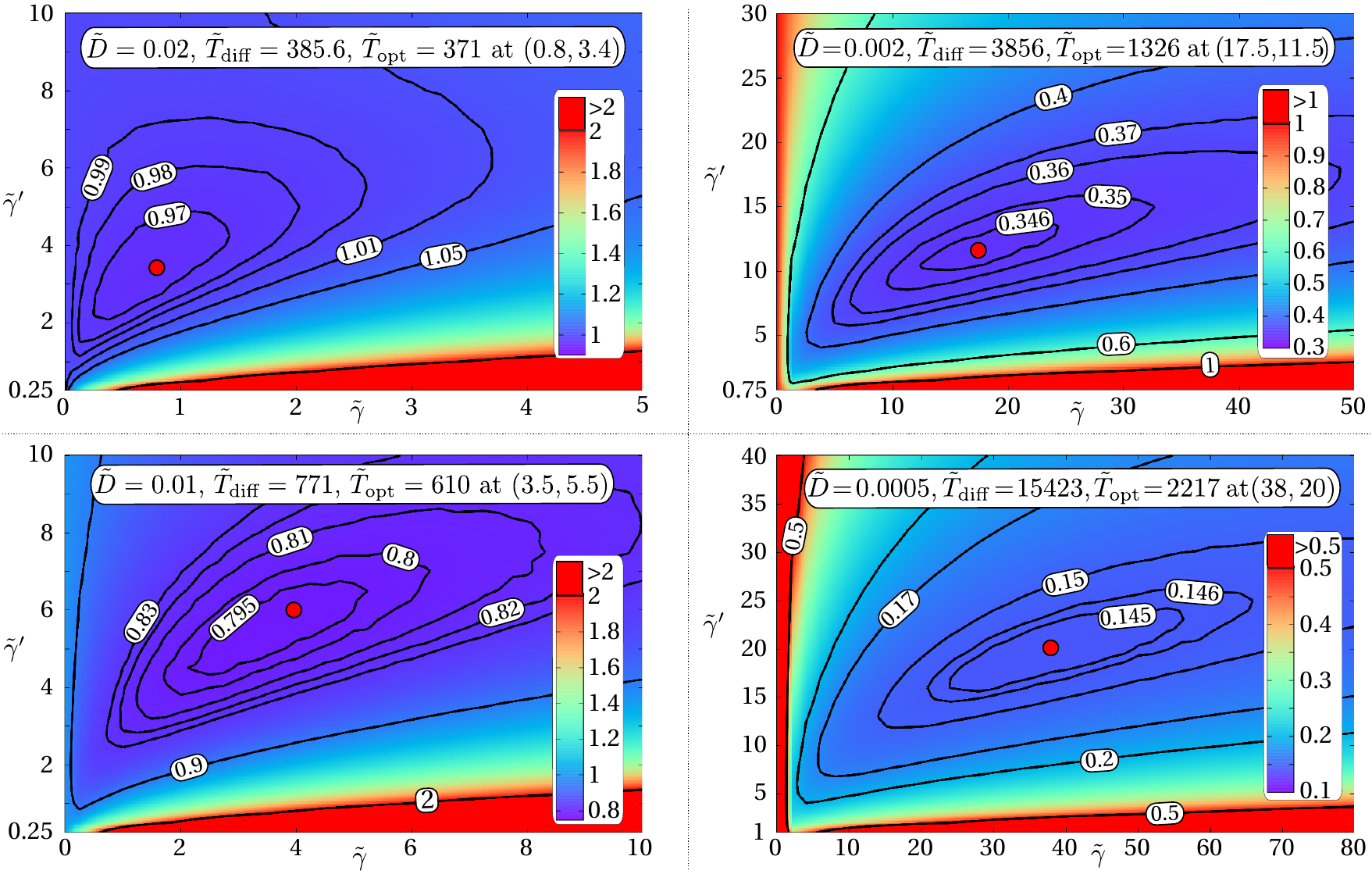} 
\caption{{\bf narrow escape, BB, $\rho_\text{hom}^\alpha$}; The normalized MFPT $\skaltime$ (Eq.$\;$\ref{def_gain}) is color-coded as a function of the rates $\tilde\gamma$ and $\tilde\gamma'$ for 
different values of $\tilde{D}$ and $\vartheta_\text{abso}=\text{arcsin}(1/7)$ (interpolation from a grid of $41 \times 41$ data points each). $2-5\cdot 10^6$ samples have been done for
each pair ($\tilde \gamma$ , $\tilde \gamma'$), which leads to relative stochastic fluctuations of $\skaltime$, which are smaller than 0.2\%. The position of the 
minimum $(\tilde\gamma_\text{opt},\tilde\gamma'_\text{opt})$  is always shown with a red dot. } 
\label{ball_stay_ball_hom}
\end{figure*}

  A comparison of $\tilde T_\text{diff}^\text{(appro)}$ and $\tilde T_\text{v}^\text{(appro)}$ points out an important difference in the behavior of divergence of $\tilde T_\text{diff}$ and $\tilde T_\text{v}$ for small escape areas:
\begin{eqnarray} 
 \tilde T_\text{diff}(\vartheta_{abso})&\propto & \frac{1}{\tilde D \cdot \vartheta_{abso}}\nnn
 \tilde T_\text{v}(\vartheta_{abso})&\propto & \left\{
\begin{array}{ccc}
\infty\quad &,& \text{BB boundary cond.}  \\
\frac{1}{(\vartheta_{abso})^2} &,& \text{BD boundary cond.}
\end{array}\right. \;.\non
\end{eqnarray}


After having studied the two possible extreme cases in search behavior, which is necessary for understanding the later discussed $\vartheta_\text{abso}$ dependence, 
we now face intermittent strategies and analyze their efficiency. \\

In subsection \ref{stayballistic} the condition BB is studied, i.e. a 
ballistically moving particle, which hits the boundary of the simulation sphere, stays in its ballistic mode with the reflected velocity direction. The arrival at the escape area 
of the sphere will only be detected if the particle is in the diffusive mode, otherwise it is reflected. We compare the problem of the homogeneously distributed 
direction density $\rho_\text{hom}^\alpha$ to the inhomogeneous scenario of $\rho_x^\alpha$.\\
In subsection \ref{switchdiffusive} the condition BD is studied, i.e. a
ballistically moving particle, which hits the boundary of the simulation sphere, switches immediately to the diffusive mode, i.e. if this switch happens at
the escape area, the particle immediately recognizes the exit. 
Here, the homogeneous case is compared to the inhomogeneous scenarios of $\rho_x^\alpha$ and $\rho_{p,\tilde\Delta}^\alpha$.
\subsection{BB}
\label{stayballistic}
For the BB condition, the searcher will start in the center of the sphere and the escape area is given by a spherical cab with angle $\vartheta_\text{abso}=\text{arcsin}(1/7)\approx 0.1433$ within this subsection, i.e. the radius of the 
absorbing spherical cab is seven times smaller then the radius of the sphere. In consequence, the area the particle searches,
is about $0.51\%$ of the total spherical surface, i.e. we are in the limit of a small escape area. In this setup, the reference time (taken from the MC data of 
Fig. \ref{MFPT_diff_general_plot}) is given by
\begin{eqnarray} 
 \tilde T_\text{diff}=\frac{7.71}{\tilde D}\;. \label{MFPT_diff_absogl0143}
\end{eqnarray}
\subsubsection{homogeneous distribution $\rho^\alpha_\text{hom}$}
\label{homrefl}
\begin{figure} [htb]
\incgrapw{0.9}{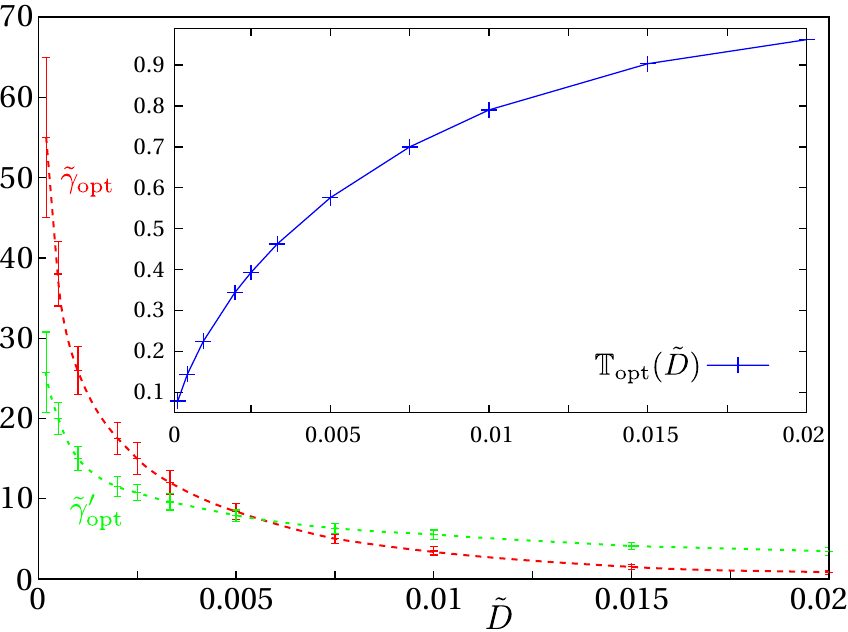} 
\caption{{\bf narrow escape, BB, $\rho_\text{hom}^\alpha$}; The optimal transition rates $\tilde\gamma_\text{opt}$, $\tilde\gamma'_\text{opt}$ and the resulting normalized MFPT $\skaltime_\text{opt}$ (inset) as a function of $\tilde D$. For the diffusion coefficients shown 
in Fig. \ref{ball_stay_ball_hom}, the data coincides with the coordinates of the red dots and its value of $\skaltime$.} 
\label{skal_min_rates_ball_stay_ball_at_bound}
\end{figure}

For different values of $\tilde{D}$, we look for the best strategy to search for the absorbing area as a function of the switching parameters $\tilde\gamma$ 
and $\tilde\gamma'$. $\skaltime$ as a function of $\tilde\gamma$ and $\tilde\gamma'$ is shown in Fig. \ref{ball_stay_ball_hom} for four different examples of $\tilde D$. 
For $\tilde{D}=\frac{D}{v R}$ larger than about 0.025 there is no benefit in a mixed strategy. Here, a purely diffusive particle is on average the better searcher 
as diffusive motion is faster on these scales. As $\tilde{D}$ decreases, phases of ballistic displacement become more and more efficient, as the diffusive displacement per time unit
shrinks. Hence, a global minimum $\skaltime_\text{opt}(\tilde D)=\tilde T_\text{opt}(\tilde D)/\tilde T_\text{diff}(\tilde D)<1$ occurs in the  ($\tilde\gamma,\tilde\gamma'$) space, 
i.e. there is a benefit in an intermittent search strategy. 
As expected, this benefit further increases with decreasing $\tilde D$, i.e. $\skaltime_\text{opt}(\tilde D)$ increases monotonically and
\begin{equation}
\lim_{\tilde D\rightarrow 0^+} \skaltime_\text{opt}(\tilde D)=0
\end{equation}
holds, although $\lim_{\tilde D\rightarrow 0^+} \tilde T_\text{opt}(\tilde D)=\infty$.
Surprisingly, the efficiency of the strategy changes only very little in a quite large (relative to the absolute values) surrounding of the optimal 
solution $(\tilde\gamma_\text{opt},\tilde\gamma'_\text{opt}$) for all diffusivities $\tilde D$. This can be seen by having a closer look to the values of the isolines in Fig. \ref{ball_stay_ball_hom}.
In consequence, due to stochastic fluctuations, the relative error in the optimal values of $\tilde\gamma_\text{opt}$ and $\tilde\gamma'_\text{opt}$ is much larger than 
the relative error in the value of $\tilde T_\text{opt}$ and $\skaltime_\text{opt}$. Fig. \ref{skal_min_rates_ball_stay_ball_at_bound} 
shows $\tilde\gamma_\text{opt}$, $\tilde\gamma'_\text{opt}$ and (in the inset) $\skaltime_\text{opt}$ as a function of the diffusivity $\tilde D$.
The corresponding values of $\tilde T_\text{diff}$ and $\tilde T_\text{opt}$ are also listed in 
Table \ref{tabelle_ball_ball} and plotted in Fig. \ref{all_MFPT_in_one} for a comparison to the later treated inhomogeneous search scenarios. 
$\tilde\gamma_\text{opt}$ and $\tilde\gamma'_\text{opt}$ decrease monotonically in $\tilde D$. As this happens faster for $\tilde\gamma$ than for $\tilde\gamma'$, 
the fraction of time spend in the diffusive mode $\tilde\gamma'/(\tilde\gamma + \tilde\gamma')$ increases with $\tilde D$. \\

Due to the enormous numerical effort, it is not possible to vary $\vartheta_\text{abso}$ systematically here. Nevertheless, we exemplarily investigated also some
values of $\tilde D$ for smaller and larger values of $\vartheta_\text{abso}$. Similar to the results in the following chapters, we found, that a decrease in target size
results in an increase in both transition rates.

\subsubsection{inhomogeneous distribution $\rho_x^\alpha$}
\label{inhomrefl}
\begin{figure}[tb!]
\incgrapw{0.9}{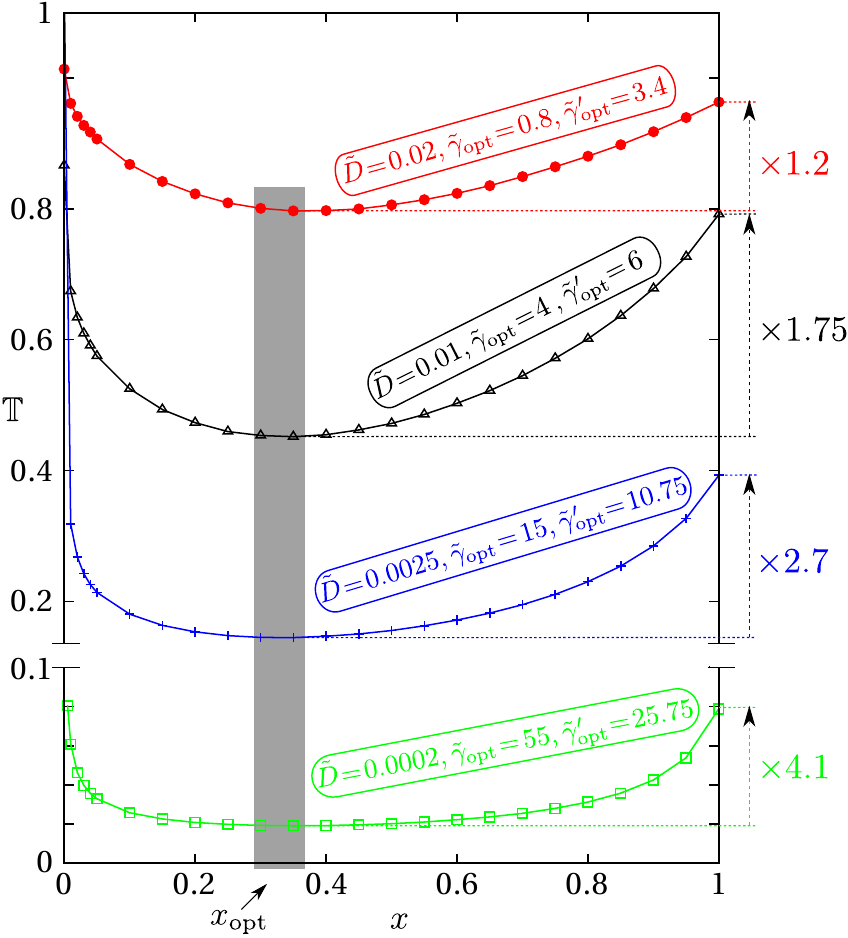} 
\caption{{\bf narrow escape, BB, $\rho_x^\alpha$}; $\skaltime$ as a function of the spreading parameter $x$ for the transition rates $(\tilde\gamma_\text{opt},\tilde\gamma'_\text{opt})$ of the optimal solution of 
Fig. \ref{skal_min_rates_ball_stay_ball_at_bound}. The minimum position $x_\text{opt}\approx 0.35$ (gray bar) of all curves is almost identical. The colored numbers on the 
right show the ratio $\skaltime_\text{min}/\skaltime_\text{opt}$.}
\label{x_vs_MFPT_for_ideal_homogen}
\end{figure}

Within this subsubsection, we study the influence of $\rho_x^\alpha$ on the search strategy and the transition rates.
Fig. \ref{x_vs_MFPT_for_ideal_homogen} shows the dependence of $\skaltime$ on $x$ for representatively selected values of $\tilde D$ and the corresponding optimal parameters $\tilde\gamma_\text{opt}(\tilde D),\;\tilde\gamma'_\text{opt}(\tilde D)$ of 
the homogeneous scenario, shown in Fig.  \ref{skal_min_rates_ball_stay_ball_at_bound}. 

\begin{figure} [!htb]
\incgrapw{0.9}{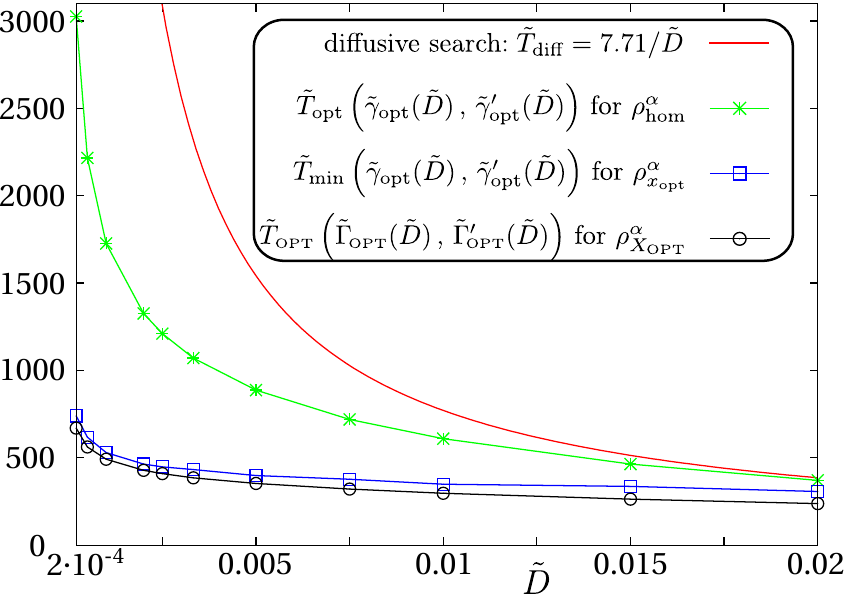} 
\caption{{\bf narrow escape, BB}; MFPT for purely diffusive search (Eq. \ref{MFPT_diff_absogl0143}, red line); optimal search with homogeneously distributed velocity direction (green); optimal search with 
inhomogeneously distributed velocity direction for the fixed rates $\tilde\gamma_\text{opt}$, $\tilde\gamma'_\text{opt}$ for the homogeneous scenario (blue) ; optimal search with 
inhomogeneously distributed velocity direction for rates $\tilde\Gamma_\text{\hspace*{-0.1cm}\textsubscript{\tiny OPT}}$, $\tilde\Gamma'_\text{\hspace*{-0.1cm}\textsubscript{\tiny OPT}}$ (black).}
\label{all_MFPT_in_one}
\end{figure}

The global minimum for each $\tilde D$ will be denoted $\tilde T_\text{min}(\tilde D)$ in the following. 
At $x=1$ the values of $\skaltime$ coincide with the corresponding $\skaltime_\text{opt}$ of Fig. \ref{skal_min_rates_ball_stay_ball_at_bound}.
In none of the cases, the value at $x=1$ is the minimum. It follows, that an anisotropic velocity direction distribution increases the efficiency of the search 
strategy significantly. For small values of $\tilde D$, $\tilde T_\text{min}$ is much smaller than $\tilde T_\text{opt}$, which can also be seen by comparing the blue and the green 
curves of Fig. \ref{all_MFPT_in_one} and the corresponding  values in Table \ref{tabelle_ball_ball}. As $\tilde D$ increases the benefit of an inhomogeneous strategy becomes less 
pronounced. It is remarkable, that the degree of inhomogeneity $x_\text{opt}\approx 0.35$ is constant for all $\tilde D$.\\

Nevertheless it is even possible to decrease $\skaltime$ further: In Fig. \ref{x_vs_MFPT_for_ideal_homogen} the transition rates were chosen as the optimal solution 
for the homogeneous case. There is no reason, that this is also the optimal choice in the inhomogeneous case. In consequence, we varied $\tilde\gamma,\;\tilde\gamma'$ and $x$ 
simultaneously for finding the optimal parameters $\tilde\Gamma\text{\hspace*{-0.05cm}\textsubscript{\tiny OPT}},\;\tilde\Gamma'\text{\hspace*{-0.15cm}\textsubscript{\tiny OPT}}$ 
and $X\text{\hspace*{-0.05cm}\textsubscript{\tiny OPT}}$ for the MFPT $\tilde T\text{\hs{-0.8}\textsubscript{\tiny OPT}}$ (be aware of the different meaning of the index "opt`` and ''OPT``). 
The results are shown in Table \ref{tabelle_ball_ball} and $\tilde T\text{\hs{-0.8}\textsubscript{\tiny OPT}}$
is plotted in Fig. \ref{all_MFPT_in_one}.
\begin{table}[htb]
\begin{tabular}{|l||l|l|l|l||l|l|l|l||l|l|} 
\hline
 $\tilde D$ & $\hs{-0.4}\tilde T_\text{diff}\hs{-0.4}$& $\tilde T_\text{opt}$ & \hs{-0.4}$\tilde T_\text{\hs{-0.6}min}$\hs{-0.4}& \hs{-0.5}$\tilde T\text{\hs{-0.8}\textsubscript{\tiny OPT}}\hs{-0.6}$ & $\tilde\gamma_\text{opt}$& $\tilde\gamma'_\text{opt}$& $\hs{-0.4}\tilde\Gamma\text{\hs{-0.8}\textsubscript{\tiny OPT}}\hs{-0.4}$ & $\hs{-0.4}\tilde\Gamma'\text{\hs{-1.8}\textsubscript{\tiny OPT}}\hs{-0.4}$ & $x_\text{opt}$& $X\text{\hs{-0.8}\textsubscript{\tiny OPT}}$\hs{-0.4}   \\
\hline
\hline
 \hs{-0.4}0.02   &386& 371 &307 & 238 &0.8 &3.4    &11.5 & 8   &0.35& \hs{-0.4}0.325\hs{-0.4}   \\
\hline
 \hs{-0.4}0.015  &514& 465 &337 &  264     &1.5 &4.1    &  18   & 8.5   &0.35& \hs{-0.4}0.325\hs{-0.4}  \\ 
\hline
 \hs{-0.4}0.01   &771& 610 &349 & 297 &4 &6    &25   & 9.5 &0.35& \hs{-0.4}0.325\hs{-0.4}   \\
\hline
 \hs{-0.4}0.0075\hs{-0.4} &1028& 720 & 377&  321     &5   &6.3    &  30   & 10    &0.35&\hs{-0.4}0.325\hs{-0.4}  \\
\hline
 \hs{-0.4}0.005  &1542& 888 & 398& 353 &8.4 &7.9    &36   & 11  & 0.35& \hs{-0.4}0.325\hs{-0.4}    \\
\hline
 \hs{-0.4}1/300  &2314& 1071& 433& 386 &12  &9.6    & 42  & 12  &0.35& \hs{-0.4}0.325\hs{-0.4}    \\
\hline
 \hs{-0.4}$0.0025\hs{-0.4}$&3085& 1211& 448& 410 &15  &\hs{-0.4}1\hs{-0.1}0\hs{-0.1}.\hs{-0.1}7\hs{-0.1}5\hs{-0.4}  &48   & 13  &0.35& \hs{-0.4}0.325\hs{-0.4}   \\
\hline
 \hs{-0.4}0.002  &3856& 1326& 466& 429 &17.5&11.5   &50   &\hs{-0.4}13.5\hs{-0.4}&0.35& \hs{-0.4}0.325\hs{-0.4}   \\
\hline
 \hs{-0.4}0.001  &7712& 1727& 530& 492 &26  & 15    &60   & 15  &0.35& 0.3    \\
\hline
 \hs{-0.4}$0.0005\hs{-0.4}$&\hs{-0.4}1\hs{-0.1}5\hs{-0.1}4\hs{-0.1}2\hs{-0.1}0\hs{-0.4}& 2217& 618& 562 &38  &20     &75   & 18  &0.35& 0.3    \\
\hline
 \hs{-0.4}$0.0002\hs{-0.4}$&\hs{-0.4}3\hs{-0.1}8\hs{-0.1}5\hs{-0.1}6\hs{-0.1}0\hs{-0.4}& 3026 &740& 670 &55  & \hs{-0.4}2\hs{-0.1}5\hs{-0.1}.\hs{-0.1}7\hs{-0.1}5\hs{-0.4} & 95  & 21  &0.35& 0.3   \\
\hline
\end{tabular}
\caption{$\tilde T_\text{diff}$: purely diffusive MFPT; $\tilde T_\text{opt}$: optimized intermittent MFPT for 
$\rho_\text{hom}^\alpha$ with optimal rates $\tilde\gamma_\text{opt}$ and $\tilde\gamma'_\text{opt}$; $\tilde T_\text{min}$: optimized intermittent MFPT for 
$\rho_x^\alpha$ with optimal inhomogeneity coefficient $x_\text{opt}$ and fixed rates $\tilde\gamma_\text{opt}$ and $\tilde\gamma'_\text{opt}$; 
$\tilde T\text{\hspace*{-0.05cm}\textsubscript{\tiny OPT}}$: optimized intermittent MFPT for 
$\rho_x^\alpha$ with optimal inhomogeneity coefficient $X\text{\hs{-0.8}\textsubscript{\tiny OPT}}$ and 
corresponding optimal rates $\tilde\Gamma\text{\hs{-0.8}\textsubscript{\tiny OPT}}$ and $\tilde\Gamma'\text{\hs{-1.8}\textsubscript{\tiny OPT}}$.}
\label{tabelle_ball_ball}
\end{table}
Table \ref{tabelle_ball_ball} delivers some remarkable results:  
\begin{itemize}[align=left,labelwidth=\widthof{i},leftmargin=\labelwidth+\labelsep]
\item The optimal value of $x$ seems to be almost constant in all cases. For the rates of the homogeneous optimization $(\tilde\gamma_\text{opt},\tilde\gamma'_\text{opt})$ and for 
the rates of the inhomogeneous optimization $(\tilde\Gamma\text{\hspace*{-0.05cm}\textsubscript{\tiny OPT}},\tilde\Gamma'\text{\hspace*{-0.15cm}\textsubscript{\tiny OPT}})$ 
the best solution is always given by $x\approx0.325\pm0.025$. Hence, the degree of inhomoegeneity for an optimal solution does not seem to depend much on the diffusion coefficient 
and the transition rates, which is quite surprising. 
\item Comparing the values of $(\tilde\gamma_\text{opt},\tilde\gamma'_\text{opt})$ with 
$(\tilde\Gamma\text{\hspace*{-0.05cm}\textsubscript{\tiny OPT}},\tilde\Gamma'\text{\hspace*{-0.15cm}\textsubscript{\tiny OPT}})$, one recognizes remarkable changes in the 
transition rates. For large values of $\tilde D$, the change is more than a factor of 10. 
\item Like in the homogeneous case, the efficiency of the inhomogeneous strategy changes only very little in a quite large surrounding of the optimal solution 
$(\tilde\Gamma\text{\hspace*{-0.05cm}\textsubscript{\tiny OPT}},\tilde\Gamma'\text{\hspace*{-0.15cm}\textsubscript{\tiny OPT}})$.
\end{itemize}
For concluding this subsection, Fig. \ref{all_MFPT_in_one} shows the optimal MFPTs for the different discussed scenarios. 
Compared to a purely diffusive searcher (red), an intermittent search strategy with a homogeneous velocity direction distribution (green) optimizes the search 
process especially  for small $\tilde D$ significantly, which has already been shown in the inset of Fig. \ref{skal_min_rates_ball_stay_ball_at_bound}. In the next 
step, we introduced an inhomogeneity in the velocity direction distribution (blue), but kept the optimal rates of the homogeneous case. Again, the largest benefit 
can be seen for small $\tilde D$ (see Fig. \ref{x_vs_MFPT_for_ideal_homogen}). In the last step, we varied the transition rates and the degree of inhomogeneity 
simultaneously (black). Although the optimal rates changed dramatically, 
the additional benefit is much smaller than in the optimization steps before. But this time it increases with $\tilde D$.

\subsection{BD}
\label{switchdiffusive}
\begin{figure} [htb]
\begin{center}
\incgrapw{0.9}{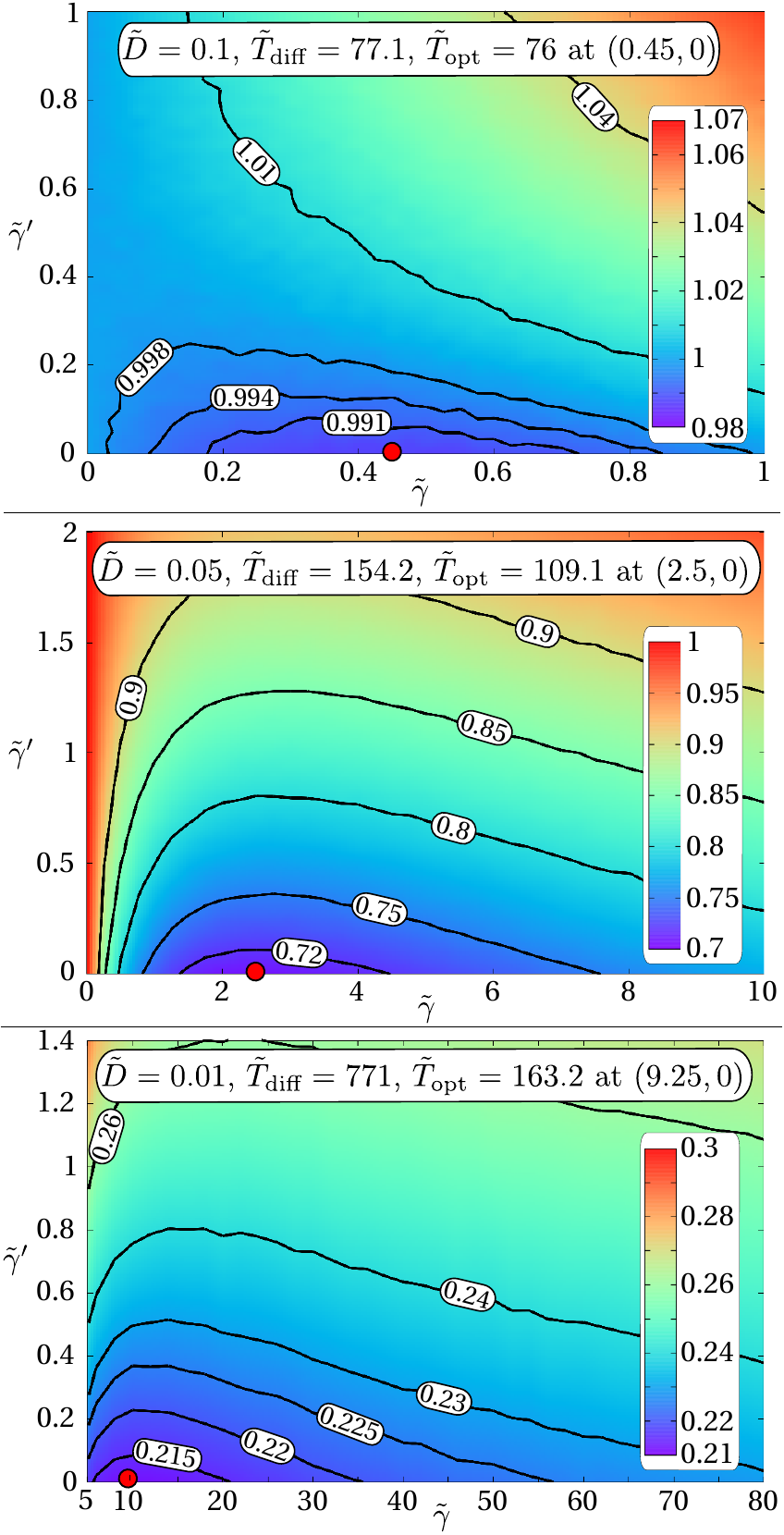} 
\caption{{\bf narrow escape, BD, $\rho_\text{hom}^\alpha$}; The normalized MFPT $\skaltime$ (Eq.$\;$\ref{def_gain}) is color-coded as a function of the parameters $\tilde\gamma$ and $\tilde\gamma'$ for 
different values of $\tilde{D}$ and $\vartheta_\text{abso}=\text{arcsin}(1/7)$ (interpolation from a grid of $41 \times 41$ data points each). Between $2\cdot 10^6$ and $5\cdot 10^6$ samples have been done for each 
pair ($\tilde \gamma$ , $\tilde \gamma'$), which leads to relative stochastic fluctuations of $\skaltime$ which are smaller than 0.2\% for. The position of the 
minimum $(\tilde\gamma_\text{opt},\tilde\gamma'_\text{opt})$  is always shown with a red dot. } 
\label{ball_to_diff_hom}
\end{center}
\end{figure}

For all investigated direction distributions ($\rho_\text{hom}^\alpha$ and both inhomogeneous scenarios $\rho_x^\alpha$, $\rho_{p,\tilde \Delta}^\alpha$ ) in this subsection the optimal search strategy 
is either a purely diffusive one (for $\tilde D$ large) or the simulations yield $\tilde \gamma'_\text{opt}=0$.  
Exemplarily, this is shown in Fig. \ref{ball_to_diff_hom}.
For $\vartheta_{abso}=\text{arcsin}(1/7)$ and different values of $\tilde D$ the figure shows 
$\skaltime$ as a function of the transition rates for the case of $\rho_\text{hom}^\alpha$ and the initial position in the origin. 
A comparison to Fig. \ref{ball_stay_ball_hom} shows the different behaviour of the optimal solution for the two boundary conditions.
We verified $\tilde \gamma'_\text{opt}=0$ also for smaller values of $\vartheta_{abso}$ and larger ones ($0.025<\vartheta_{abso}<\pi$).  
Consequently, the numerical effort of finding the best strategy is dramatically reduced, as there is one parameter less to vary. Due to this 
reduced effort, the variation of the absorbing angle $\vartheta_\text{abso}$ will also be studied in the case of a $\rho_\text{hom}^\alpha$.\\
Apart from this additional study, the beginning of the subsection is organized identically to the one before: 
We start with the case of $\rho_\text{hom}^\alpha$ , followed by the inhomogeneous scenario $\rho_x^\alpha$ for $\vartheta_{abso}=\text{arcsin}(1/7)$. 
In both cases the initial position is the origin.
Afterwards we study the case $\rho_{p,\tilde\Delta}^\alpha$ for a homogeneously distributed initial position $\tilde {\mb{r}}_0$ and $\vartheta_{abso}=\text{arcsin}(1/7)$.
\subsubsection{homogeneous distribution  $\rho_\text{hom}^\alpha$}
\begin{figure} [htb]
\incgrapw{0.9}{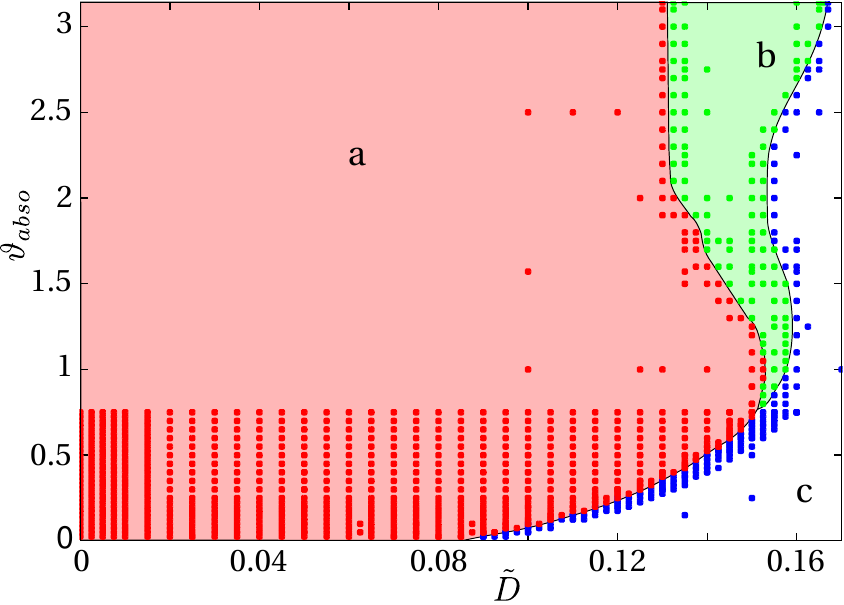} 
\caption{{\bf narrow escape, BD, $\rho_\text{hom}^\alpha$}; Diagram for the choice of the best search strategy as a function of $\tilde D$ and $\vartheta_{abso}$ : In the red (a) and the green (b) domain, an 
intermittent search strategy is preferable, whereas in the white domain (c) pure diffusion is the best strategy. For the construction of the diagram, the behaviour 
at the position of the dots was investigated.}
\label{phase_diagram_hom}
\end{figure}

At first, we study whether an intermittent search strategy or a purely diffusive strategy is better for a given pair of parameters ($\tilde D,\,\vartheta_{abso}$). 
For the reason of completeness we faced this question for all values of $\vartheta_{abso}\in ]0;\pi]$ and not only for a "narrow" escape area.
The result for $\tilde{\mb{r}}_0=0$ is shown in Fig. \ref{phase_diagram_hom}. 

\begin{figure} 
\incgrapw{0.9}{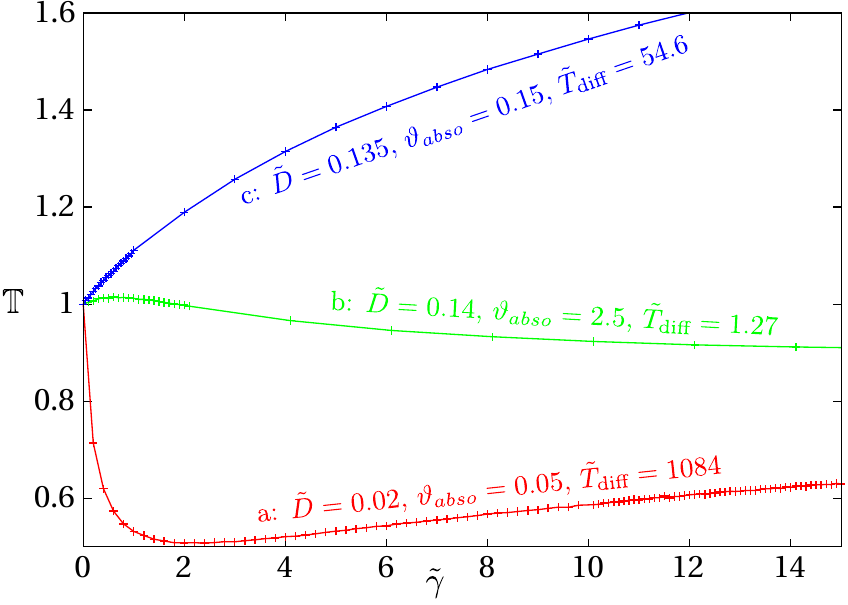} 
\caption{{\bf narrow escape, BD, $\rho_\text{hom}^\alpha$}; Examples of the function $\skaltime(\tilde\gamma, \tilde\gamma'_\text{opt}=0$) for the three different colored areas in Fig. \ref{phase_diagram_hom}.}
\label{phase_diagram_abc_examples}
\end{figure}
In the red (a) domain an intermittent search strategy is preferable. $\skaltime(\tilde\gamma)$ starts monotonically decreasing at $\tilde\gamma=0$. It follows the 
global optimum at $\tilde \gamma_\text{opt}>0$. An example for this behavior for $\tilde D=0.02$, $ \vartheta_{abso}=0.05$ is given in Fig. \ref{phase_diagram_abc_examples}.\\
In the green domain (b), intermittent search is also preferable. Although $\skaltime(\tilde\gamma)$ starts monotonically increasing, it decreases to $\skaltime<1$ for some values of $\tilde\gamma$. 
Again, an example for this behavior for $\tilde D=0.14$, $ \vartheta_{abso}=2.5$ is given in Fig. \ref{phase_diagram_abc_examples}.\\ 
Finally, in the white domain (c) $\tilde \gamma_\text{opt}=0$ 
holds, hence a diffusive search is the best strategy. An example for this behavior for $\tilde D=0.135$, $ \vartheta_{abso}=0.15$ is also given in Fig. \ref{phase_diagram_abc_examples}.\\

Fig. \ref{phase_diagram_hom} only answers the question about the best strategy in principle, it is neither quantifying the transition rate $\tilde \gamma_\text{opt}(\tilde D, \tilde \vartheta_{abso})$ 
nor the MFPTs $\tilde T_\text{opt}(\tilde D, \tilde \vartheta_{abso})$ and $\skaltime_\text{opt}(\tilde D, \tilde \vartheta_{abso})$. A quantification has only been done in the case of small escape 
areas ($\vartheta_{abso}<0.75$) due to the following reasons:\\ 
If the escape area is large, the searcher will find it soon, hence there is no need for a special strategy. The largest impact of $\tilde \gamma$ on the efficiency of the strategy is given 
for small values of $\tilde \vartheta_{abso}$, i.e. for large $\tilde \vartheta_{abso}$ either a purely diffusive searcher or a random velocity model ($\tilde\gamma=\infty$) is always close to the optimal strategy.          
Additionally, for small values of $\vartheta_{abso}$ the optimal strategy is almost independent of the starting position of the searcher, i.e. the shown results for a searcher starting at the origin will
remain true in the more general context of an arbitrary initial position. 
For the angle $\tilde \vartheta_{abso}=\text{arcsin}(1/7)$ this independence is shown explicitly later.

\begin{figure} [htb]
\incgrapw{0.9}{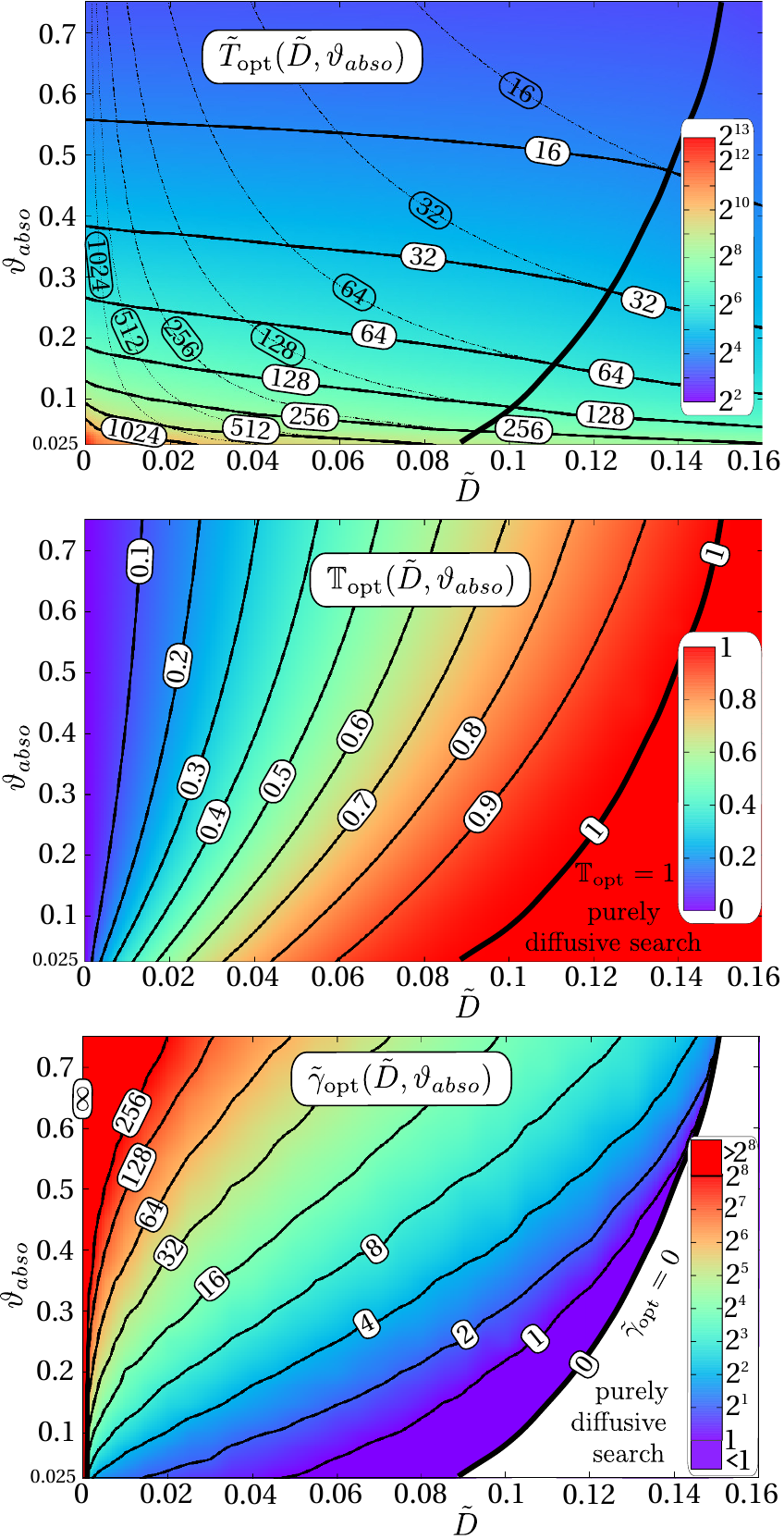}
\caption{{\bf narrow escape, BD, $\rho_\text{hom}^\alpha$};
$\tilde T_\text{opt}$, $\skaltime_\text{opt}$ and the corresponding $\tilde\gamma_\text{opt}$ as a function of $\tilde D$ and $\vartheta_{abso}<0.75$ 
(interpolation from the non equidistant grid shown at the bottom of Fig. \ref{phase_diagram_hom}). The thick black line in each subfigure separates the area of intermittent search 
and purely diffusive search. It coincides with the boundary line between the area a and c in Fig. \ref{phase_diagram_hom}. {\bf top}: $\tilde T_\text{opt}$ in a logscale color plot, 
the dashed lines with transparent label show isolines of the purely diffusive search scenario (Eq. (\ref{approx_purely_diff_narrow_escape})) for the reason of comparison. {\bf middle}: $\skaltime_\text{opt}$ 
is color-coded plotted  {\bf bottom}: $\tilde\gamma_\text{opt}$ in a logscale color plot.
}
\label{3d_ball_to_diff_theta_D_vs_G_and_gamma}
\end{figure}

Fig. \ref{3d_ball_to_diff_theta_D_vs_G_and_gamma} quantifies the values of $\tilde\gamma$, $\tilde T_\text{opt}$ and $\skaltime_\text{opt}$ for $0.025<\vartheta_{abso}<0.75$.
The corresponding curves, from which the optimal values of $\tilde\gamma_\text{opt}$ and $\tilde T_\text{opt}$ were taken for each data point, qualitatively all look like the red 
curve in Fig. \ref{phase_diagram_abc_examples}.
Depending on $\tilde D$ and $\vartheta_{abso}$, $500.000$ up to $10^{10}$ samples have been performed for each parameter triple ($\tilde\gamma$, $\tilde D$, $\vartheta_{abso}$).
As the depth of the minimum position is differently strong pronounced this is necessary to control the stochastic fluctuations in the value of $\tilde\gamma_\text{opt}$.\\
For $\tilde D=0$ the optimal strategy is trivially given by $\tilde\gamma_\text{opt}=\infty$ for all $\vartheta_{abso}$, i.e. the optimal strategy is the random velocity model with 
MFPT $\tilde T_\text{v}$, which is very well approximated by $\tilde T_\text{v}^\text{(appro)}$ in Eq. (\ref{hatqv}), shown in Fig. \ref{T_v_and_q_v}.   
For small diffusivities $\tilde D > 0$ the transition rate $\tilde\gamma_\text{opt}$ is finite. Its value strongly depends on the size of the escape 
area, i.e. on the value of $\vartheta_{abso}$. The thick black line in Fig. \ref{3d_ball_to_diff_theta_D_vs_G_and_gamma} shows the ''break-even`` diffusivity $\tilde D_\text{be}(\vartheta_{abso})$, 
where the optimal strategy changes from intermittent search to purely diffusive search. $\tilde D_\text{be}$ increases monotonically in $\vartheta_{abso}$. It rises the interesting
question about the limit of $\tilde D_\text{be}$ for $\vartheta_{abso}\rightarrow 0$ (be aware of $0.025<\vartheta_{abso}$ in Fig. \ref{3d_ball_to_diff_theta_D_vs_G_and_gamma}). 
If $\lim_{\vartheta_{abso}\rightarrow 0} \tilde D_\text{be}=0$ held, for every $\tilde D$ there would be a threshold value $\vartheta_{thres}$ below which pure diffusion would be the best
strategy. In the opposite case of a positive limit $\tilde D_\text{be0}$, i.e. $\lim_{\vartheta_{abso}\rightarrow 0} \tilde D_\text{be}=\tilde D_\text{be0}>0$, intermittent search would be more 
efficient for all $\tilde D<\tilde D_\text{be0}$, no matter how small $\vartheta_{abso}$ becomes. Due to the divergence of the MFPT for $\vartheta_{abso}\rightarrow 0$ it is not possible to
face this limit numerically for the reason of running time. Nevertheless, there are clear arguments for a limit $\tilde D_\text{be0}>0$: The second derivatives of the isolines of $\skaltime_\text{opt}$ 
in the second subfigure of Fig. \ref{3d_ball_to_diff_theta_D_vs_G_and_gamma} seems to vanish for small $\vartheta_{abso}$. Hence, they were expected to reach the x-axis in a straight line at positions 
larger than zero. Due to the enormous running time for very small angles we verified this hypothesis of affine extrapolation only partially at $\vartheta_{abso}=0.0125$ for some $\tilde D$. \\

\begin{figure}[htb]
\incgrapw{0.9}{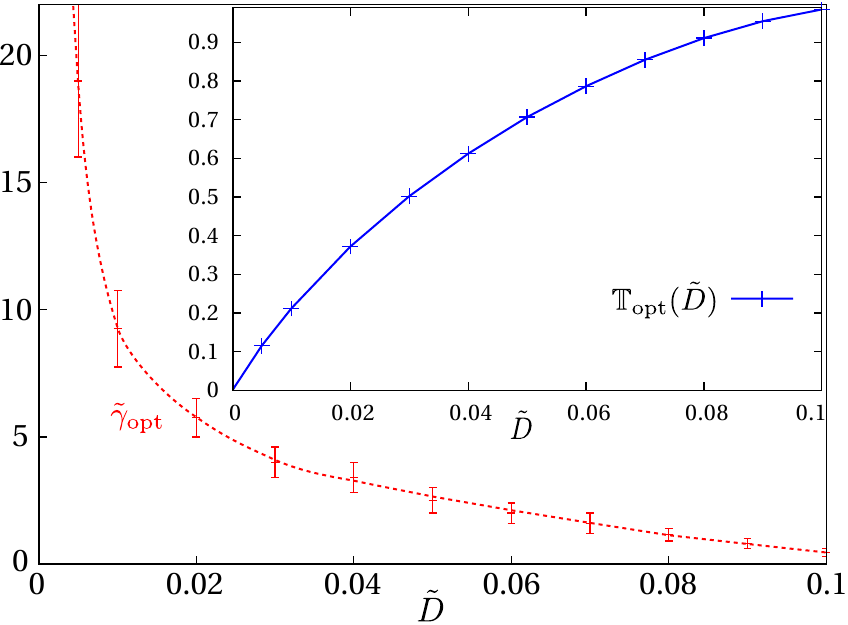} 
\caption{{\bf narrow escape, BD, $\rho_\text{hom}^\alpha$}; $\tilde\gamma_\text{opt}$ and $\skaltime_\text{opt}$ (inset) as a function of $\tilde D$ for $\thetab=\text{arcsin}(1/7)$: 
The corresponding curves $\skaltime(\tilde\gamma)$ from which the minima are taken qualitatively all belong to 
case (a) in the diagram. For $2 \cdot 10^6$ samples for each investigated $\tilde D$ 
the position of the minimum $\tilde\gamma_\text{opt}$ and its value $\skaltime_\text{opt}$ are shown.} 
\label{skal_min_rates_ball_to_diff_at_bound}
\end{figure}

For comparison to the boundary condition BB and the later investigated inhomogeneous search scenarios, the angle $\vartheta_{abso}=\text{arcsin}(1/7)$ is shown separately in Fig.   
\ref{skal_min_rates_ball_to_diff_at_bound} and the corresponding values of $\tilde T_\text{opt}$ are shown in Fig. \ref{all_MFPT_in_one_ball_to_diff} and Table \ref{tabelle_ball_diff}.
Qualitatively, Fig. \ref{skal_min_rates_ball_to_diff_at_bound} does not differ from the result of Fig. \ref{skal_min_rates_ball_stay_ball_at_bound} (except for $\tilde\gamma'=0$), but quantitatively it differs a lot.
The interval where an intermittent search strategy is preferable ($\tilde D<0.11$) is almost five times larger compared to the boundary condition BB. 
For the BD condition the benefit of an intermittent search strategy is always larger, for the following reason: A ballistically moving particle detects the target area immediately after 
switching to diffusive mode at the boundary. In the subsection before, the particle was simply reflected without recognizing the target area. In consequence, the status 
of the ballistic mode is enhanced here, which can also be seen by comparing  the values of $\tilde\gamma_\text{opt}$ in the common interval of Fig. 
\ref{skal_min_rates_ball_stay_ball_at_bound} and Fig. \ref{skal_min_rates_ball_to_diff_at_bound}. In case of the BD condition of this subsection the searcher stays on average shorter in 
the diffusive mode before switching back to ballistic motion again compared to the BB scenario. 

\subsubsection{inhomogeneous distribution $\rho_x^\alpha$}
\begin{figure} [htb]
\incgrapw{0.9}{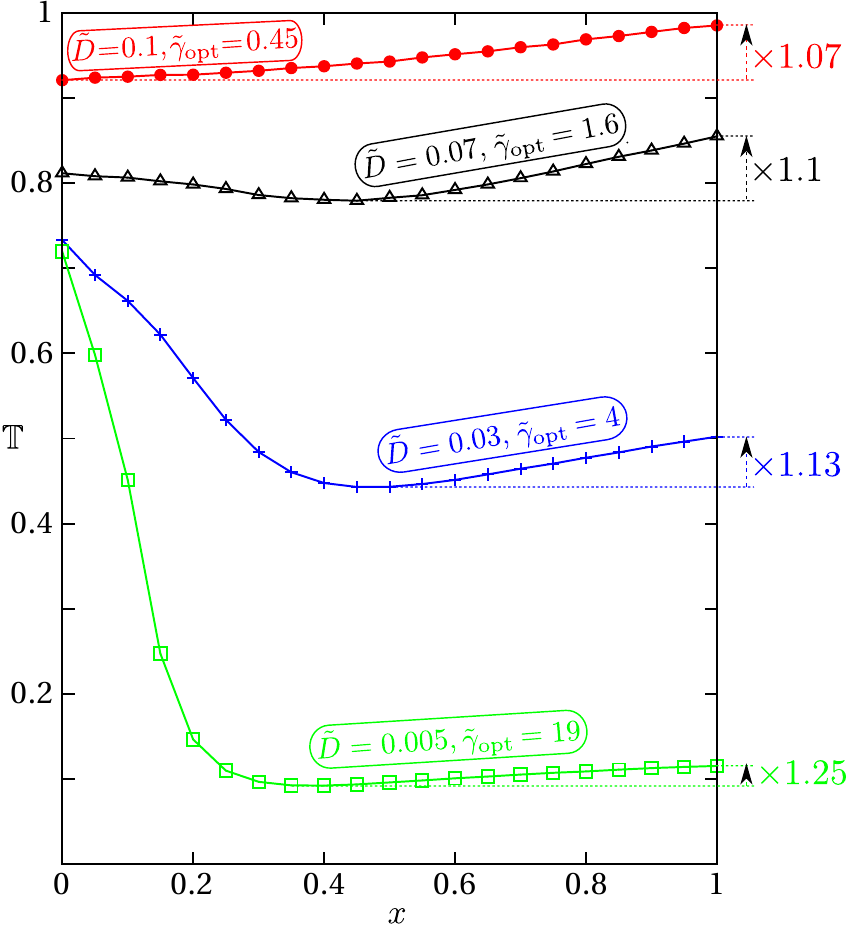} 
\caption{{\bf narrow escape, BD, $\rho_x^\alpha$}; $\skaltime$ as a function of the spreading parameter $x$ for the transition rate $\tilde\gamma_\text{opt}$ of the optimal solution of 
Fig. \ref{skal_min_rates_ball_to_diff_at_bound}. The colored numbers on the right show the ratio $\skaltime_\text{min}/\skaltime_\text{opt}$.} 
\label{skal_with_x_ball_to_diff}
\end{figure}

\begin{figure} [tb]
\incgrapw{0.9}{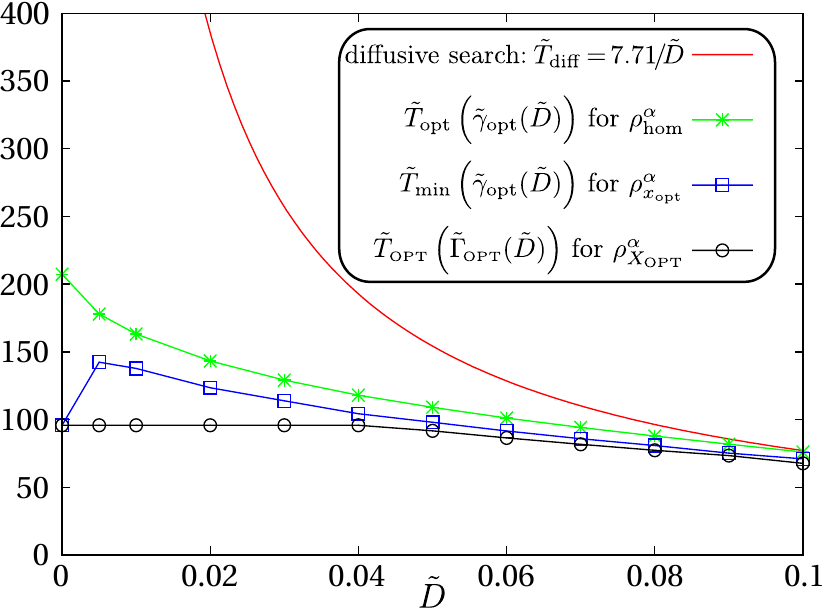}
\caption{{\bf narrow escape, BD};  MFPT for purely diffusive search (Eq. \ref{MFPT_diff_absogl0143}, red line); search with homogeneously distributed velocity direction (green); search with 
inhomogeneously distributed velocity direction for the fixed rates $\gamma_\text{opt}$, $\gamma'_\text{opt}=0$ (blue) ; search with 
inhomogeneously distributed velocity direction for rates $\Gamma_\text{\hspace*{-0.1cm}\textsubscript{\tiny OPT}}$, $\Gamma'_\text{\hspace*{-0.1cm}\textsubscript{\tiny OPT}}=0$ (black).
The points at $\tilde D=0$ are based on random velocity direction simulations, as this is the limit 
for $\tilde D\rightarrow0$.}

\label{all_MFPT_in_one_ball_to_diff}
\end{figure}

Fig. \ref{skal_with_x_ball_to_diff} shows $\skaltime$ as a function of $x$ for the optimal parameters $\tilde\gamma_\text{opt}$ of 
Fig. \ref{skal_min_rates_ball_to_diff_at_bound} for different values of $\tilde D$. For each $\tilde D$ the minimal MFPT $\tilde T_\text{min}$ is plotted in Fig. \ref{all_MFPT_in_one_ball_to_diff} and listed in Table 
\ref{tabelle_ball_diff}.    
For large values of $\tilde D$, $x_\text{opt}=0$ holds, meaning the optimal velocity direction is always radially to the outward. As $\tilde D$ decreases, 
the minimum $x_\text{opt}$ switches to the interior of the interval $[0,1]$. A comparison to Fig. \ref{x_vs_MFPT_for_ideal_homogen} shows the following differences 
between the two boundary conditions: The position of $x_\text{opt}$ is not constant any more, here it depends strongly on $\tilde D$. The value of $\skaltime_{min}$ at $x_\text{opt}$ 
differs less from the value of the homogeneous velocity direction distribution (x=1). Hence, the additional benefit of an inhomogeneous velocity direction is less 
than in the case of the previous subsection. This can also be seen by comparing the gap between the green and blue lines of Fig. \ref{all_MFPT_in_one} and 
Fig. \ref{all_MFPT_in_one_ball_to_diff}. \\
Similar to the BB boundary condition before, we varied $\tilde \gamma$, $\tilde \gamma'$ and $x$ simultaneously to find the optimal parameters 
$\tilde \Gamma_\text{OPT}$ and $X\text{\hspace*{-0.05cm}\textsubscript{\tiny OPT}}$ for the MFPT $\tilde T\text{\hspace*{-0.05cm}\textsubscript{\tiny OPT}}$. The optimal $\gamma'$ again vanishes,
i.e. $\tilde \Gamma'_\text{OPT}=0$. The other results are shown in Table \ref{tabelle_ball_diff}.
\begin{table} [htb]
\begin{tabular}{|l||l|l|l|l||l|l||l|l|l|} 
\hline
 $\tilde D$ & $\tilde T_\text{diff}$ & $\tilde T_\text{opt}$ & $\tilde T_\text{min}$ &$\tilde T\text{\hspace*{-0.05cm}\textsubscript{\tiny OPT}}$ & $\tilde\gamma_\text{opt}$ & $\tilde\Gamma\text{\hspace*{-0.05cm}\textsubscript{\tiny OPT}}$ & $x_\text{opt}$ & $X\text{\hspace*{-0.05cm}\textsubscript{\tiny OPT}}$  \\
\hline
\hline
 0.1  & 77.1 & 76   & 71  & 67.6 & 0.45 & 1.7 & 0 & 0  \\
\hline
 0.09 & 85.7 & 81.8 & 75.2& 73.4 & 0.8  & 1.6 & 0 & 0   \\
\hline
 0.08 & 96.4 & 87.8 & 80.8& 77.3 & 1.15 & 4  & 0.35 & 0.375   \\
\hline
 0.07 & 110  & 94.2 & 85.9& 81.8 & 1.6  & 5.5 & 0.45& 0.4   \\
\hline
 0.06 & 129  & 101 & 91.6 & 86.5 & 2    & 7.5 & 0.45& 0.4   \\
\hline
 0.05 & 154  & 109 & 98   & 91.7 & 2.5  & 14  & 0.45& 0.375  \\
\hline
 0.04 & 193  & 118 & 104  & 95.8 & 3.4  & -  & 0.45 &0.12   \\
\hline
 0.03 & 257  & 129 & 114  & 95.8 & 4    & -  & 0.45&0.12   \\
\hline
 0.02 & 386  & 143 & 124  & 95.8 & 5.75 & -  & 0.45& 0.12   \\
\hline
 0.01 & 771  & 163 & 138  & 95.8 & 9.25 & -  & 0.45&0.12 \\
 \hline
 0.005& 1542 & 178 & 142  & 95.8 & 19   & -  & 0.4 &0.12  \\
\hline
\end{tabular}
\caption{$\tilde T_\text{diff}$: purely diffusive MFPT; $\tilde T_\text{opt}$: optimized intermittent MFPT for  
$\rho_\text{hom}^\alpha$ with optimal rate $\tilde\gamma_\text{opt}$; $\tilde T_\text{min}$: optimized intermittent MFPT  for 
$\rho_x^\alpha$ with optimal inhomogeneity coefficient $x_\text{opt}$ and fixed 
rate $\tilde\gamma_\text{opt}$; $\tilde T\text{\hspace*{-0.05cm}\textsubscript{\tiny OPT}}$: optimized intermittent MFPT  for 
$\rho_x^\alpha$ with optimal inhomogeneity coefficient $X\text{\hs{-0.8}\textsubscript{\tiny OPT}}$ and 
corresponding optimal rate $\tilde\Gamma\text{\hs{-0.8}\textsubscript{\tiny OPT}}$}
\label{tabelle_ball_diff}
\end{table}

A comparison of the table with Fig. \ref{skal_with_x_ball_to_diff} and the rates of Fig. \ref{skal_min_rates_ball_to_diff_at_bound} delivers some remarkable results:  
\begin{itemize}[align=left,labelwidth=\widthof{i},leftmargin=\labelwidth+\labelsep]
\item In contrast to the case of the BB condition, $x_\text{opt}$ varies a lot in the different optimization scenarios.
\item Comparing the values of $\tilde\gamma_\text{opt}$ with $\tilde\Gamma\text{\hspace*{-0.05cm}\textsubscript{\tiny OPT}}$, one recognizes remarkable changes in the 
transition rates. Especially for $\tilde D<0.05$, it is not possible to find $\tilde\Gamma\text{\hspace*{-0.05cm}\textsubscript{\tiny OPT}}$ as it tends to infinity, 
i.e. the best strategy here is a random velocity search. A particle reaching the boundary, immediately switches to ballistic motion again. The velocity direction 
distribution is a renormalization of $\rho^\alpha_{0.12}$ to the interval $[\pi/2,\pi]$, as $\alpha<\pi/2$ is not possible for particles at the boundary. 
\end{itemize}
For concluding this subsection, Fig. \ref{all_MFPT_in_one_ball_to_diff} shows the optimal MFPTs for the different discussed scenarios. 

\begin{figure*}
\incgraptw{1}{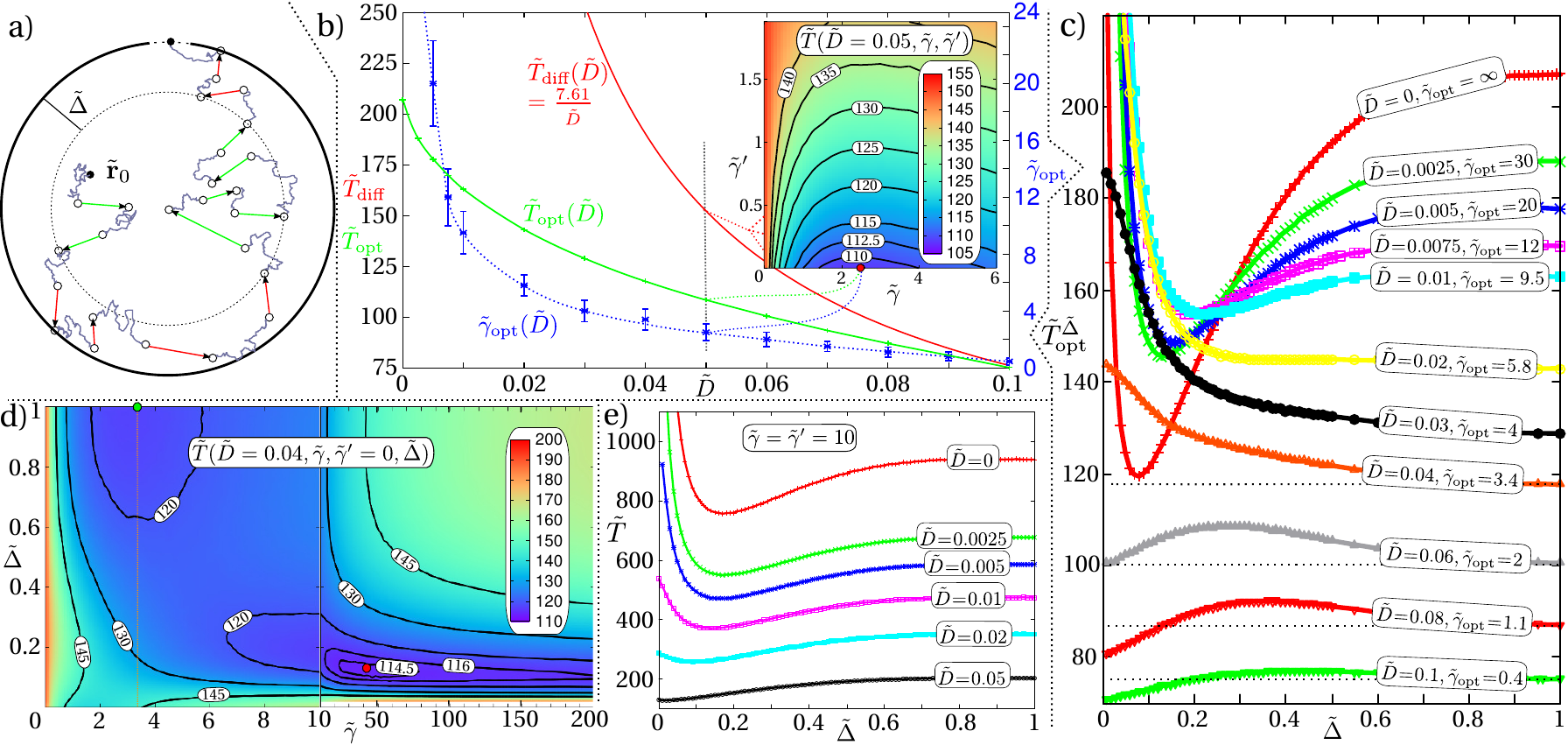}
\caption{
{\bf narrow escape, BD, $\rho_\text{hom}^\alpha$ and $\rho_{p=1,\tilde D}^\alpha$}; 
{\bf{a)}} 
Sketch of the process with the direction distribution $\rho_{p,\tilde\Delta}^\alpha$
with a narrow escape region represented by the dotted segment on the surface of
the spherical search volume (full black circle). Trajectories of the searcher are represented as in Fig. \ref{sketch_cytoskeleton}b.
{\bf{b)}} MFPT for the purely diffusive, ${\tilde T}_{\rm diff}$ (red line),
and intermittent search, ${\tilde T}_{\rm opt}$ (green line), 
with $\rho_\text{hom}^\alpha$ and optimal transition rates, as functions of the
diffusion constant $\tilde{D}$ for $\thetab=\text{arcsin(1/7)}$. The optimal rate 
$\tilde \gamma_\text{opt}(\tilde D)$ is represented by the blue line
(and right y-axis), ${\tilde{\gamma}'}_\text{opt}$ vanishes for all $\tilde D$. 
The inset shows as an example for $\tilde{D}=0.05$ a color plot of the MFPT in 
dependence of the attachment and detachment rates, 
$\tilde{\gamma}$ and $\tilde{\gamma}'$ respectively,
the red dot indicates the optimal values (yielding the minimal 
MFPT) $\tilde \gamma_\text{opt}(\tilde D=0.05)$ and 
${\tilde{\gamma}'_\text{opt}(\tilde D=0.05)}$. 
Optimal attachment / detachment rates like this are used in the main plot 
for varying diffusion constants.
{\bf{c)}} MFPT for the inhomogeneous scenario $\rho_{1,\tilde \Delta}^\alpha$
as a function of $\tilde{\Delta}$ for different diffusion constants $\tilde{D}$
using the optimal rates for the homogeneous scenario. The values at $\tilde{\Delta}=1$ coincide with the data from the blue curve in Fig. b) , as $\rho_{p,\tilde \Delta=1}^\alpha=\rho_\text{hom}^\alpha$.
{\bf{d)}} MFPT as a function of $\tilde \gamma$ and $\tilde \Delta$ for $\tilde D=0.04$, $\tilde{\gamma}'=0$.
The red dot indicates the global minimum ($\tilde T=113.7$) at $\tilde \gamma=42$ and $\tilde \Delta=0.12$. 
the green dot (top left) indicates the minimum for the homogeneous case ($\tilde\Delta=1$, $\tilde T=117.6$).
{\bf{e)}} MFPT as a function of $\tilde \Delta$ for fixed rates 
$\tilde \gamma=\tilde \gamma'=10$ for different $\tilde D$ and $p=1$. 
} 
\label{narrow_escape_letter_scenario}
\end{figure*}

Compared to a purely diffusive searcher (red), an intermittent search strategy with a homogeneous velocity direction distribution (green) optimizes the search process 
especially for small $\tilde D$ significantly, which has already been shown in the inset of Fig. \ref{skal_min_rates_ball_to_diff_at_bound}. This benefit is even more 
pronounced than in the case of BB boundary conditions. Again, in the next step, we introduced an inhomoegeneity in the velocity direction distribution (blue), but kept the optimal 
rate $\tilde\gamma$ of the homogeneous case. The additional benefit is much smaller than it was in the BB case, although the total benefit is still larger. 
In the last step, we varied the transition rate and the degree of inhomoegeneity simultaneously (black). Although the optimal rate again changes dramatically, the 
additional benefit is as small as in the case of the BB condition.

\subsubsection{inhomogeneous distribution $\rho_{p,\tilde\Delta}^\alpha$}
Fig. \ref{narrow_escape_letter_scenario}a) shows a sketch of the class of stochastic first passage processes with direction distribution $\rho_{p,\tilde\Delta}^\alpha$. The initial
position ${\tilde{\mb{r}}_0}$ of the searcher is now homogeneously distributed within the unit sphere. As already mentioned, the reference time $\tilde T_\text{diff}$ is expected to 
decreases slightly by $0.1/\tilde D$ and the optimal rates $\tilde \gamma_\text{opt}$, $\tilde \gamma'_\text{opt}$ are expected to be almost identical to the case of $\tilde{\mb{r}}_0=0$. Hence, in order to avoid 
long repetition of almost identical data, the result for the homogeneous scenario is summarized in Fig. \ref{narrow_escape_letter_scenario}b): $\tilde T_\text{diff}(\tilde D)$ and 
$\tilde T_\text{opt}(\tilde D)$ are almost identical to the corresponding curves of Fig. \ref{all_MFPT_in_one_ball_to_diff}. $\tilde \gamma'_\text{opt}=0$ is also true for a homogeneously
chosen initial position, which shows the inset of the plot exemplarily for $\tilde D=0.05$ (compare Fig. \ref{ball_to_diff_hom}, middle subfigure ) and the values of $\tilde \gamma_\text{opt}$ are identical (within stochastic fluctuations)
to those of Fig. \ref{skal_min_rates_ball_to_diff_at_bound}.  \\

Similar to the procedure in the sections before, the MFPT for the optimal values of $\tilde \gamma_\text{opt}(D)$ is now minimized according to the class parameters $p$ and $\tilde \Delta$. \\

Unsurprisingly, $p_\text{opt}=1$ holds for all values of $\tilde D$. For $p_\text{opt}=1$, the dependence of the MFPT on $\tilde \Delta$ is shown in 
Fig. \ref{narrow_escape_letter_scenario}c) for different values of $\tilde D$. For small values of $\tilde D<0.02$, there is always a minimum for $\tilde\Delta_\text{opt}\in[0.1;0.2]$, i.e. an
inhomogeneous strategy is favorable. 
For $\tilde D>0.06$, $\tilde\Delta_\text{opt}=0$ holds, i.e. the velocity direction of the ballistic motion should always be chosen radially to the outside for all switching positions.
For $0.02<\tilde D<0.06$ the minimum is at $\tilde\Delta=1$, i.e. a homogeneous strategy seems to be optimal in this interval. In order 
to verify this statement, we varied $\tilde \gamma$ and $\tilde \Delta$ simultaneously. Exemplarily, the result for $\tilde\Delta=0.04$ is shown in Fig. \ref{narrow_escape_letter_scenario}d).
The dotted line corresponds to the orange ($\tilde D=0.04$) curve of subfigure c), i.e. the green dot indicates the minimum at $\tilde \Delta=1$. However, there is a global minimum for $\tilde\Delta_\text{OPT}\approx0.15$ and $\tilde\gamma_\text{OPT}\approx45$,
indicated by the red dot. Consequently, the most efficient strategy is again inhomogeneous.\\

Up to now, we always minimized according to the rates $\tilde\gamma$ and $\tilde\gamma'$ first in order to demonstrate the efficiency of an inhomogeneous strategy
afterwards for these optimal rates. In real search, however, these rates might be restricted, for example by an upper value for the allowed energy consumption 
or the number of available motor proteins in the case of intracellular search. Consequently, systematic studies on the direction distribution for fixed non-optimal rates, 
motivated by biological data, will also be of interest in further research, but it will go beyond the scope of this publication. However, it should be at least mentioned, that there 
are robust (here: according to changes in $\tilde D$) inhomogeneous strategies, which minimize the MFPT, thus \ref{narrow_escape_letter_scenario}e) shows $\tilde T$ as a function
of $\tilde \Delta$ for $\tilde\gamma=\tilde\gamma'=10$ and different values of $\tilde D$.

\section{reaction kinetics}
\label{target_in_sphere}
Within this section, the efficiency of intermittent search strategies for an immobile target at the interior of the simulation sphere will be studied. The search process will succeed, 
if the distance between the diffusive searcher and the target becomes smaller than a reaction distance $d$ for the first time. This introduces a second length scale to the system (in addition to the radius 
$R$ of the sphere). As the following will stick to the dimensionless units, introduced in the equations (\ref{entdimensionalisierung}) and (\ref{skaling_of_D_gamma}), we 
additionally define 
\begin{equation}
 \tilde d=\frac{d}{R},
\end{equation}
which is the reaction distance in the dimensionless units.  Within this section we will again study BB and BD boundary conditions for a ballistically moving particle. 
For the reason of comparison to other publications, we take BB boundary conditions. On the other hand the studies in case of the inhomogeneity $\rho_{p,\tilde\Delta}^\alpha$
appear more meaningful with BD conditions. But for small $\tilde d$ the results differ only very less. Thus the results are almost independent on the applied boundary condition,
which is in contrast to the narrow escape problem.\\
We will study and compare different scenarios for the target position. In subsection \ref{target_in_center_subsection} the target is centered 
in the middle of the simulation sphere and the boundary conditions BB are applied for the reason of comparison. Afterwards, subsection \ref{target_hom_verteilt_subsection} faces the problem of a homogeneously randomly chosen target position, again with the boundary conditions BB. 
Finally, in subsection \ref{inhomogeneously distributed random target position} the scenario of an inhomogeneously distribution of the target position is discussed for the
direction distribution $\rho_{p,\tilde\Delta}^\alpha$ and BD boundary conditions.
\subsection{target in the center of the sphere}
\label{target_in_center_subsection}
Due to the radial symmetry of the problem an analytic expression for the reference time $\tilde T_\text{diff}$ can easily be derived for a searcher, 
starting at radius $\tilde r_0> \tilde d$ by solving the boundary value problem:
\begin{eqnarray}
 \frac{1}{\tilde r_0^2}\pd{\tilde r_0} \left( \tilde r_0^2 \pd{\tilde r_0} \tilde T_\text{diff}(\tilde d,\tilde r_0)\right)=\frac{-1}{\tilde D}\quad\text{ with}\quad\quad\\
 T_\text{diff}(\tilde d,\tilde d)=0\quad\text{and}\quad \pd{\tilde r_0} \tilde T_\text{diff}(\tilde d,\tilde r_0)|_{\tilde r_0=1}=0\;\quad \\
 \Rightarrow\quad \tilde T_\text{diff}(\tilde d,\tilde r_0)=\frac{- \tilde d \tilde r_0^3  +  \kr{2+\tilde d^3}\tilde r_0  -2 \tilde d}{6\tilde D\tilde d \tilde r_0}\;.\quad\quad\quad \label{loverdo_r_0_start}
\end{eqnarray}
In this section, the initial position of the searcher will always be homogeneously distributed in the spherical shell given by $\tilde d<\tilde r_0<1$. The reference MFPT $\tilde T_\text{diff}$ of the purely
diffusive searcher $\tilde T_\text{diff}$ will then be given by
\begin{eqnarray}
 \tilde T_\text{diff}(\hs{-0.1}\tilde d\hs{0.2})\hs{-0.5}=\hs{-1.5}\int_{\tilde d}^1 \hs{-2} d{\tilde r}_0\frac{3{\tilde r_0}^2\cdot \tilde T_\text{diff}(\tilde d,\tilde r_0\hs{-0.5})}{1-\tilde d^3}\hs{-0.2}=\hs{-0.2}\frac{5\hs{-0.2}-\hs{-0.2}9\tilde d\hs{-0.2}+\hs{-0.2}5\tilde d^3\hs{-0.2}-\hs{-0.2}\tilde d^6}{15 \tilde D\kr{1-{\tilde{d}}^3}\tilde d}\;. \label{loverdo_hom_start}
\end{eqnarray}
It is plotted in Fig. \ref{Loverdo_not_center_only_diff_two_plot} (red line). In order to check and prove the accuracy of our numerical method for this 
scenario we simulated $\tilde T_\text{diff}$ for $\tilde d=0.2$ and $\tilde d=0.025$, as these values of $\tilde d$ will be used in the following:
\begin{align}
\tilde T_\text{diff}(0.2)&=\frac{12656}{11625 \tilde D}&\approx \frac{1.08869}{\tilde D}& \nnn
\tilde T_\text{diff}^\text{num}(0.2)&=&\frac{1.08857}{\tilde D}&\quad (10^7 \text{samples)} \nnn
\tilde T_\text{diff}(0.025)&=\frac{55722849}{4376000 \tilde D}&\approx \frac{12.7337}{\tilde D} \nnn
\tilde T_\text{diff}^\text{num}(0.025)&=&\frac{12.7331}{\tilde D}&\quad (5\cdot10^6 \text{samples)} \non
\end{align}
In both cases the relative deviation is smaller than 0.02 \%, which is in the range of the statistical error. We expect the results reported below to have the same numerical accuracy.

\begin{figure*}
\incgraptw{1}{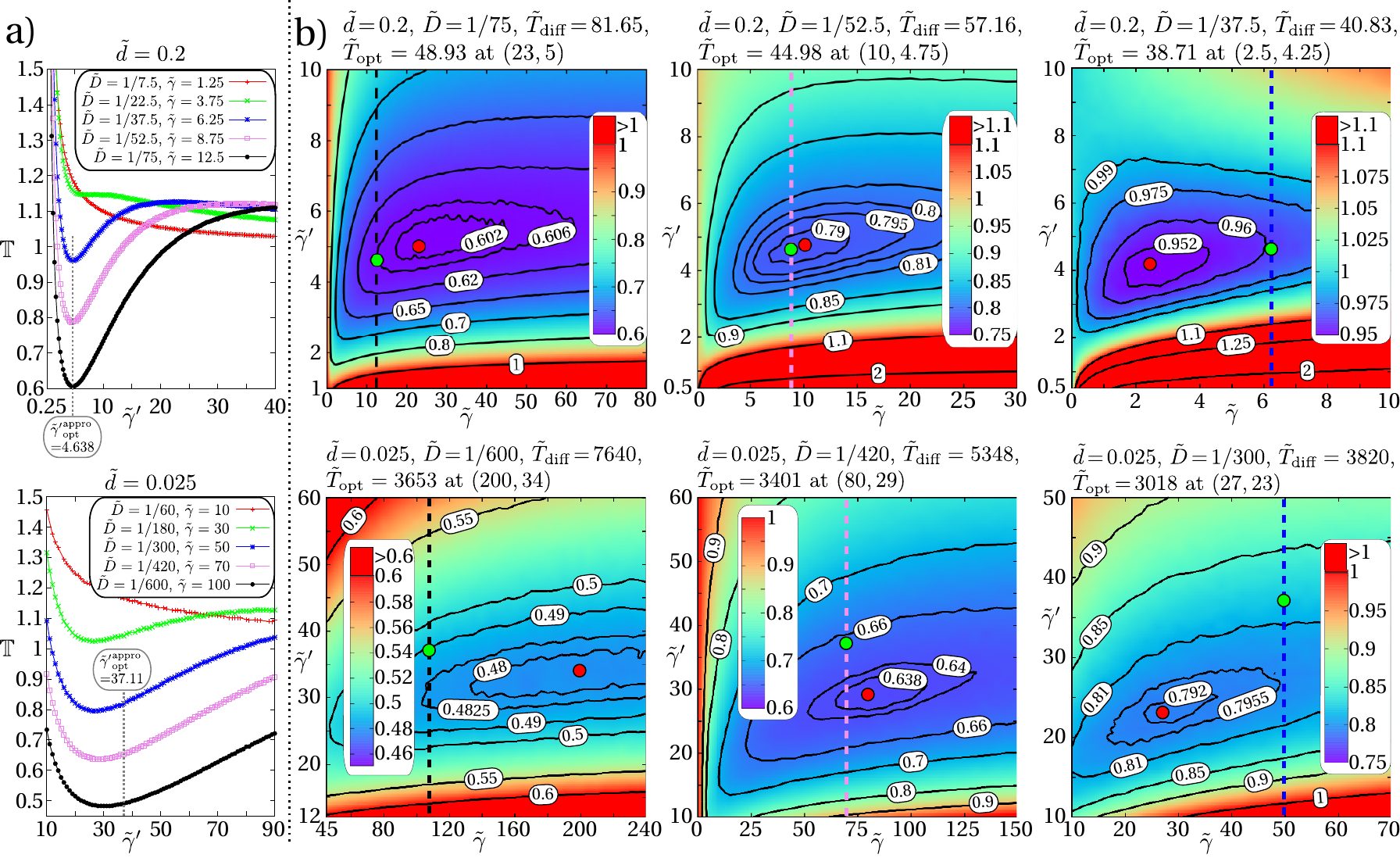}
\caption{{\bf reaction kinetics, $\tilde{\mb{r}}_\text{tar}=0$, $\rho_\text{hom}^\alpha$}; The upper line shows data for $\tilde d=0.2$ ($2\cdot 10^6$ samples per data point) , the lower line for $\tilde d=0.025$ ($5\cdot 10^5$ - $2\cdot 10^6$ 
samples per data point). {\bf a)} $\skaltime$ as a function of $\tilde\gamma'$ for different values of $\tilde D$ with $\tilde\gamma=\tilde\gamma_\text{opt}^\text{appro}=1/(6 \tilde D)$.
The dotted gray lines show the position of the suggested minimum $\tilde\gamma{'}^\text{appro}_\text{opt}$.
{\bf b)} $\skaltime$ is color-coded as a function of $\tilde\gamma$ and $\tilde\gamma'$ for the three 
smallest values of $\tilde D$ in subfigure a). The position of the global minimum $(\tilde\gamma_\text{opt},\tilde\gamma'_\text{opt})$ is always shown with a red dot. 
Each colored vertical dotted line corresponds to the curve of subfigure a) in the same color. The green dot shows the position of the proposed optimal rates 
$\tilde\gamma_\text{opt}^\text{appro}, \tilde\gamma{'}^\text{appro}_\text{opt}$ by 
\cite{Benichou2011, Loverdo2008, Loverdo2009}.
} 
\label{Loverdobestaetigen}
\end{figure*}

\subsubsection{homogeneous distribution  $\rho_\text{hom}^\alpha$}
The studies of \cite{Benichou2011, Loverdo2008, Loverdo2009} already considered the intermittent search problem for the homogeneously distributed 
velocity direction distribution $\rho_\text{hom}^\alpha$ and a target centered in the middle of the sphere. Approximating expressions  
for the transition rates $\gamma_\text{opt}$ and $\gamma'_\text{opt}$ of the search problem were derived there: The dependence of the MFPT 
$T$ on the rate $\gamma$ is claimed to be very weak and $\gamma_\text{opt}^\text{appro}=v^2/(6 D)$ to be a good guess for the optimal switching rate 
from diffusive to ballistic motion. For the optimal rate from ballistic to diffusive motion, their approximative calculations deliver 
${\gamma'}_\text{opt}^\text{appro}=v/(1.078 d)$. In the nondimensional coordinates of this article, this relations are transformed to 
\begin{equation}
\tilde\gamma{'}^\text{appro}_\text{opt}=1/(1.078 \tilde d)\quad\text{and}\quad \tilde\gamma_\text{opt}^\text{appro}=1/(6 \tilde D)\;. \label{loverdo_predict}
\end{equation}
The numeric simulations of \cite{Benichou2011, Loverdo2008, Loverdo2009} do not show a simultaneous variation of the two rates, as $\gamma$
is always set to the assumed optimal value $\gamma_\text{opt}^\text{appro}$. \\

We now study this scenario more extensively. For the reason of comparison to their results, we investigate the cases 
$\tilde d=0.2$ ($\tilde D=$1/7.5, 1/22.5, 1/37.5, 1/52.5, 1/75) and $\tilde d=0.025$ ($\tilde D=$1/60, 1/180, 1/300, 1/420, 1/600) 
as these nondimensional values correspond to the geometry parameters of their studies. \\ 
Fig. \ref{Loverdobestaetigen}a shows $\skaltime$ as a function of $\tilde\gamma'$ for $\tilde d=0.2$ and  $\tilde d=0.025$ and $\tilde\gamma=\tilde\gamma_\text{opt}^\text{appro}$.
These results agree with the numerical results of \cite{Benichou2011, Loverdo2008, Loverdo2009}, when rescaling our plots and plotting them in the same manner than the data of 
their publications. 
For $\tilde d=0.2$ the position of the minimum is in agreement to $\tilde\gamma{'}^\text{appro}_\text{opt}$. For $\tilde d=0.025$ there are already deviations visible. 
Next, we varied the rates simultaneously. Our simulations confirm the very weak dependence on $\tilde\gamma$. Nevertheless, $\tilde\gamma_\text{opt}$ and $\tilde\gamma_\text{opt}'$ do 
not seem to scale exactly like predicted in Eq. \ref{loverdo_predict}. Fig. \ref{Loverdobestaetigen}b show this for the three smallest diffusion coefficients that 
we have studied for the same values of $\tilde{d}$ as in Fig. \ref{Loverdobestaetigen}a. Although $\skaltime_\text{opt}$ (red dots) is less than 2$\%$ smaller than the suggested 
minima (green dots), it is nevertheless stochastically significant enough to claim a deviation in the optimal rates. 
For small values of $\tilde D$, $\tilde\gamma_\text{opt}$ is larger than $\tilde\gamma^\text{appro}_\text{opt}$, for large values of $\tilde D$ it becomes smaller. Furthermore, the optimal
value of $\tilde\gamma_\text{opt}$ seems not to be independent of $\tilde D$, as it slightly decreases with increasing diffusivity. \\
Nevertheless, although the approximations $\tilde\gamma{'}^\text{appro}_\text{opt}$ and $\tilde\gamma_\text{opt}^\text{appro}$ sometimes differ essentially from simulated minima, they always 
define a very good search strategy, which is close to the optimal one, as the corresponding MFPT is always very close to $\tilde T_\text{opt}$. 
\subsubsection{optimal inhomogeneous distribution of $\rho^\alpha$}
Similar to the narrow escape problem in section \ref{narrow_escape}, there are more efficient velocity direction distributions than the homogeneous distribution. 
For a target located in the center of the sphere the optimal intermittent search strategy is obvious: The starting direction of a ballistically moving particle 
is always chosen to point to the origin, i.e.
\begin{eqnarray}
 \rho^\alpha(\alpha)=\rho_{p=0,\tilde \Delta=0}^\alpha=\delta(\alpha-\pi).
\end{eqnarray}

For this setup, there are no finite values for $\tilde \gamma_\text{opt}$ and $\tilde \gamma'_\text{opt}$. As the ballistic motion happens only radially and always directed to 
the center, it is possible to construct a ballistic motion with target detection. For $\tilde \gamma \rightarrow \infty$, $\tilde \gamma' \rightarrow \infty$ with 
$\tilde \gamma/\tilde \gamma'\rightarrow 0$ the particle switches infinitely often between diffusion and ballistic motion within every time period. Nevertheless, it moves 
like a ballistic particle. In consequence, for a fixed starting radius $\tilde r_0$, we simply get $\tilde T(\tilde r_0)=\tilde r_0-\tilde d$. With the help of the 
Eqs. (\ref{loverdo_r_0_start}-\ref{loverdo_hom_start}) analytic expressions for $\skaltime$ and the break-even value $\tilde D_\text{be}$ for a switch from a purely diffusive search to an 
intermittent search (here: ballistic search) can be derived, but will be skipped here. 
\subsection{homogeneously distributed random target position }
\label{target_hom_verteilt_subsection}
\begin{figure}[htb!]
\incgrapw{0.9}{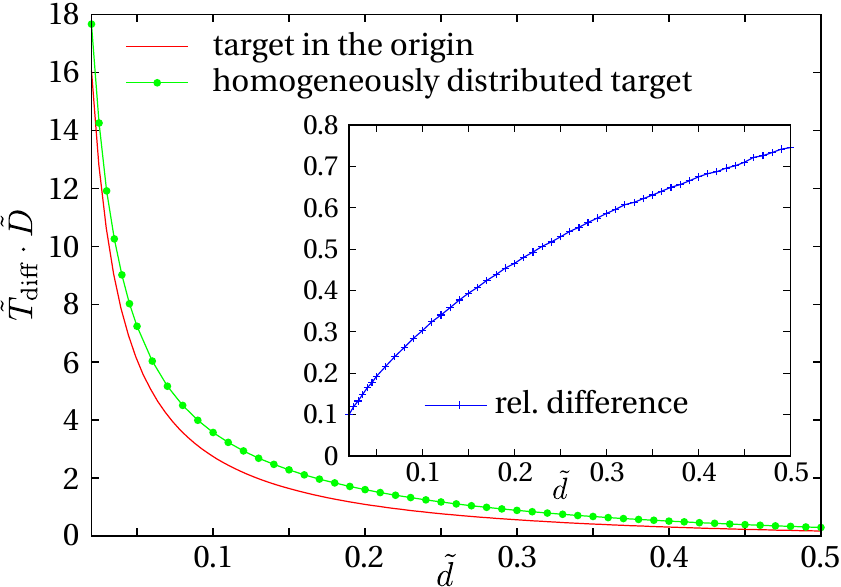} 
\caption{$\tilde T_\text{diff}\cdot\tilde D$ as a function of the reaction distance $\tilde {d}$. {\bf{Red:}} Target in the center of the 
sphere,  plot of Eq. \ref{loverdo_hom_start}. {\bf{Green:}} Target homogeneously distributed, each green dot is the average value over $2\cdot 10^6$ 
Monte Carlo samples. {\bf{Blue}} (inset): relative difference of the green and the red curve.}
\label{Loverdo_not_center_only_diff_two_plot}
\end{figure}

\begin{figure}[htb!]
\incgrapw{0.9}{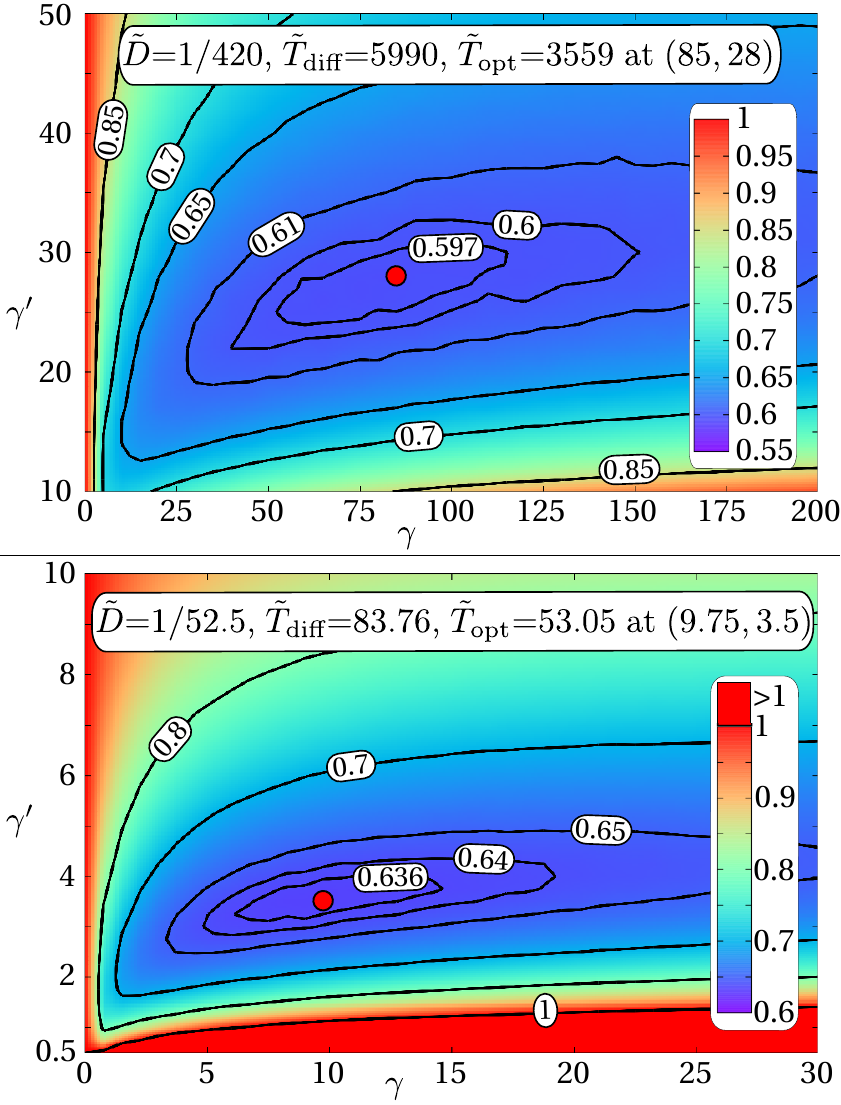}
\caption{{\bf reaction kinetics, ${\tilde{\mb{r}}}_\text{tar}$ eq. distributed, $\rho_\text{hom}^\alpha$}; $\skaltime$ is color-coded plotted as a function of $\tilde\gamma$ and $\tilde\gamma'$  
 for the same $\tilde D$ and $\tilde d$ than in the middle column of Fig. \ref{Loverdobestaetigen}b. ({\bf top}: $\tilde D=1/420$, $\tilde d=0.025$,
 {\bf bottom}: $\tilde D=1/52.5$, $\tilde d=0.2$). $2-4\cdot 10^6$ samples have been done for each data point.
 The position of the global minimum $(\tilde\gamma_\text{opt},\tilde\gamma'_\text{opt})$ is always shown with a red dot.}
\label{Loverdo_not_center_example}
\end{figure}

The position $\tilde {\bm r}_\text{tar}$ of the target is homogeneously distributed in a sphere of radius $1-\tilde d$. The initial position $\tilde{\bm r}_0$ of the searcher 
is homogeneously distributed in the unit sphere with the restriction $|| {\bm r}_\text{tar}- \tilde{\bm r}_0||>\tilde d$. Compared to the situation of a target in the center 
of the sphere, the reference time $\tilde T_\text{diff}$ slightly increases and the relative difference increases monotonically with $\tilde d$, which can both be seen 
in Fig. \ref{Loverdo_not_center_only_diff_two_plot}. 
For the reason of comparison to the subsection before, we analyzed the same parameters $\tilde d$ and $\tilde D$ as in Fig. \ref{Loverdobestaetigen}b. 
Exemplarily the results for $\tilde d=0.2$, $\tilde D=1/52.5$ and $\tilde d=0.025$, $\tilde D=1/420$ are shown in Fig. \ref{Loverdo_not_center_example}. 
For all investigated cases, the distribution of the target position changes the value of $\tilde\gamma_{\text{opt}}(\tilde D)$ only very less and within the stochastic 
fluctuations, i.e. $\tilde\gamma_{\text{opt}}(\tilde D)$ seems to depend only on $\tilde D$ and $\tilde d$. For small values of $\tilde d$, this is also true for  
$\tilde\gamma'_{\text{opt}}(\tilde D )$. For larger values of $\tilde d$, $\tilde\gamma'_{\text{opt}}(\tilde D )$ decreases in the case of a homogeneously distributed target 
position. Compared to Fig. \ref{Loverdobestaetigen}, $\tilde T_{\text{opt}}$ is larger for all
$\tilde d$ and $\tilde D$. As this increase is smaller than the increase in $\tilde T_\text{diff}$, $\skaltime_\text{opt}$ decreases. For $\tilde d=0.025$ this 
decrease is only about 7$\%$, for $\tilde d=0.2$ it is already about 20 $\%$. \\

We find that for a homogeneously distributed target position, there is no gain in an inhomogeneously distributed $\rho_\mb{v}(\Omega|\tilde{\mb{r}})$.

\begin{figure*}[t]
\includegraphics[width=\textwidth]{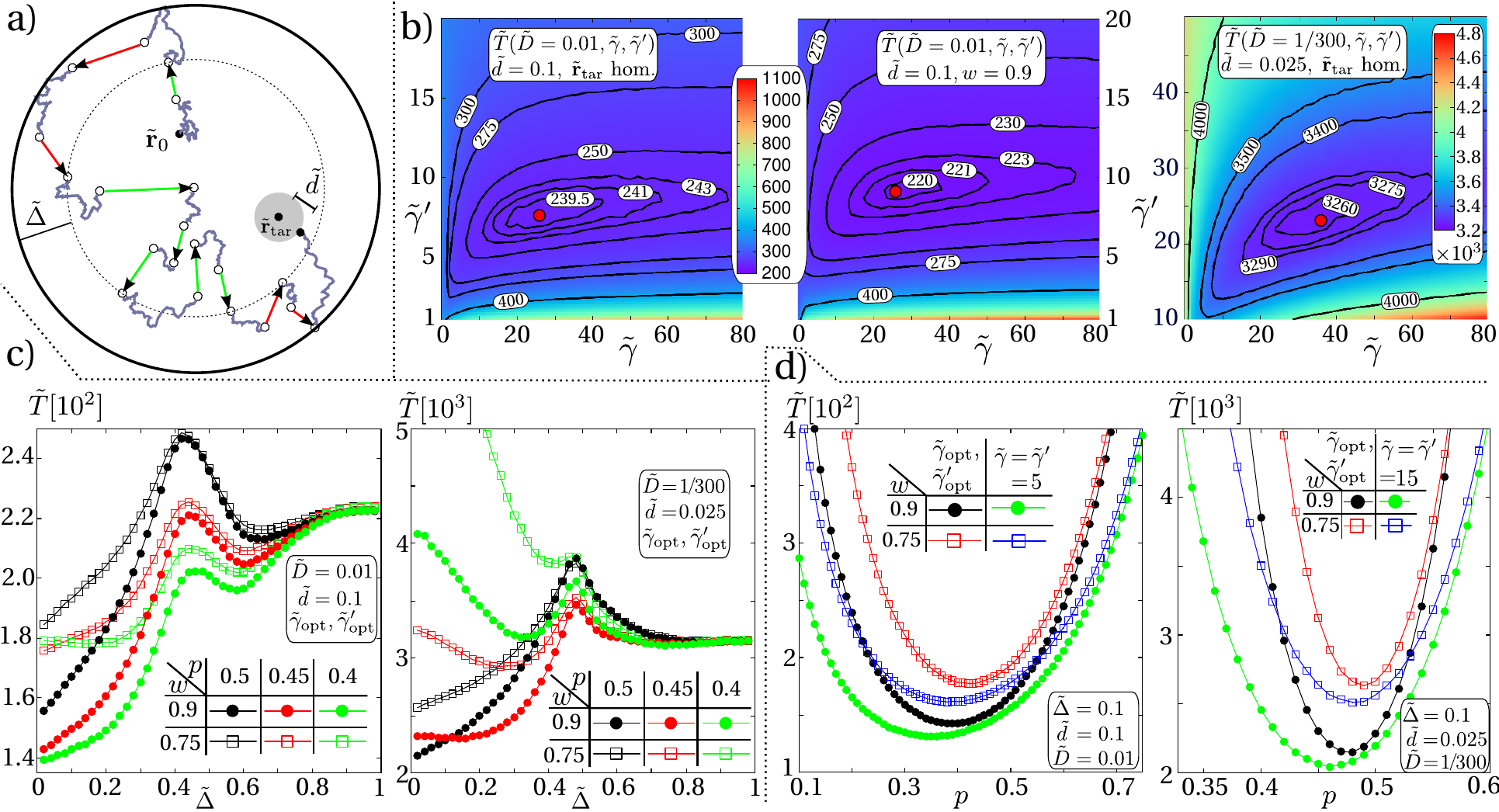} 
\caption{
{\bf reaction kinetics, ${\tilde{\mb{r}}}_\text{tar}$ according to $\rho_w^\text{init}(\tilde r_\text{tar})$ , $\rho_\text{hom}^\alpha$ and $\rho_{p,\tilde\Delta}^\alpha$}; {\bf{a)}}
Sketch of the reaction process for $\rho_{p,\tilde \Delta}^\alpha$ with an immobile target of
diameter $\tilde{d}$ indicated by the gray circle. Trajectories of the 
mobile particle (searcher), starting at ${\tilde {\bf r}_0}$, are represented  
as in Fig.\ref{sketch_cytoskeleton}b. 
{\bf{b)}} Homogeneous direction distribution:
MFPTs as a function of $\tilde \gamma$ and $\tilde \gamma'$ for a spatially homogeneous scenario  $\rho_\text{hom}^\alpha=\rho_{p,1}^\alpha$, for
({\it left}) $\tilde D=0.01, \tilde d=0.1$, homogeneously distributed target position $\tilde {\bf r}_\text{tar}$; 
({\it middle}) $\tilde D=0.01, \tilde d=0.1$, 
 target position $\tilde {\bf r}_\text{tar}\leq 0.5$ with probability $w=0.9$;  ({\it right}) $\tilde D=1/300, \tilde d=0.025$,  
 homogeneously distributed target position.  
{\bf{c)}} MFPT for the inhomogeneous distribution $\rho_{p,\tilde\Delta}^\alpha$
with the optimal rates $\tilde \gamma_\text{opt}(\tilde D, \tilde d)$, $\tilde \gamma'_\text{opt}(\tilde D, \tilde d)$
from the homogeneous case $\tilde{\Delta}=1$ as function of 
$\tilde{\Delta}$ for different values of the forward radial transport $p$
and different $w$.
{\bf{d)}} MFPT as in c) but now with fixed width $\tilde{\Delta}=0.1$ 
as function of the forward probability $p$ for different 
fixed rates $\tilde\gamma$ and $\tilde\gamma'$ and different values of $w$.
} 
\label{Fig_enhanc_reac_kin_rho_p_delta}
\end{figure*}

\subsection{inhomogeneously distributed random target position}
\label{inhomogeneously distributed random target position}
In the interesting case of a small area where the immobile target is predominantly placed, the best search strategy in not obvious anymore. On the on hand, the searcher should
be prevalent in the surrounding of this area. On the other hand, it can't stay there exclusively, as the target might be somewhere else with a non vanishing probability. 
This situation shall be studied now for the following distribution of $\tilde r_\text{tar}=||\tilde {\bm r}_\text{tar}||$:
\begin{flalign}
&\rho_w^\text{init}(\tilde r_\text{tar})=\left\lbrace{\small \begin{matrix} 
 24w\cdot \tilde r_\text{tar}^2&,&  0\leq \tilde r_\text{tar}\leq\frac{1}{2} \vspace*{0.1cm}
\\ \frac{24(1-w)}{8\kr{1-\tilde d}^3-1}\tilde r_\text{tar}^2&,&  \frac{1}{2}< \tilde r_\text{tar}\leq 1-\tilde d \vspace*{0.1cm}
\end{matrix}} \right.\;, \label{inhomrtar}
\end{flalign}
i.e. with probability $w$, the particle is homogeneously distributed in a sphere of radius $1/2$ around the origin, with  probability $1-w$, the particle is homogeneously
distributed in the outer region.\\
The initial position  $\tilde{\bm r}_0$ of the searcher is again homogeneously distributed in the unit sphere with the restriction $|| \tilde{\bm r}_\text{tar}- \tilde{\bm r}_0||>\tilde d$.
Fig. \ref{Fig_enhanc_reac_kin_rho_p_delta}a) shows a sketch of the resulting stochastic first passage process for the direction distribution $\rho_{p, \tilde \Delta}^\alpha$. 
Within this section, we exemplarily study the parameter sets $\tilde d=0.1,\tilde D=0.01$ and $\tilde d=0.025,\tilde D=1/300$. Like before, we first face the scenario of a
homogeneous search strategy in order to quantify the values of $\tilde\gamma_\text{opt}$ and $\tilde\gamma'_\text{opt}$. For small values of $\tilde d$ these rates are almost independent
on $w$, which can be seen by comparing the left and the middle subfigure of Fig. \ref{Fig_enhanc_reac_kin_rho_p_delta}b) for $\tilde d=0.1$. For $\tilde d=0.025$ the differences in the rates
totally vanishes within the stochastic fluctuations, hence, only the scenario of a homogeneous initial target position is shown for this case in the right subfigure. \\

Like before, the influence of the inhomogeneous direction distribution $\rho_{p,\tilde \Delta}^\alpha$ is studied for the optimal values 
$\tilde \gamma_\text{opt}(\tilde D, \tilde d)$ and $\tilde \gamma'_\text{opt}(\tilde D, \tilde d)$. But due to the computational effort we did not
minimize according to $p$ and $\tilde \Delta$ in parallel. First, $\tilde T$ is minimized according to $\tilde \Delta$ for three different values of $p$
for two different $w$. The corresponding plots are shown in Fig. \ref{Fig_enhanc_reac_kin_rho_p_delta}c). For $\tilde d=0.1$ an inhomogeneous strategy is more efficient
for all investigated values of $0.4\leq p\leq 0.5$. For the smaller detection distance $\tilde d=0.025$ an inhomogeneous strategy is also preferable, but only in the range of 
$p\approx 0.5$. Like in the section before, the optimal values of $\tilde \gamma_\text{OPT}$ and $\tilde \gamma_\text{OPT}'$ in case of an inhomogeneous strategy might strongly differ from 
$\tilde \gamma_\text{opt}$ and $\tilde \gamma'_\text{opt}$. We did not calculate the optimal strategy  $\tilde \gamma_\text{OPT}$, $\tilde \gamma_\text{OPT}'$, $p_\text{OPT}$, $\tilde \Delta_\text{OPT}$ 
explicitly due to the enormous numerical effort of minimizing according to four parameters. Instead Fig. \ref{Fig_enhanc_reac_kin_rho_p_delta}d) exemplarily shows the dependence 
on $p$ for a fixed value of $\tilde \Delta=0.1$ for the rates $\tilde \gamma_\text{opt}(\tilde D, \tilde d)$ and $\tilde \gamma'_\text{opt}(\tilde D, \tilde d)$ and 
chosen transition rates. For both $\tilde d$ and both values of $w$ the MFPT of the chosen parameters is always beneath the MFPT for $\tilde \gamma_\text{opt}$, $\tilde \gamma'_\text{opt}$. 
As already seen in subfigure c), the dependence on $p$ increases for smaller $\tilde d$, i.e. the minima are stronger pronounced and the optimal value of $p$ tends to 0.5 (independent on $w$).

\begin{figure*}[t]
\includegraphics[width=\textwidth]{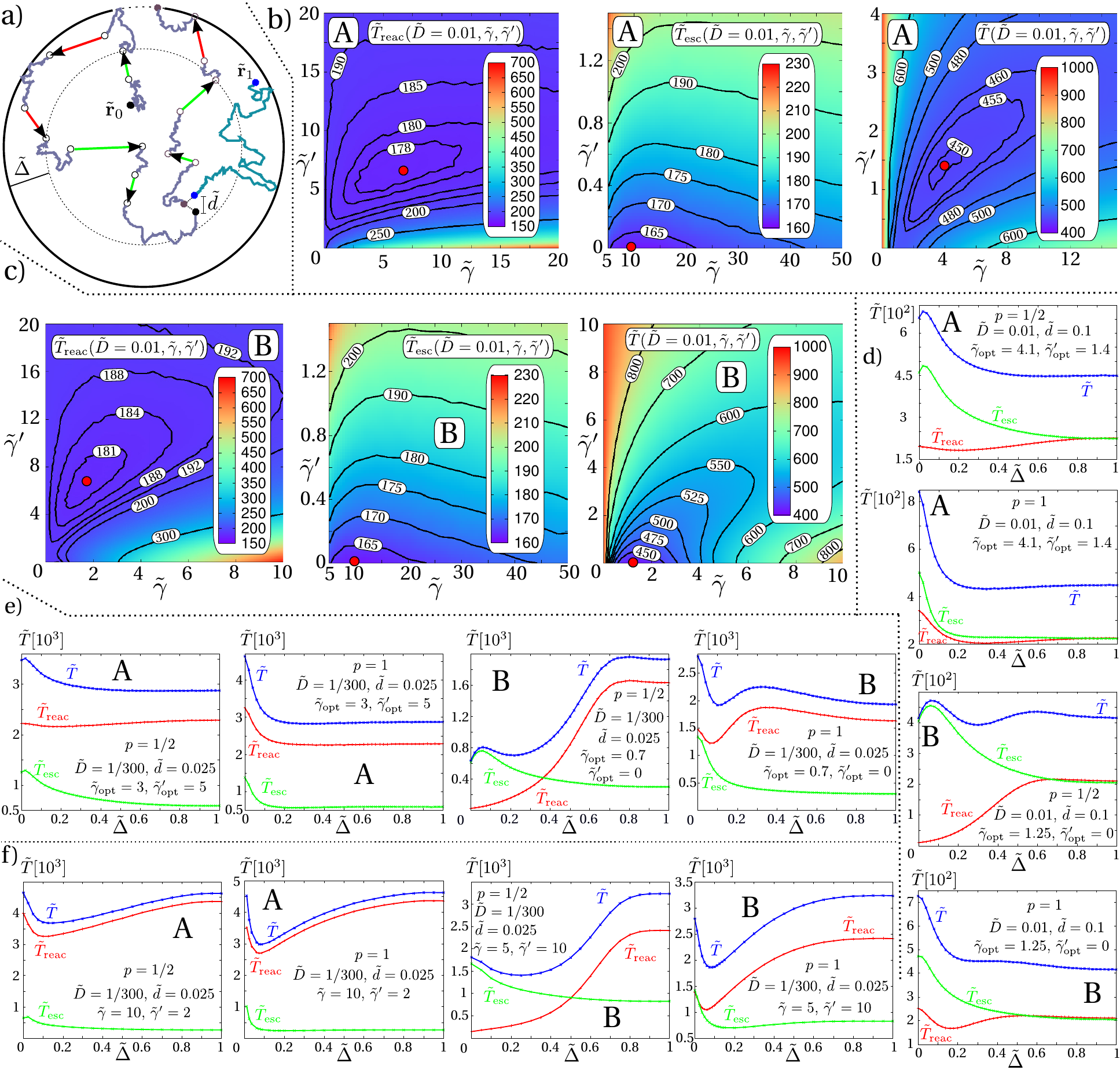} 
\caption{
\textbf{reaction-escape, BD, $\rho_\text{hom}^\alpha$ and $\rho_{p,\tilde\Delta}^\alpha$}; {\bf{a)}}
Sketch of the reaction-escape process, involving
two particles, an intermittently searching particle (black) and a diffusive target (blue particle), both starting diffusively at random positions $\tilde{\mb{r}}_0$  and $\tilde{\mb{r}}_1$. 
Absorption at the narrow escape region is only possible for searcher-target pair, and the two particles react when coming closer than a distance $\tilde d$ and form a pair (brown
particle), which will be absorbed at the escape region represented by the dotted segment on the boundary.
{\bf{b)}}
The MFPTs $\tilde T_\text{reac}, \tilde T_\text{esc}$ and the resulting $\tilde T$ are color coded for the homogeneous direction distribution $\rho_\text{hom}^\alpha$ for situation A
and the parameters $\tilde d=0.1$ $\tilde D=0.01$.
{\bf{c)}}
The MFPTs $\tilde T_\text{reac}, \tilde T_\text{esc}$ and the resulting $\tilde T$ are color coded for the homogeneous direction distribution $\rho_\text{hom}^\alpha$ for situation B
and the parameters $\tilde d=0.1$ $\tilde D=0.01$.
{\bf{d)}}
The MFPTs $\tilde T_\text{reac}, \tilde T_\text{esc}$ and the resulting $\tilde T$ as a function of $\tilde \Delta$ in case of the inhomogeneous 
direction distributions $\rho_{1/2,\tilde \Delta}^\alpha$ and $\rho_{1,\tilde \Delta}^\alpha$ for the situations A and B with parameters $\tilde d=0.1$ $\tilde D=0.01$
and the optimal transition rates belonging to the corresponding homogeneous direction distribution scenario (position of red dots in the right figures of b) and c) ).  
{\bf{e)}}
The MFPTs $\tilde T_\text{reac}, \tilde T_\text{esc}$ and the resulting $\tilde T$ as a function of $\tilde \Delta$ in case of the inhomogeneous 
direction distributions $\rho_{1/2,\tilde \Delta}^\alpha$ and $\rho_{1,\tilde \Delta}^\alpha$ for the situations A and B with parameters $\tilde d=0.025$ $\tilde D=1/300$
and the optimal transition rates belonging to the corresponding homogeneous direction distribution scenario (data not shown).  
{\bf{f)}}
The MFPTs $\tilde T_\text{reac}, \tilde T_\text{esc}$ and the resulting $\tilde T$ as a function of $\tilde \Delta$ in case of the inhomogeneous 
direction distributions $\rho_{1/2,\tilde \Delta}^\alpha$ and $\rho_{1,\tilde \Delta}^\alpha$ with parameters $\tilde d=0.025$ $\tilde D=1/300$
and the transition rates $\tilde\gamma=10$,$\tilde\gamma'=2$  (A) , respectively $\tilde\gamma=5$,$\tilde\gamma'=10$  (B).
}
\label{Fig_reac_esc_290416}
\end{figure*}

\section{Reaction-Escape Problem}
\label{pred-prey}
Finally, we study the influence of an inhomogeneous search strategy to a combination of a reaction- and an escape problem for the BD boundary condition. An intermittently 
searching particle is looking for a mobile particle, which will be found if the searcher and the mobile particle are in the diffusive phase and the particles distance is smaller or 
equal $\tilde d$. Afterwards the particle complex has to solve the narrow escape problem ($\thetab=\text{arcsin}(1/7)$) (see section \ref{narrow_escape}) with the same search strategy. \\
A
In the following, two possibilities for the target particle will be studied. In the first case, the target particle moves only diffusively with $\tilde D$. Fig. \ref{Fig_reac_esc_290416}a) 
sketches this situation of the resulting stochastic first passage process for the direction distribution $\rho_{p, \tilde \Delta}^\alpha$. This scenario is always denoted by 
situation "A'' in the following. In the second case, there is no difference between the searcher and the target. Both are intermittent searchers. This situation is denoted by "B''.\\

The total MFPT $\tilde T$ to the escape area at the boundary is the sum of the mean reaction time $\tilde T_\text{reac}$ and the mean escape time $\tilde T_\text{esc}$
for the final narrow escape problem:
\begin{equation}
\tilde T=\tilde T_\text{reac}+\tilde T_\text{esc}\;.
\end{equation}
It is no surprise, that we are dealing with a frustrated problem, i.e. the optimal rates for the reaction problem differ from the optimal rates of the narrow escape scenario.
Exemplarily  Fig. \ref{Fig_reac_esc_290416}b) shows this in case of the homogeneous direction distribution for $\tilde d=0.1$, $\tilde D=0.01$ in situation A.  
Furthermore, the optimal rates depend on the considered situation (A or B) , Fig. \ref{Fig_reac_esc_290416}c) shows the same data as subfigure b), but this time for the situation B.   
However the plots for $\tilde T_\text{esc}$ (middle in subfigure b) and c), see also Fig. \ref{ball_to_diff_hom} at bottom) are almost identical. The only difference in the 
investigated narrow escape processes in situation A
and B is the distribution of the starting position, as the spatial likelihood of the reaction position differs from A to B. But it has already been shown, that 
this influence is neglectable for $\vartheta_{abso}=\text{arcsin}(1/7)$. In both situations, for the chosen parameters $\tilde d=0.1$ and $\tilde D=0.01$ the addends 
$\tilde T_\text{reac}$ and $\tilde T_\text{esc}$ contribute roughly equal to the sum $\tilde T_\text{opt}$.\\
Like in the sections before, we now varied $\tilde \Delta$ for the optimal rates of the homogeneous scenario for $p=1/2\, ,\, 1$ in the situations A and B. The four results are shown in Fig.
\ref{Fig_reac_esc_290416}d). $\tilde T_\text{reac}$ is always minimized by an inhomogeneous strategy (especially in situation B).
Surprisingly, there is no or almost no a gain in
an inhomogeneous strategy for the MFPT $T$, as the MFPT $\tilde T_\text{esc}$ is always minimized by a homogeneous strategy for the chosen rates.\\
It raises the question whether an inhomogeneous strategy might be more favorable in scenarios, where $\tilde T_\text{reac}$ is much larger than $\tilde T_\text{esc}$.
For answering it, we decreased $\tilde d$ and investigated the parameters $\tilde D=1/300$, $\tilde d=0.025$. First, the optimal rates of the homogeneous scenario were 
determined in the situations A and B. As the corresponding figures qualitatively look like Fig. \ref{Fig_reac_esc_290416}b) and c), these plots are skipped here. 
Similar to the parameter set before, we now varied $\tilde \Delta$ for these optimal rates of the homogeneous scenario for $p=1/2\, ,\, 1$ in the situations A and B, 
the result is shown in \ref{Fig_reac_esc_290416}e). In situation A, there is again no or only little gain in an inhomogeneous strategy. For B, there is an enormous 
gain for p=1/2, and a small one for p=1. \\    
Due to the number of simulations, we did not minimize the rates $\tilde \gamma, \tilde \gamma'$ and the inhomoegeneity parameters 
$p$ and $\tilde D$ simultaneously. Instead, in Fig \ref{Fig_reac_esc_290416}f) we show examples for the variation of $\tilde \Delta$ for rates, which do not optimize 
the homogeneous scenario. In all cases, the MFPT for small $\tilde\Delta$ is significantly less than in the homogeneous scenario ($\tilde\Delta=1$). For p=1 in situation B, 
the value of the inhomogeneous minimum is even a little bit smaller than the optimal value of the scenario with the rates $\tilde\gamma_\text{opt}$ and $\tilde\gamma_\text{opt}$,
compare right figures of subfigure e) and f) . In consequence, the optimal strategy is for sure also an inhomogeneous one, at least in this scenario.

\section{Summary}
In this work we have studied the efficiency of spatially homogeneous and inhomogeneous intermittent search strategies for three paradigmatic search problems in
spheres: narrow escape problem, reaction kinetics and the reaction-escape problem. Our results are obtained by an event driven Monte Carlo algorithm, which has recently been 
published \cite{schwarz2013}. The working horses of this algorithm are sampling routines which depend on the geometry of the search domain under consideration. We developed 
highly efficient sampling routines for spherical domains, which are much faster than routines published so far.   
Since the potential applications of these routines are universal in the field of First Passage Kinetic Monte Carlo algorithms, they are explained 
in the appendix in detail. \\

Before summarizing each of the three search problems individually, some general remarks, relevant for all studied scenarios, are in appropriate:: \\
The break-even diffusivity $\tilde D_\text{be}$ (the value of $\tilde D$ where the best strategy changes from 
intermittent to purely diffusive search) increases with the target size $s\in\lbrace\thetab$, $\tilde d\rbrace$. Consequently, if intermittent search is the best strategy, the fraction 
of time spend in the diffusive mode will be monotonically increasing in $\tilde D$ and decreasing in $s$. \\ 
For small targets, the MFPT does almost not depend on the distribution of the initial position $\mb{r}_0$ or the initial mode (diffusive or ballistic), as the time for the 
searcher to lose its memory about the initial position is much shorter than the MFPT. \\
Furthermore, we observed, that the MFPT $\tilde T$ as a function of the transition rates $\tilde\gamma$ and $\tilde\gamma'$ seems always to be convex. 
However, the positions of the minima $(\tilde\gamma_\text{opt},\tilde\gamma'_\text{opt})$ and 
$(\tilde\Gamma\text{\hspace*{-0.05cm}\textsubscript{\tiny OPT}},\tilde\Gamma'\text{\hspace*{-0.15cm}\textsubscript{\tiny OPT}})$ 
are never sharp, neither in $\tilde\gamma$, nor in $\tilde\gamma$, which can be seen by comparing the values of neighbored isolines in the color coded plots.
Hence, in a quite large surrounding (relative to the absolute values), the MFPT 
$\tilde T$ is only slightly larger than the optimal value. This is remarkable for real search, as this fact offers the opportunity to optimize the search 
process also according to other criteria (e.g. energy consumption or usage of limited resources, for instance fuel necessary for ballistic motion, like ATP 
for motor proteins in the biological context) without increasing the MFPT significantly. \\
For the inhomogeneous search strategies that we studied, the behavior of the MFPT as a function of the inhomogeneity parameters $x$ and $(p,\tilde\Delta)$ sometimes differs. 
Especially for small diffusion constants $\tilde D$ and a large transition rate $\tilde\gamma$, the optimal searching strategy depends strongly on the chosen inhomogeneity parameters.
In addition, more than one local minimum of $\tilde T$ might occur in dependence of the tunable parameters (see Fig. \ref{narrow_escape_letter_scenario}d). \\ 
The dependence on the applied boundary conditions at the border of the searching domain varies strongly in the scenarios that we analyzed. If the target is predominantly located 
very close (compared to $1/\tilde \gamma'$ $\hat{=}$  average covered distance in the ballistic state ) to the boundary or even part of it (narrow escape problem), the MFPT and 
the optimal strategy will strongly be influenced by the boundary condition. If $\tilde\gamma'$ and/or the average distance from the target to the boundary increases, this influence 
shrinks rapidly. \\

The first search scenario that we considered is the so called narrow escape problem. It is well understood for a purely diffusive particle, but apart from \cite{schwarz2016} 
there are no studies for intermittent search available in literature. Thus, before studying inhomogeneous strategies, we analyzed first the homogeneous scenarios for the 
reason of comparison. 
The value of the break-even diffusivity $\tilde D_\text{be}$ depends strongly on the considered boundary conditions. For the exemplarily chosen small opening angle $\thetab=$arcsin(1/7) 
it is about 4 times smaller for the BB (ballistic-ballistic) boundary condition ($\tilde D_\text{be}\approx0.025$) than for the BD  (ballistic-diffusive) scenario ($\tilde D_\text{be}\approx0.1$). Furthermore, there is a qualitative
difference in the behaviour of the optimal transition rates $\tilde\gamma_\text{opt}$ and $\tilde\gamma^{'}_\text{opt}$ as a function of $\tilde D<\tilde D_\text{be}$.
For the BB boundary condition, the optimal transition rates both decrease with $\tilde D$. For BD, only $\tilde\gamma_\text{opt}(\tilde D)$ decreases, while 
$\tilde\gamma^{'}_\text{opt}(\tilde D)=0$ holds for all $\tilde D$. Thus, it is always part of the best strategy to end the ballistic phase only when being forced by the BD condition at 
the boundary of the simulation sphere. As $\tilde\gamma^{'}_\text{opt}$ vanishes for all $\thetab$, the numerical effort for finding the best strategy is essentially reduced.  
Hence, in addition to $\tilde D$, we also varied $\thetab$ systematically. Fig. \ref{3d_ball_to_diff_theta_D_vs_G_and_gamma} (in combination with the nondimensionalisation relations) 
offers a full numeric solution for the best homogeneous search strategy in the BD case as a function all parameters, which is the diffusivity $D$, the radius $R$, 
the velocity $v$ (ballistic mode) and the target area with polar angle $\thetab$.\\  
For both boundary conditions, the MFPT can be significantly reduced by the usage of inhomogeneous searching strategies, which has been shown for the direction distributions 
$\rho_x^\alpha$ and $\rho_{p,\tilde \Delta}^\alpha$. The family $\rho_x^\alpha$ has exclusively been
designed by us for optimizing the narrow escape problem, it is not efficient for targets at the interior of the searching domain. Surprisingly, 
for the rates $\tilde\gamma_\text{opt}$ and $\tilde\gamma^{'}_\text{opt}$ the optimal strategies of the 
biologically inspired family $\rho_{p,\tilde \Delta}^\alpha$ can almost compete with the results of $\rho_x^\alpha$ for small $\tilde D$, which can be seen by comparing the 
values of $\tilde T_\text{min}$ in Table \ref{tabelle_ball_diff} with the minima of Fig. \ref{narrow_escape_letter_scenario}c). This is remarkable, as $\rho_{p,\tilde \Delta}^\alpha$
was not specially designed for the narrow escape problem and is also a good strategy for the other search scenarios.\\
Furthermore, the optimal transition rates depend strongly on the considered direction distribution (up to a factor of 10) in all investigated scenarios, 
which can be seen by comparing $\tilde\gamma_\text{opt}$ vs $\tilde \Gamma_\text{OPT}$, $\tilde\gamma^{'}_\text{opt}$ vs $\tilde \Gamma^{'}_\text{OPT}$ in the Tables \ref{tabelle_ball_ball}, \ref{tabelle_ball_diff} and the $\tilde\gamma$-coordinates of the homogeneous and inhomogeneous minima in 
Fig. \ref{narrow_escape_letter_scenario}d).\\ 

Next, we focused on the problem an immobile target in the interior of the sphere, called reaction kinetics. 
For a target at the origin, it has been shown that the analytic approximations of \cite{Benichou2011, Loverdo2008, Loverdo2009} for the optimal rates $\tilde\gamma_\text{opt}$ 
and $\tilde\gamma'_\text{opt}$ slightly (but systematically) differ from the minima position. Nevertheless, these approximations define almost perfect searching strategies, as
the MFPT is almost nearly insensitive to a variation of the rates in quite a large surrounding of the optimal values. \\ 
In case of a homogeneous searching strategy, the influence of the distribution of the target position is rather small. If the target position is homogeneously 
randomly chosen, $\tilde T_\text{diff}$ and $\tilde T_\text{opt}$  
slightly increase compared to a centered target. Nevertheless, for small target sizes the optimal transition rates 
$\tilde\gamma_\text{opt}$ and $\tilde\gamma'_\text{opt}$ turned out to be independent of the target distribution within the sphere. For larger target sizes $\tilde\gamma_\text{opt}$ 
remains independent, only $\tilde\gamma'_\text{opt}$ slightly increases in the cases that we investigated. Furthermore, if there is no predominantly chosen target position, a homogeneous 
searching strategy will be the optimal solution.\\ 
Things change, when the immobile target is predominantly (but not exclusively) placed in a specific area. Within the family $\rho_{p,\tilde \Delta}^\alpha$ there 
are inhomogeneous strategies which are essentially more efficient than a homogeneous one (Fig. \ref{Fig_enhanc_reac_kin_rho_p_delta}). \\

Finally, we considered the combination of two search processes, called reaction-escape problem. An intermittently searching particle has first to find an either purely diffusive (A) 
or also intermittently moving particle (B) before finding a narrow escape. Dealing with a frustrated problem, the optimal rates for the MFPT $\tilde T_\text{reac}$ of the 
particle-particle binding differ from the optimal rates of the narrow escape problem $\tilde T_\text{esc}$. In consequence, the overall 
optimal strategy, i.e. transition rates which minimize $\tilde T=\tilde T_\text{reac}+\tilde T_\text{esc}$ are a compromise in between. Depending on the ratio of the absolute 
values of $\tilde T_\text{reac}$ and $\tilde T_\text{esc}$ (mostly controlled via the size of the reaction distance $\tilde d$ in comparison to the opening angle $\thetab$), the total
influence on the best strategy varies. \\
The gain of an inhomogeneous searching scenario depends strongly on the investigated parameters and states of motion for the target particle (A or B). However, there is a large
parameter regime in which an inhomogeneous searching strategy is most efficient. \\

To conclude we have demonstrated the efficiency of spatially inhomogeneous search strategies, which were introduced by us recently \cite{schwarz2016}. 
The space of possible spatial inhomogeneities is large and we confined our study only to two parameterized families of search strategies, which already 
turned out to be more efficient than homogeneous strategies. Most probably even more efficient strategies exist outside the families studied here, 
and it would be highly desirable to explore the space of possible strategies, in particular direction distributions, with alternative, possibly more powerful 
tools than brute force numerical studies. Currently the quest for {\it the} optimal inhomogeneous search strategy remains a challenge for future work. 
Potential applications comprise the spatial organization of cytoskeleton in living cells \cite{schwarz2016}, but also the wide field of search 
in spatially inhomogeneous domains and/or with spatially inhomogeneous target distributions.

\subsection*{Acknowledgement}
This work was financially supported by the German Research Foundation 
(DFG) within the Collaborative Research Center SFB 1027. 

\appendix

\section{Fast generation of random numbers}

\label{Samplingappendix}

Based on the Green's functions $P_S$ (sphere) and $P_C$ (cone) for a diffusive particle starting at the origin of a sphere or spherical cone with polar angle 
$\Theta$ ($0\leq \vartheta \leq \Theta$) respectively, this appendix presents efficient methods for sampling the occurring densities, needed within the 
simulations of this paper, in detail:
\begin{itemize}[align=left,labelwidth=\widthof{i},leftmargin=\labelwidth+\labelsep] 
\item[$\rho_b(t)$:] probability density for reaching the absorbing radius $R_{pro}$ of a sphere/cone for the first time at time $t$ at an arbitrary solid angle when starting at the origin.
\item[$\rho_n(r|t)$:] probability density for being at radius $r$ within a sphere/cone at time $t$  at an arbitrary solid angle under the condition of not having reached radius 
$R_{pro}$ before and having started at the origin.  
\end{itemize}
Due to the radial symmetry of these problems, the density of the solid angle is homogeneously distributed on the surface of either a sphere or a spherical cone for all $t$ 
and $r\in [0;R_{pro}]$. Thus its sampling in case of a position update can be done very fast, which is among others shown in chapter V. 4 of \cite{devroye1986}. \\
In the following, the index "$pro$" is skipped at the radius $R_{pro}$. It shortens the notation and there is no danger of mixing it up with the radius of the simulation 
sphere within this appendix.\\  
Although ways of sampling the densities $\rho_b$, $\rho_n$ have already been published in \cite{donev2010}, we want to present significantly faster methods to sample them 
by picking up their idea (\cite{donev2010}) of different representations for large and small times, 
but avoiding their bottleneck: the numerical inversion of cumulative distribution functions, which results in the calculation of many exponential and trigonometrical 
functions, slowing the algorithm down. Especially a fast sampling of $\rho_b$ is very important as a random number according to this distribution is always needed after 
the creation of a protection sphere/cone, which is part of the innermost loop of the simulation.
    
\subsection{The analytic solutions $P_S$ and $P_C$ of the diffusion equation}
\subsubsection{The sphere}
The diffusion problem within a totally absorbing sphere of radius $R$ is given by  
\begin{eqnarray}
\pdou{P(r,\varphi,\vartheta,t)}{t}&=&D\Delta P(r,\varphi,\vartheta,t)\;, \text{ with} \label{iniprobsphere}\\
 P(R,\varphi,\vartheta,t)&=&0\;\;\;\forall\; \varphi\in[0;2\pi[,\vartheta\in[0;\pi]\;.\nonumber
\end{eqnarray} 
In \cite{carslaw1959} the solution for a particle at an arbitrary starting radius $r_0<R$ is derived. With the help of l'Hospital's rule ($r_0 \rightarrow 0$) the following expression for the radial symmetric probability density $P_S$ can be obtained:  
\begin{eqnarray}
P_S(r,t)=\frac{1}{2 R^2 r}\sum_{n=1}^\infty e^{-\pi^2 n^2\frac{Dt}{R^2}}\,n\,\sin\left(\frac{n\pi r}{R}\right)\;.\label{appendix1}
\end{eqnarray}

For the sake of computational efficiency it is useful to derive a second expression for $P_S$ by applying Poisson's summation formula to Eq. (\ref{appendix1}).
\begin{eqnarray}
P_S(r,t)=\frac{1}{8 \left( \pi D t \right)^{\frac{3}{2}}}\,\times\hs{45} \label{appendix2} \\
\hs{2}\left\lbrace e^{\frac{-r^2}{4Dt}}\hs{-0.5}+\hs{-1}\sum_{k=1}^\infty \hs{-1}\left(\frac{2kR\hs{-0.5}+\hs{-0.5}r}{r} e^{\frac{-(2kR+r)^2}{4Dt}}\hs{-0.5}-\frac{2kR\hs{-0.5}-\hs{-0.5}r}{r} e^{\frac{-(2kR-r)^2}{4Dt}}\hs{-0.5}\right) \hs{-1}\right\rbrace\hs{-5}\nonumber 
\end{eqnarray}

For large $t$ the series of Eq. (\ref{appendix1}) converges very fast, whereas for small $t$ the series of Eq. (\ref{appendix2}) does. 

\subsubsection{The spherical cone}
The diffusion problem within a spherical cone of radius $R$ and polar angle $\Theta$, reflecting at the conical boundary and absorbing at the spherical cap is given by       
\begin{eqnarray}
\pdou{P(r,\varphi,\vartheta,t)}{t}&=&D\Delta P(r,\varphi,\vartheta,t),\;\; \text{with} \label{iniprobcone}\\
 P(R,\varphi,\vartheta,t)&=&0\;\;\;\forall\; \varphi\in[0;2\pi[,\vartheta\in[0;\Theta]\; ,\nnn
          \left.\pdou{P(r,\varphi,\vartheta,t)}{\vartheta}\right|_{\vartheta=\Theta}&=&0\;\;\;\forall\; r\in[0;R],\varphi\in[0;2\pi[\;.\nonumber
\end{eqnarray}
In the general case of an arbitrary starting position, its solution looks very complex, but for the case of interest ($r_0=0$), we simply obtain a radial symmetric probability 
density, which is proportional to the solution of the subsection above:  
\begin{eqnarray}
P_C(r,t)=\frac{2}{1-\cos(\Theta)} P_S(r,t) 
\end{eqnarray}

\subsection{The probability densities $\rho_b(t)$, $\rho_n(r|t)$}
Based on the formulas (\ref{appendix1}) and (\ref{appendix2}), two different expressions for each of the probability densities $\rho_b$ and $\rho_n$ can be derived. 
The upper index ``$>$'' will always mark the series, which converges fast for large $t$, while the upper index ``$<$'' will mark the series, which converges fast for 
small $t$. If there is no upper index, it is a general statement, i.e. independent of the series representation.       
The probability density $\rho_b(t)$  is identical for the sphere and the cone of radius $R$ and given by
\begin{eqnarray}
\rho_b(t)\hspace*{-0.1cm}&=&\hspace*{-0.1cm}-\frac{d}{dt}\left[\int_0^R\hspace*{-0.15cm} dr \int_0^{2\pi}\hspace*{-0.15cm} d\varphi \int_0^{\pi}\hspace*{-0.15cm} d\vartheta\,r^2\sin(\vartheta)\,P_S(r,t)\right]\quad\quad\label{defrho_b}\\
         &=&\hspace*{-0.1cm}-\frac{d}{dt}\left[\int_0^R\hspace*{-0.15cm} dr \int_0^{2\pi}\hspace*{-0.15cm} d\varphi \int_0^{\Theta}\hspace*{-0.15cm} d\vartheta\,r^2\sin(\vartheta)\,P_C(r,t)\right]\;.\nonumber
\end{eqnarray}
Applying this to the series (\ref{appendix1}) and (\ref{appendix2}), we get: 

\begin{eqnarray}
\rho_b^>(t)\hspace*{-0.15cm}&=&\hspace*{-0.15cm}\frac{2\pi^2 D}{R^2}\sum_{n=1}^\infty e^{-\pi^2 n^2\frac{Dt}{R^2}} (-1)^{n+1} n^2\; , \nnn
\rho_b^<(t)\hspace*{-0.15cm}&=&\hspace*{-0.15cm}\frac{R^3}{2\sqrt{\pi D^3 t^5}}\hspace*{-0.1cm}\sum_{k=1}^\infty e^{-\frac{R^2(2k-1)^2}{4 D t}}\hspace*{-0.12cm} \left(\hspace*{-0.1cm}(2k-1)^2\hspace*{-0.1cm}-2\frac{Dt}{R^2}\hspace*{-0.05cm}\right)\;.\non
\end{eqnarray}

The probability density $\rho_n(r|t)$ is also identical for the sphere and the cone of radius $R$ and given by

\begin{eqnarray}
\rho_n(r|t)&=&\dfrac{\int_0^{2\pi} d\varphi \int_0^{\pi} d\vartheta\,r^2\sin(\vartheta)\,P_S(r,t)}{\int_0^R dr \int_0^{2\pi} d\varphi \int_0^{\pi} d\vartheta\,r^2\sin(\vartheta)\,P_S(r,t)}\label{defrhor}\\
&=&\dfrac{\int_0^{2\pi} d\varphi \int_0^{\Theta} d\vartheta\,r^2\sin(\vartheta)\,P_C(r,t)}{\int_0^R dr \int_0^{2\pi} d\varphi \int_0^{\Theta} d\vartheta\,r^2\sin(\vartheta)\,P_C(r,t)}\; .\non
\end{eqnarray}

In consequence, we get: 

\begin{eqnarray}
\rho_n^>(r|t)&=&\frac{\pi r}{R^2}\dfrac{\sum\limits_{n=1}^\infty e^{-\pi^2 n^2\frac{Dt}{R^2}} n\sin\left(\frac{n\pi r}{R}\right)}{\sum\limits_{n=1}^\infty e^{-\pi^2 n^2\frac{Dt}{R^2}} (-1)^{n+1}}\; ,\hs{50}\nnn
\rho_n^<(r|t)&=&\non
\end{eqnarray}
\vspace*{-0.5cm}
\begin{equation}
\dfrac{r^2 e^{\frac{-r^2}{4Dt}}\hspace*{-0.08cm}+\hspace*{-0.06cm}r\hspace*{-0.1cm}\sum\limits_{k=1}^\infty\hspace*{-0.12cm}\left(\hs{-1}(2kR\hs{-0.5}+\hs{-0.5}r) e^{\frac{-(2kR+r)^2}{4Dt}}\hspace*{-0.06cm}-\hspace*{-0.06cm}(2kR\hs{-0.5}-\hs{-0.5}r) e^{\frac{-(2kR-r)^2}{4Dt}} \hs{-0.3}\right)\hs{-1} }{2Dt\left(\sqrt{\pi Dt}-2R\sum\limits_{k=1}^\infty e^{-\frac{R^2 (2k-1)^2}{4Dt}}\right)\hs{-1}}\;.\nonumber
\end{equation}

\subsection{Efficient sampling of $\rho_b(t)$}
Instead of designing a sampling routine including the parameters $R$ and $D$, it is computationally more efficient to sample the dimensionless random number 
\begin{equation}
\tau=\frac{D}{R^2}t \; ,\non
\end{equation}
as $R^2/D$ is the characteristic timescale. For its probability density $\tilde\rho_b$, we get 
\begin{equation}
\tilde\rho_b(\tau)=\frac{R^2}{D}\,\,\rho_b\left(\frac{R^2}{D}\tau\right)\; ,\non
\end{equation}
which leads to the following series representations
\begin{eqnarray}
\tilde\rho_b^>(\tau)\hspace*{-0.1cm}&=&\hspace*{-0.1cm}2\pi^2 \sum_{n=1}^\infty e^{-\pi^2 n^2\tau} \left(-1\right)^{n+1} n^2\label{dens_dimless_one}\\
\tilde\rho_b^<(\tau)\hspace*{-0.1cm}&=&\hspace*{-0.1cm}\frac{1}{2\sqrt{\pi  \tau}\tau^2}\sum_{k=1}^\infty e^{-\frac{(2k-1)^2}{4 \tau}} \left(\left(2k-1\right)^2\hspace*{-0.1cm}-2\tau\right)\,\, . \quad\quad\label{dens_dimless_two}
\end{eqnarray}
The corresponding distribution function can be expressed via the Jacobi-Theta function $\vartheta_4$: 
\begin{equation}
\tilde{F}_b(\tau)=\vartheta_4\left(0,e^{-\pi^2\tau}\right) \; .\non
\end{equation}
Although this function is well studied in mathematics, there seems to be no way to invert it analytically. In consequence, sampling via the Inversion method would require a numerical 
inversion tool, which slows the algorithm dramatically down.  

Hence, we decided to use a fast way of rejection sampling, which is described below:\\\hspace*{+0.5cm}

Depending on the needed numerical accuracy, the questions of where to truncate the sums (\ref{dens_dimless_one}) , (\ref{dens_dimless_two}) and when to switch between 
these two representations has to be answered. With the following choice one is on the safe side for all practical purposes:
\begin{eqnarray}
\tilde\rho_b^\text{num}(\tau)\hspace*{-0.05cm}=\hspace*{-0.1cm}\left\lbrace{\small \begin{matrix} 
2\pi^2 \sum\limits_{n=1}^{n_\text{max}(\tau)} e^{-\pi^2 n^2\tau} \left(-1\right)^{n+1} n^2&,& \tau\hs{-0.5}\geq\hs{-0.5}\tau_c \vspace*{0.2cm}\\ 
\hs{-0.6}\frac{1}{2\sqrt{\pi  \tau}\tau^2}\hspace*{-0.2cm}\sum\limits_{k=1}^{k_\text{max}(\tau)}\hspace*{-0.1cm} e^{-\frac{(2k-1)^2}{4 \tau}} \hspace*{-0.1cm}\left(\left(2k\hs{-0.5}-\hs{-0.5}1\right)^2\hspace*{-0.1cm}-2\tau\right)&,&\tau\hs{-0.5}<\hs{-0.5}\tau_c \end{matrix}} \right.\;. \nonumber
\end{eqnarray}
with $\tau_c=0.25$ and the piece-wise constant integer functions


\begin{eqnarray}
n_\text{max}(\tau)=\left\lbrace{\small \begin{matrix} 
4 &,& \tau \in\left[0.25;0.3 \right[ \vspace*{0.1cm}\\ 
3 &,& \tau \in\left[0.3;0.6 \right[ \vspace*{0.1cm}\\ 
2 &,& \tau \in\left[0.6;1.5 \right[ \vspace*{0.1cm}\\ 
1 &,& \tau \in\left[1.5;\infty \right[
\end{matrix}} \right.\;, \nonumber
\end{eqnarray}

\begin{eqnarray}
k_\text{max}(\tau)=\left\lbrace{\small \begin{matrix} 
3 &,& \tau \in\left[0.125;0.25\right[ \vspace*{0.1cm}\\ 
2 &,& \tau \in\left[0.04;0.125\right[ \vspace*{0.1cm}\\ 
1 &,& \tau \in\left]0;0.04\right[
\end{matrix}} \right.\;. \nonumber
\end{eqnarray}

Using this choice, the relative deviation fulfills
\begin{eqnarray}
\frac{|\tilde\rho_b(\tau)-\tilde\rho_b^\text{num}(\tau)|}{\tilde\rho_b(\tau)}<10^{-18}\;\; \forall\; \tau>0\;, \text{i.e.}\nonumber
\end{eqnarray}
we never have to add more than four addends to calculate $\tilde\rho_b(\tau)$ within double precision. \\\hspace*{+0.5cm}

For using rejection sampling, a helping probability density $\rho_h(\tau)$ and a scaling constant $k>1$ (as small as possible) with 
$k\geq\frac{\tilde\rho_b(\tau)}{\rho_h(\tau)} \;\forall\; \tau$ have to be found \cite{devroye1986}. Furthermore there must be the possibility to generate random 
numbers according to $\rho_h(\tau)$ very fast. For these purposes we divide $\mathbbm{R}^+_0$ into $N$ intervals $[\tau_i,\tau_{i+1}[$ with $\tau_0=0$ and $\tau_{N}=\infty$. 
For the first $N-1$ intervals we want $\rho_h(\tau)$ to be piece-wise constant:
\begin{eqnarray}
\rho_h(\tau)=\frac{p_{i}}{k} \quad \forall\; \tau\in[\tau_i,\tau_{i+1}]\;,\; i\in\left\lbrace 0,...,N-2\right\rbrace\;, \quad\quad 
\end{eqnarray}
with $p_i=\maxi\limits_{[\tau_i,\tau_{i+1}]}\tilde\rho_b(\tau)$. Having chosen the length of the first interval $[\tau_0,\tau_{1}]$, we define a constant 
$Q=(\tau_{1}-\tau_{0})\cdot p_{0}$ and the length of all the other intervals is determined via
\begin{equation}
(\tau_{i+1}-\tau_{i})\cdot p_{i}=Q \quad\; \forall\; i\in\left\lbrace 1,...,N-2\right\rbrace\;.
\end{equation}
These calculations cannot be done in a straightforward manner, as $p_{i}$ is also a function of $\tau_{i+1}$. Nevertheless it is possible to iterate it as exact, 
as wanted (more than double precision) by using floating data-types with an arbitrary exactness. Since the calculation of all $\tau_i$ and $p_{i}$ have to be done 
once only, in order to include them in our implementation, we don't have to care about the running time of it. 

\begin{figure}[htb]
\incgrapw{0.9}{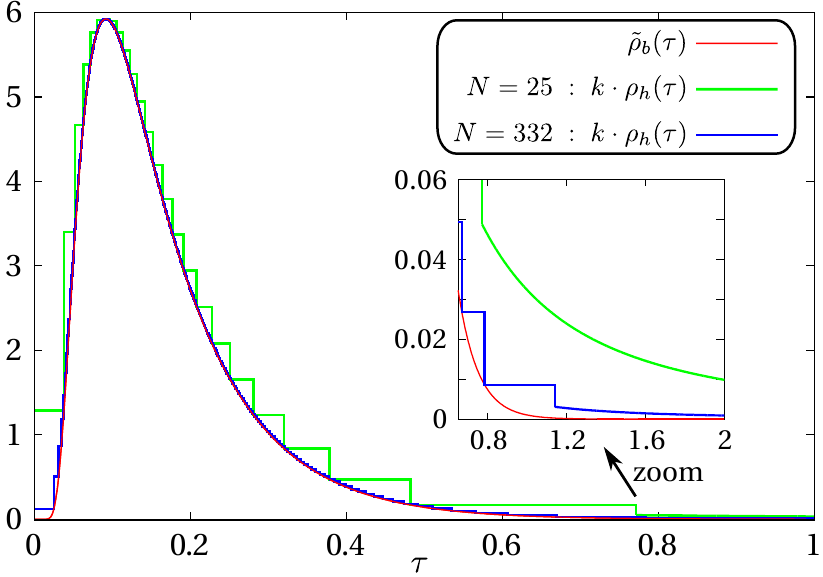}
\caption{Illustration for the construction of $\rho_h(\tau)$ for the cases $\tau_1=0.038$ ($\rightarrow N=25$) and $\tau_1=0.025$ ($\rightarrow N=332$). The case N=25 is created 
only to clarify the construction mechanism. The case N=332 is used in our implementation.}
\label{rho_b_sampling}
\end{figure}

\begin{figure}[htb]
\incgrapw{0.9}{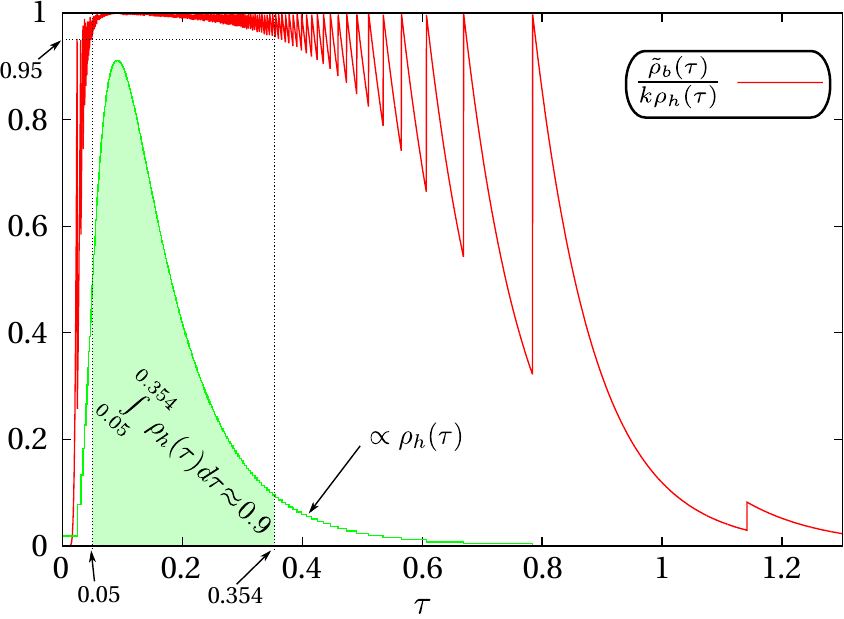} 
\caption{$\tilde\rho_b(\tau)/(k\,\rho_h(\tau))$: Around 90 $\%$ of all random numbers $\tau_\text{cand}$ will be produced in the interval $[0.05;0.354]$. The ratio $\tilde\rho_b/(k\,\rho_h)$ is bigger 
than 0.95 within this interval.}
\label{rho_n_smaller0_054_sampling_ratio}
\end{figure}

For the tail interval $[\tau_{N-1};\infty[$ we choose $\rho_h(\tau)=\frac{Q}{k(\tau+1-\tau_{N-1})^2}$.
Thus, the cumulative probability of each interval is the same: $Q/k$. Keeping in mind, that there are $N$ intervals, we get: 
\begin{equation}
k=N\cdot Q \; , 
\end{equation}
and in consequence:
\begin{eqnarray}
\rho_h(\tau)&=&\left\lbrace{\small \begin{matrix} 
 \frac{p_0}{k}&,& \tau \in [\tau_0;\tau_1[ \vspace*{0.1cm}
\\ \frac{p_1}{k}&,& \tau \in [\tau_1;\tau_2[ \vspace*{0.1cm}
\\ \vdots&& \vdots
\\ \frac{p_{N-2}}{k}&,& \tau \in [\tau_{N-2};{\tau_{N-1}}[ \vspace*{0.1cm}
\\ \frac{1}{N(\tau+1-\tau_{N-1})^2}&,& \tau\geq\tau_{N-1} 
\end{matrix}} \right. \;.\quad\quad\quad
\end{eqnarray}
An illustration of this procedure for a rather rough $\rho_h$ by choosing $\tau_1=0.038$ ($\rightarrow N=25$) and the case of $\tau_1=0.025$ ($\rightarrow N=332$), 
which was used in our simulations is shown in Fig. \ref{rho_b_sampling}.  
The corresponding distribution function is given by
\begin{eqnarray}
F_h(\tau)\hspace{-0.08cm}=\hspace{-0.08cm}\left\lbrace{\small \begin{matrix} 
\frac{p_0}{k}\tau&,& \tau \in [\tau_0;\tau_1[
\\ \frac{1}{N}+\frac{p_1}{k}(\tau-\tau_1)&,& \tau \in [\tau_1;\tau_2[ 
\\ \frac{2}{N}+\frac{p_2}{k}(\tau-\tau_2)&,& \tau \in [\tau_2;\tau_3[ 
\\ \vdots&& \vdots
\\ \frac{N-2}{N}+ \frac{p_{N-2}}{k} (\tau-\tau_{N-2})&,& \tau \in [\tau_{N-2};\tau_{N-1}[ 
\\ 1-\frac{1}{N(\tau+1-\tau_{N-1})}&,& \tau\geq\tau_{N-1} 
\label{as}
\end{matrix}} \right.
\end{eqnarray}
As every interval has the same probability mass $1/N$, a random number $\tau_\text{cand}$ according to $\rho_h$ can be generated very fast, as $F_h$ can be 
inverted very fast straightforwardly without the usage of a bisection method: \\
The integer number $m\in\left\lbrace 0,...,N-1 \right\rbrace$, given by  
\begin{equation}
m=\lfloor r_\text{cand}\cdot N \rfloor\;, \label{mgleich}
\end{equation}
immediately indicates the candidate's interval $[\tau_m;\tau_{m+1}]$ of $F_h(\tau)$, where $r_\text{cand}$ is a random number from a uniform 
distribution in the interval $]0,1[$. 
Solving for $\tau$ in the $m$-th interval of Eq. (\ref{as}) yields
\begin{eqnarray}
\tau_\text{cand}\hspace{-0.08cm}=\hspace{-0.08cm}\left\lbrace{\small \begin{matrix} 
\left(r_\text{cand}-\frac{m}{N}\right)\cdot\frac{k}{p_m}+\tau_m&,&m\hspace{-0.08cm}\in\hspace{-0.04cm}\left\lbrace 0,...,N\hspace{-0.04cm}-\hspace{-0.08cm}2\right\rbrace
\\ \tau_{N-1}-1+\frac{1}{N (1-r_\text{cand})}&,&m\hspace{-0.04cm}=\hspace{-0.04cm}N-1
\end{matrix}} \right..\quad\quad \label{inv_F}
\end{eqnarray}
The choice of $\tau_1=0.025$ for the construction of $\rho_h(\tau)$ (see Fig. \ref{rho_b_sampling}) results in $Q=0.003078...$, $N=332$ and $k=1.02...\,$. 
In consequence we are dealing with a very efficient way of rejection sampling, as the rejection rate is about 2 $\%$. \\
Nevertheless it is possible to improve this sampling significantly by having a closer look at the procedure of rejection sampling: At first, a candidate 
random number $\tau_\text{cand}$ according to $\rho_h(\cdot)$ is chosen. As explained above, this is possible very fast. $\tau_\text{cand}$ is accepted, 
if the quotient $\tilde\rho_b(\tau_\text{cand})/(k\rho_h(\tau_\text{cand}))$ is bigger than a uniformly distributed random number $r_\text{rej}$ in the 
interval $[0;1]$. Using the method above, this happens in around $1/k\approx98$ $\%$ of the cases. The most time consuming part is the calculation of 
$\rho_b(\tau_\text{cand})$, as it includes 1-4 addends. This calculation can often be avoided by formulating one (or more) precondition for acceptance:
About 90$\%$ of the $\tau_\text{cand}$ will be in the interval $[0.05;0.354]$. There the quotient $\tilde\rho_b/(k\,\rho_h)$ is bigger than 0.95 everywhere, 
which is illustrated in Fig. \ref{rho_n_smaller0_054_sampling_ratio} 
Hence, for $r_\text{rej}<0.95$, we can accept without evaluating $\tilde\rho_b$. In consequence, only in $1-0.95\cdot 0.9\approx 15$ $\%$ of the cases we really have 
to evaluate $\tilde\rho_b$. This fraction can be further reduced by formulating more preconditions for $\tau<0.05$ and $\tau>0.354$. But the gain won't be large, 
as already 90$\%$ of the $\tau_\text{cand}$ are covered.\\
Let's summarize the average effort: Per candidate $\tau_\text{cand}$ we need only one equally distributed random number $r_\text{cand}$ and 6 trivial 
operations ($+,-,\cdot,/$) for the calculations of the Eqs. (\ref{mgleich}, \ref{inv_F}).  The rejection rate $\frac{k-1}{k}$ is only about 2 $\%$ and the 
computational price for the acceptance decision is the generation of an equally distributed random number $r_\text{rej}$ and on average less 
than 0.5 addends in the formulas (\ref{dens_dimless_one}), (\ref{dens_dimless_two}). A C++-implementation of the described method on a single CPU-core 
with 3.4 GHz takes around 35 sec for $10^9$ random $\tau$. About 14 sec of this time have been used to generate high quality uniformly distributed random numbers 
for $r_\text{cand}$ and $r_\text{rej}$ with the help of the gsl routine ``gsl\_rng\_mt19937'' based on the Mersenne Twister \cite{mersenne1998}.\\ 

\subsection{Efficient sampling of $\rho_n(r|t)$}
In addition to the characteristic time scale, we want to use the characteristic length scale $R$ and define the dimensionless length
\begin{equation}
x=\frac{1}{R}r \quad .
\end{equation}
For its probability density, we get:
\begin{equation}
\tilde\rho_n(x|\tau)=R \rho_n\left(xR\left|\frac{R^2}{D}\tau\right.\right)\; ,\non
\end{equation}
which leads to the following series representations
\begin{eqnarray}
\tilde\rho_n^>(x|\tau)&=&
\pi x\dfrac{\sin\left(\pi x\right)+\sum\limits_{n=2}^\infty e^{-\pi^2 \left(n^2-1\right)\tau} n\sin\left(n\pi x\right)}{1-\sum\limits_{n=2}^\infty e^{-\pi^2 \left(n^2-1\right)\tau} \left(-1\right)^n}\;,\quad\quad\label{dens_dimless_one_x}\nnn
\tilde\rho_n^<(x|\tau)&=&\non\label{dens_dimless_two_x}
\end{eqnarray}
\vspace*{-0.5cm}
\begin{eqnarray}
\dfrac{x^2 e^{\frac{-x^2}{4\tau}}\hspace*{-0.06cm}+\hspace*{-0.06cm}x\hspace*{-0.06cm}\sum\limits_{k=1}^\infty\hspace*{-0.08cm}\left((2k+x) e^{\frac{-(2k+x)^2}{4\tau}}\hspace*{-0.06cm}-\hspace*{-0.06cm}(2k-x) e^{\frac{-(2k-x)^2}{4\tau}} \right) }{2\tau\left(\sqrt{\pi \tau}-2\sum\limits_{k=1}^\infty e^{-\frac{(2k-1)^2}{4\tau}}\right)}\;.\non
\end{eqnarray}
For later usage, we want to decompose $\tilde\rho_n^<(x|\tau)$ in the not normalized probability density $g(x|\tau)=4\pi R^3 x^2 P_S\left(xR\left|\frac{R^2}{D}\tau\right.\right)$ 
of being at radius $x$ and the probability $S(\tau)=1-\tilde F_b(\tau)$ of not reaching the radius $R$ until time $\tau$, i.e. $\tilde\rho_n^<(x|\tau)=g(x|\tau)/S(\tau)$ with 
\begin{eqnarray}
g(x|\tau)\hspace*{-0.1cm}=\hspace*{-0.08cm}\frac{x^2 e^{\frac{-x^2}{4\tau}}\hspace*{-0.1cm}+\hspace*{-0.06cm}x\hspace*{-0.12cm}\sum\limits_{k=1}^\infty\hspace*{-0.15cm}\left(\hspace*{-0.08cm}(2k\hspace*{-0.06cm}+\hspace*{-0.06cm}x) e^{\frac{-(2k+x)^2}{4\tau}}\hspace*{-0.12cm}-\hspace*{-0.08cm}(2k\hspace*{-0.06cm}-\hspace*{-0.06cm}x) e^{\frac{-(2k-x)^2}{4\tau}}\hspace*{-0.06cm}\right) }{2\tau\sqrt{\pi\tau}} \nnn
\text{and} \quad S(\tau)=1-\frac{2}{\sqrt{\pi \tau}}\sum\limits_{k=1}^\infty e^{-\frac{(2k-1)^2}{4\tau}}=1-\tilde F_b(\tau)\;.\quad\quad \non
\end{eqnarray}
  
Again, we have to answer the questions, when to switch between the different series representations and where to truncate them. Choosing the same values for 
$\tau_c$ , $n_\text{max}$, $k_\text{max}$ like in the previous section one is again on the safe side for all practical purposes.\\
As the density $\tilde\rho_n$ contains the parameter $\tau$, which influences the shape of $\rho_n$ dramatically, we decomposed our algorithm in a sampling procedure for 
short times ($\tau<0.054$, a) and a procedure for long times ($\tau\geq 0.054$, b). This decomposition is not connected to the switch in the series representations for 
evaluating the density, its origin is the result of a purely empiric optimization process. 
\subsubsection{Sampling for $\tau<0.054$}
\begin{figure}[htb!]
\incgrapw{0.9}{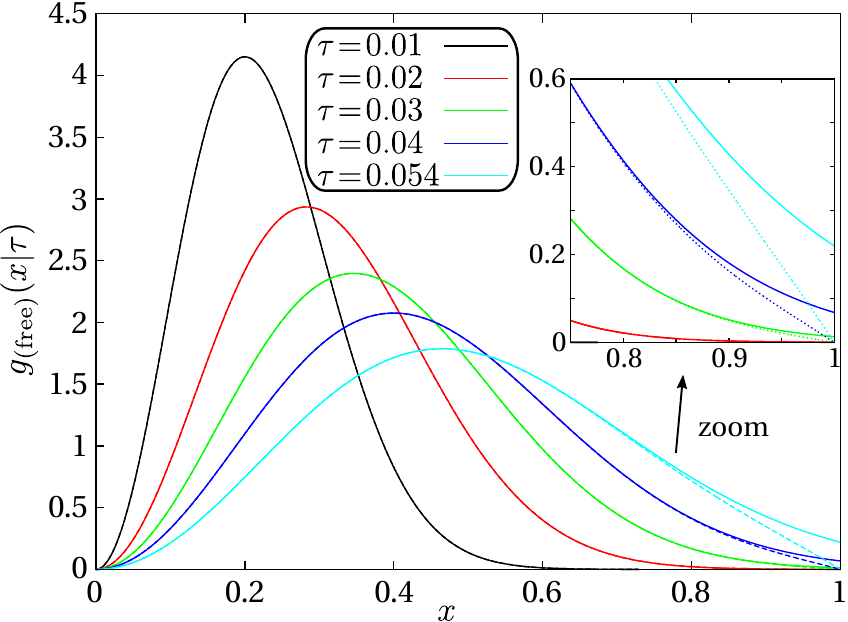} 
\caption{Illustration of the functions $g_\text{free}(x|\tau)$ and $g(x|\tau)$ for different times $t$. Full lines represent $g_\text{free}$, dotted 
ones $g$. For small $\tau$ the curves coincide almost perfectly.}
\label{rho_n_smaller0_054_sampling}
\end{figure}

For $\tau<0.054$, the not normalized probability density $g(x|\tau)$ is almost identical to the also not normalized (according to the interval [0;1]) probability density 
\begin{equation}
g_\text{free}(x|\tau)=\frac{1}{2\tau\sqrt{\pi\tau}}x^2e^{-\frac{x^2}{4\tau}}  
\end{equation}
of a freely diffusing particle in $\mathbbm{R}^3$, where $x$ is the distance to the origin, which is illustrated in Fig. \ref{rho_n_smaller0_054_sampling}.
As $g(x|\tau)\leq g_\text{free}(x|\tau)$ holds for all $x$ and $\tau$, the rejection sampling procedure for $\tilde\rho_n$ is the following:\\
We generate a random $x_\text{c}\in\; ]0;1[$ according to $g_\text{free}(x|\tau)$ in a way, which will be described later. Afterwards it is accepted with 
probability $p_a=g(x_\text{c}|\tau)/g_\text{free}(x_\text{c}|\tau)$, which can be simplified to
\begin{eqnarray}
p_a\hspace*{-0.04cm}(x_\text{c}|\tau\hspace*{-0.04cm})\hspace*{-0.08cm}=\hspace*{-0.08cm}1\hspace*{-0.08cm}+\hspace*{-0.08cm}\frac{\hspace*{-0.12cm} \sum\limits_{k=1}^{\hspace*{-0.3cm}k_\text{max}(\tau)}\hspace*{-0.2cm}\left(\hspace*{-0.08cm}(2k\hs{-0.5}+\hs{-0.5}x_\text{c})e^{\frac{-k(k+x_\text{c})}{\tau}}\hspace*{-0.1cm}-\hspace*{-0.1cm}(2k-x_\text{c}) e^{\frac{-k(k-x_\text{c})}{\tau}}\hspace*{-0.05cm} \right)}{x_c}\;.\non
\end{eqnarray}



\begin{figure}[htb!]
\incgrapw{0.9}{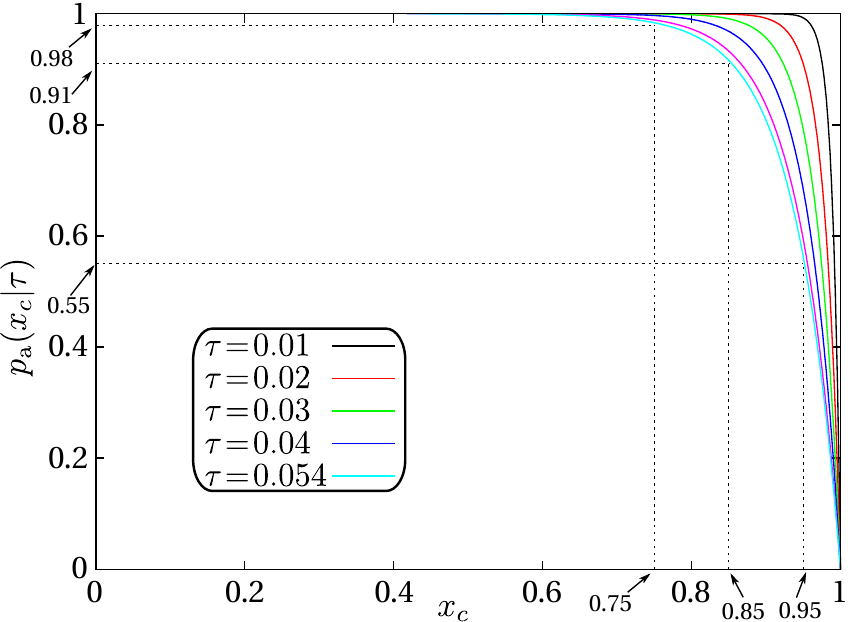} 
\caption{Illustration of the empirically chosen preconditions: For more than 95 $\%$ of the sampled $x_c$, the sampled rejection random number 
$r_\text{rej}$ will be within one of the three black dotted rectangles. There $x_c$ can be accepted without a calculation of $p_a(x_\text{c}|\tau)$.}
\label{rho_n_smaller0_054_sampling_reject_ratio}
\end{figure}

Similar to the method in the previous subsection, it is possible to save a lot of computation time by having a closer look at the function 
$p_a(x_\text{c}|\tau)$ in order to formulate some preconditions for acceptance. For all $\tau\in[0;0.054]$ the following inequalities hold:
\begin{itemize}[align=left,labelwidth=\widthof{i},leftmargin=\labelwidth+\labelsep] 
\item $p_a(x_\text{c}|\tau)>0.98\;\forall x\in[0;0.75]$
\item $p_a(x_\text{c}|\tau)>0.91\;\forall x\in[0.75;0.85]$
\item $p_a(x_\text{c}|\tau)>0.55\;\forall x\in[0.85;0.95]$\;,
\end{itemize}
which is illustrated in Fig \ref{rho_n_smaller0_054_sampling_reject_ratio}.
Due to these chosen relations, in the worst case ($\tau=0.054$), for less than 5 $\%$ of the sampled $x_c$ it is necessary to calculate $p_a(x_\text{c}|\tau)$ 
for the decision whether to accept or to deny $x_c$.\\
Sampling the candidate $x_c$ is also done via rejection sampling:
At first, we sample three gaussian random numbers $z_1,\,z_2,\,z_3$ with variance $\sigma^2=2\tau$ and calculate
\begin{eqnarray}
x_c=\sqrt{z_1^2+z_2^2+z_3^2}\; . \non
\end{eqnarray}
For $x_c\leq 1$, which happens (again in the worst case) in more than 97 $\%$, we go on, otherwise we resample $x_c$ until it is smaller than 1.\\
A C++-implementation of the described method on a single CPU-core with 3.4 GHz takes for the worst case scenario ($\tau=0.054$) 160 sec for $10^9$ random $x$.

At first sight, this procedure consisting of two steps of rejection sampling might not look so fast, but apart from the very small rejection rates and its fast 
generation, there is a second argument, which speeds the procedure indirectly up: We don't need to sample the solid angle for a position update any more, taking 
$z_1,\,z_2,\,z_3$ for the replacement fulfills the right statistics in this case. 

\subsubsection{Sampling for $\tau>0.054$}
\begin{figure}[htb!]
\incgrapw{0.9}{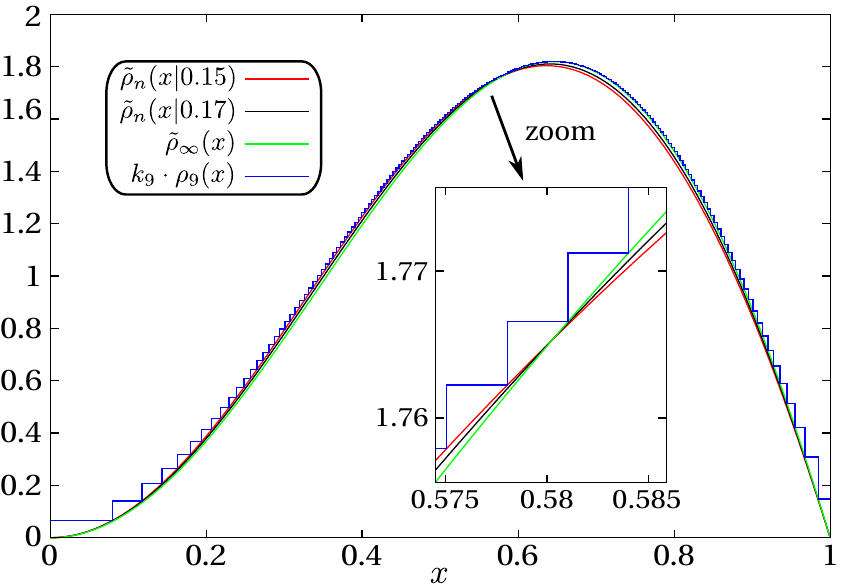} 
\caption{Illustration of the construction of $\rho_9(x)$ for the choice $x_1^9=0.08$, which results in N=194 intervals $[x_i^9,x_{i+1}^9]$ and a scaling constant $k_9=1.02$.}
\label{rho_n_greater0_15}
\end{figure}

For very small values of $\tau$, the shape of $\tilde\rho_n$ changes very fast with time, which can also be seen in Fig. \ref{rho_n_smaller0_054_sampling}. But for
$\tau\rightarrow\infty$ it converges quickly to the time independent density
\begin{eqnarray}
\tilde\rho_{\infty}(x)=\pi x\sin{\pi x}\;. \non
\end{eqnarray}
Our sampling routine makes use of this fact by dividing the time interval $[0.054;\infty[$ into a sequence of $M$ disjoint intervals $[\tau_j;\tau_{j+1}[$ with $\tau_0=0.054$ and $\tau_M=\infty$. Within every interval $[\tau_i;\tau_{i+1}[$, we want to construct an efficient rejection sampling method with a time independent helping density $\rho_j(x)$ and a scaling constant $k_j$. 
In consequence, a lower boundary $B(\tau_j,\tau_{j+1})$ for each scaling constant $k_j$ is given by 
\begin{eqnarray}
B(\tau_j,\tau_{j+1})=\int_0^1 dx \maxi\limits_{\tau\in[\tau_j;\tau_{j+1}]}(\tilde\rho_n(x,\tau)) < k_j\; .  \non
\end{eqnarray}
With the choice $M=9$ and the decomposition 
\begin{align}
I_1=[0.054;0.057[\,,\;I_2=[0.057;0.061[\,,\;I_3=[0.061;0.066[\; \nnn
I_4=[0.066;0.072[\;,\;I_5=[0.072;0.08[\;\;,\;\;I_6=[0.08;0.091[\;\; \nnn
I_7=[0.091;0.108[\;,\;I_8=[0.108;0.15[\;\;,\;\;I_9=[0.15;\infty[\;\;\;\;\;\; \non
\end{align}
we confirmed numerically 
\begin{eqnarray}
1.015< B(\tau_j,\tau_{j+1})< 1.028\quad\forall\; j\; .  \non
\end{eqnarray}
Similar to the case of $\tilde\rho_b$, we divide the spatial interval $[0;1]$ $M$ times into $N_j$ intervals $[x_i^j;x_{i+1}^j]$ with $x_0^j=0$ and $x_{N_j}^j=1$. For all $N_j$ intervals
we want $\rho_j(x)$ to be piece-wise constant:
\begin{eqnarray}
\rho_j(x)=\frac{p_{i}^j}{k_j} \quad \forall \;x\in[x_i^j,x_{i+1}^j]\; ,   \non
\end{eqnarray}
with $p_i^j=\maxi\limits_{\tau\in[\tau_j;\tau_{j+1}]}\left(\maxi\limits_{[x_i^j,x_{i+1}^j]}\tilde\rho_n(x|\tau)\right)$. 
By choosing the length of the first intervals $[x_0^j,x_{1}^j] \;\forall\;j$, we again define a set of constants 
$Q_j=(x_{1}-x_{0})\cdot p_{0}^j$ and the length of all the other intervals is determined via
\begin{equation}
(x_{i+1}^j-x_{i}^j)\cdot p_{i}^j=Q_j \quad\; \forall\; i\in\left\lbrace 1..N_j-1\right\rbrace\;. \non
\end{equation}
In order to get $x_i^j$ and $p_i^j$ within double precision, its iteration was done with much more than double precision and took some minutes. But this also has to be done only once, when implementing the routine.  
The length of the last intervals is determined by $1-x_{N_j}$, in consequence, we get:
\begin{equation}
k_j=(N_j-1)\cdot Q_j + (1-x_{N_j})p_{N_j}\; . \non
\end{equation}
Fig. \ref{rho_n_greater0_15} shows this decomposition for the case of $\tau\in I_9$ and the choice $x_1^9=0.08$. 
In consequence, we get $ N=194$ and $k_9=1.02...$, which results in a rejection rate of about $2\%$. 
Following the same argumentation as in the subsection before (same probability mass $1/N$ per interval), it is possible to generate random numbers according to all $\rho_j(x)$ very fast.
Finally we go on as in all the other cases before: By having a closer look at the quotient $\tilde\rho_n/(k_j \rho_j)$ once, we can avoid computing the 
time demanding computation of $\tilde\rho_n$ in more than $90\%$ of all cases by formulating some preconditions for accepting. 
A C++-implementation of the described method on a single CPU-core with 3.4 GHz takes around 65 sec for $10^9$ random $x$.\\
\subsection*{memory requirement}
The last two subsections introduced very fast and exact recipes for sampling random numbers according to the densities $\rho_b(t)$ and $\rho_n(r|t)$. In most cases 
these recipes were based on the precalculation of fast invertable piece wise constant density functions for rejection sampling. The total memory requirement for 
these values is less than 36 kB. 

\end{document}